\documentclass[12pt,amscd,amssymb]{report}
\usepackage[centertags]{amsmath}
\usepackage{amsthm, amsfonts}
\usepackage{graphics,color}
\usepackage{chapsty}
\usepackage{amssymb,letterspace}
\usepackage{epsfig}
\usepackage{amscd, amssymb,latexsym}
\usepackage[active]{srcltx}
\usepackage{setspace}
\usepackage{epsf}

\begin{document}
\begin{center}
\pagenumbering{roman}
\normalfont{\huge{\bfseries{\sc{\abstractname}}}} \vspace{1.75cm}

\normalfont{\large{\bfseries{\sc{LEPTON FLAVOR VIOLATING RADION
DECAYS IN THE RANDALL-SUNDRUM SCENARIO: THE THESIS}}}}
\vspace{.5cm}

Korutlu, Beste\\
M. Sc., Department of Physics\\
Supervisor: Dr. Erhan Onur \.{I}ltan \vspace{.5cm}

December 2007, \pageref{lastpage} pages.
\end{center}
\vspace{.5cm} The lepton flavor violating interactions are
worthwhile to examine since they are sensitive to physics beyond
the Standard Model. The simplest extension of the Standard Model
promoting the lepton flavor violating interactions are the so
called two Higgs doublet model which contains an additional Higgs
doublet carrying the same quantum numbers as the first one. In
this model, the lepton flavor violating interactions are induced
by new scalar Higgs bosons, scalar $h^{0}$ and pseudo scalar
$A^{0}$, and Yukawa couplings, appearing as free parameters, are
determined by using the experimental data. On the other hand, the
possible extra dimensions are interesting in the sense that they
ensure a solution to the hierarchy and cosmological constant
problems and also result in the enhancement in the physical
quantities of various processes. In the present work, we predict
the branching ratios of lepton flavor violating radion decays
$r\rightarrow e^{\pm},\mu^{\pm}$, $r\rightarrow
e^{\pm},\tau^{\pm}$ and $r\rightarrow\mu^{\pm},\tau^{\pm}$ in the
two Higgs doublet model, including a single extra dimension, in
the framework of the Randall Sundrum scenario. We observed that
the branching ratios of the processes we study are at most at the
order of $10^{-8}$ for the small values of radion mass and it
decreases with the increasing values of the radion mass. Among the
LFV decays we study, the $r\rightarrow\mu^{\pm},\tau^{\pm}$ decay
would be the most suitable one to measure its branching ratio.

\vspace{.5cm}

\\ \\
This thesis is based on the manuscript entitled "Lepton flavor
violating radion decays in the Randall-Sundrum scenario" by E.O.
Iltan, B. Korutlu (Middle East Tech. U., Ankara) hep-ph/0610147
(2006).

\newpage
\phantom{Dedication} \vspace{7.5cm}
\begin{center}
to my lovely family..
\end{center}

\newpage
\begin{center}
\normalfont{\huge{\bfseries{\sc{acknowledgments}}}}
\end{center}
\vspace{1.75cm} There are no words to describe such a great honor
that I have the chance to send my sincere gratefulness to my
supervisor Dr. Erhan Onur \.{I}ltan not only for his continuous
support and patience throughout this study, which was a great
comfort, but also for the new dimensions he brought into my life.
Without his inspiring discussions, invaluable guidance and exact
encouragement, this accomplishment would have never been
fulfilled.

\bigskip\noindent I would like to express my appreciation for
the financial support of TUBITAK, The Scientific and Technological
Research Council of Turkey, through my master education.

\bigskip\noindent I would like to thank Assist. Prof. Dr. Se\c{c}il Gerg\"{u}n
and Assist. Prof. Dr. Emre Sermutlu for their help in carrying out
some important Latex works, and \c{C}a\u{g}da\c{s} Acar who gave
me valuable support throughout this study.

\bigskip\noindent I would like to express deep gratitude to my family for their
endless care, understanding, and trust in me.

\bigskip\noindent My special thanks go to two witches, G\"{u}l Esra
B\"{u}lb\"{u}l and Nazan Kara; Esra who has brought beauty, hope,
and happiness into my life, and Nazan with whom I feel myself in a
deep confidence with her staying so close when I needed the most.

\bigskip\noindent I am thankful to our kitty \c{C}{\i}t{\i}r for
 his everlasting energy and fresh spirit.

\bigskip\noindent Last but not least, my sincere appreciation goes to
\.{I}nan\c{c} Kan{\i}k.

\newpage
\addcontentsline{toc}{chapter}{TABLE OF CONTENTS} \tableofcontents
APPENDICES . . . . . . . . . . . . . . . . . . . . . . . . . . . .
. . . . . . . .  {\large{88}}
\newpage
\addcontentsline{toc}{chapter}{LIST OF FIGURES} \listoffigures
\newpage
\addcontentsline{toc}{chapter}{LIST OF TABLES} \listoftables

\newpage
\pagenumbering{arabic}
\chapter{INTRODUCTION}

The major goal of physics has always been the simplification and
the unification of seemingly diverse and complicated natural
phenomena. In the second half of the twentieth century, as a
result of successful approaches, a significant progress has been
made in the particle physics, in the identification of fundamental
particles and the unification of their interactions.
Glashow-Weinberg-Salam \cite{WSalam, WSalam2} combined the quantum
electrodynamics (QED) and the weak interactions into the
electroweak (EW) theory and the Standard Model (SM) of elementary
particles has emerged, which can be considered as a good example
satisfying this major goal of physics. Being a quantum field
theory (QFT) based on the gauge group $SU(3)_{C}\otimes
SU(2)_{L}\otimes U(1)_{Y}$, the SM describes all of the known
elementary constituents of the universe together with the three
out of four fundamental forces: the strong force, the weak force,
and the electromagnetic force. In the QFT, all interactions are
mediated by means of force carrier particles, mediators. In the
case of electromagnetic interaction the mediator is the photon
($\gamma$), one of the four gauge bosons of the group
$SU(2)_{L}\otimes U(1)_{Y}$, and for the weak interactions, there
exist the remaining gauge bosons of the same group, the so called
$W^{\pm}$, $Z^{0}$ bosons. In addition to this, for the strong
interactions, the mediators are eight gluons $(G_{i})$, the gauge
bosons of the group $SU(3)_{C}$. The remaining force, called as
the gravity, is far too weak to be of any consequence at the
experimentally accessible energy scales that are relevant to
particle physics.
In addition to the mediator particles, the SM contains matter
particles: the Higgs boson and the fermions, namely leptons and
quarks which fall into three generations.
The first generation contains all stable stuff of which the stable
matter is composed. The second and third generations of particles
decay, therefore, they are not present in the stable matter and
physicists are still trying to understand their role in the
underlying theory. For each quark and lepton there exists a
corresponding antiquark and antilepton. This is all adding up to
an embarrassingly large number of elementary particles: 12
leptons, 36 quarks, 12 mediators, and, as we will see later,
Glashow-Weinberg-Salam theory calls for at least one Higgs
particle, so we have a minimum of 61 particles to content with. In
the next chapter, we will see how this structure leads to the
first consistent and self-contained theory. The energy range which
defines this theory extends up to several hundreds GeV. For the
details of the model construction see for example textbooks
\cite{GWeinberg, leader}, and the review \cite{abers} existing in
the literature.

The SM has been very successful in explaining many diverse
experimental results in the energy range available at present.
However, behind this energy range it possesses some conceptual
problems which motivate us to look physics beyond. The big issues
in physics beyond the SM can be conveniently grouped into four
categories. The Unification: What is the reason beyond the
hierarchy of fundamental forces? The problem of Flavor: Why are
there so many different types of quarks and leptons? The Mass
problem: What is the origin of masses of fundamental particles and
their mass hierarchies? Does the Higgs boson exist? The
cosmological constant problem. In addition to the conceptual
problems of SM, there also exist phenomenological hints obtained
from measurements of flavor changing neutral currents (FCNCs),
including lepton flavor violating (LFV) interactions which also
indicate the need for physics beyond the SM since the SM
predictions differ from the upper limits coming from current
measurements.

There are various alternative extended models proposed for solving
these problems of the SM such as the multi Higgs doublet model
(MHDM) \cite{branco, haber, gunion, Soni}, the minimal
supersymmetric model (MSSM) \cite{sohnius, barbieri, barbieri2,
barbieri3}, left-right (super) symmetric model \cite{pati}, the
Zee Model \cite{ghosal}, the see-saw model \cite{minkowski},
technicolor model \cite{yue}, extra dimensional models
\cite{akama}-\cite{RS1}; large extra dimensions \cite{antoni,
arkani, arkani1}, universal extra dimensions (UED) \cite{ued,
ued1, ued2}, non-universal extra dimensions (NUED) \cite{nued,
nued1}, split fermion scenarios \cite{split, Schmaltz, chang}, the
Randall-Sundrum model (RS model) \cite{RS, RS1}.

Based on the phenomenological hints, the violation of flavor
symmetry in the leptonic sector is of special interest to
physicists. In the SM, the FCNCs with massless neutrino, are not
allowed in the lepton sector and, in the quark sector, they are
prohibited at tree level, despite they seem not to violate any
fundamental law of nature. The negligibly small branching ratios
(BRs) of the decays based on the FCNCs stimulate one to go beyond
the SM and they are worthwhile to examine since they open a window
to test new models, to ensure considerable information about the
restrictions of the free parameters, with the help of the possible
accurate measurements. An elegant framework to open up the
possibility of the tree level FCNCs is proposed through the
general two Higgs doublet model (2HDM) (see \cite{haber,gunion,
Soni} for details), the most primitive candidate of MHDM, which is
obtained by adding a second Higgs doublet, having the same quantum
numbers as the first one. This doublet may lead to FCNCs in its
Yukawa sector, representing interactions between the Higgs fields
and fermions (see for example \cite{Soni}). In this model, the
lepton flavor (LF) violation is driven by the new scalar.
In addition, the mass hierarchy problem among third generation of
quarks, namely the top and bottom quarks, also could be solved in
the scope of 2HDM, such that, unlike in the SM where both quarks
gain mass through the interaction with the same Higgs doublet,
there is a possibility that the bottom receives its mass from one
doublet (say $\phi_{1}$) and the top from the another one (say
$\phi_{2}$). Then the hierarchy of their Yukawa couplings could be
more natural.

A theory which consists of the SM, combined with gravity, contains
two enormously different energy scales. One is the EW scale
$m_{EW}\sim 10^3$ GeV at which EW symmetry is broken, and the
other is the Planck scale $M_{Pl}\sim 10^{19}$ GeV which
determines the strength of gravitational interactions. Newton's
laws state that the strength is inversely proportional to the
second power of that energy, and because the strength of gravity
is so small, the Planck scale mass (related to the Planck scale
energy by $E=mc^{2}$) should be very large. Generally, when making
predictions in particle physics, we can ignore gravity since the
gravitational effects on particles in the EW energy scale are
completely negligible. But that is precisely a question which
particle physicists try to find an answer: Why is the gravity so
weak? A solution to this problem comes from models with extra
dimensions where gravity becomes strong and cannot be neglected.
In 1998, Nima Arkani-Hamed, Savas Dimopoulos and Gia Dvali
\cite{arkani, arkani1} proposed a model (called as ADD Model) with
$n$ compact extra spatial dimensions of large size to bring the
Planck scale down to TeV scale.
Depending on the details of their implementation, the space in
their model contains two, three or more compact extra dimensions.
For two extra dimensions, the hierarchy problem in the fundamental
scales could be solved and the true scale of quantum gravity would
be no more the Planck scale but of the order of EW scale. This is
the case that the gravity is spreading over all the volume
including the extra dimensions and thus it is diluted by a large
volume of them so much that it will be very feeble in the lower
dimensional effective theory\footnote{Effective theory is a theory
describing those elements and forces that are in principle
observable at the distance or energy scales over which it is
applied.}, although it is very strong in higher dimensions. On the
other hand, the matter fields together with the electromagnetic,
strong and weak forces are restricted in four dimensions, called
four dimensional (4D) brane. Unlike gravitational force, these
forces will not be accessible to the higher dimensions. As
mentioned above, in ADD model, the extra dimensions are compact
and their compactification leads to the appearance of towers of
heavy Kaluza- Klein (KK) modes \cite{Klein} of particles such
that, in 4D effective theory, the existence of the extra
dimensions are felt by the appearance of these KK modes. However,
since the matter fields do not travel along the extra dimensions
but bound to 4D brane, they will not carry extra dimensional
momenta. In other words, none of the SM particles will have the KK
partners. The only particle that will have KK partners is the
graviton, the force carrier particle of the gravitational force.
Since the KK partners of graviton also interact with gravitational
strength (i.e., as weakly as graviton itself), it would be no
easier to produce or detect KK partners of the graviton than to
observe the graviton itself which also has never directly seen by
anyone up to now. This means that, gravity, being the only force
which lives in higher dimensions, the existence of large extra
dimensions will not contradict with the experimental results.
Despite the success of ADD proposal in solving the hierarchy
problem, there exist also some weaknesses of the theory. In fact
their model do not actually solve the hierarchy problem, because
one still have to solve why the size of extra dimensions are so
large.

An alternative approach is introduced by Randall and Sundrum (RS1
model) \cite{RS, RS1} to explain the huge discrepancy between
$m_{EW}$ and $M_{Pl}$ without the need for a large extra
dimension, or for any arbitrary large number at all. In this
scenario, the geometry is a non-factorizable one where the gravity
is localized in a 4D brane, so called Planck brane, which is one
of the boundary of the extra dimension and away from another 4D
brane, TeV brane, which is the other boundary where we
live\footnote{The extra dimension is compactified to $S^{1}/Z_{2}$
orbifold with two 4D brane boundaries which reside at the orbifold
fixed points.}. Theory also includes a finely tuned 5D
cosmological constant $\Lambda$ which serve as sources for 5D
gravity and in 4D, with the help of opposite tensions on
boundaries, it vanishes.

The review topics we include in this thesis is, broadly, divided
into three categories:  In Chapter 2, we give a brief review of
the SM. Chapter 3 is devoted to the simplest extension of the SM,
the so called the 2HDM. In Chapter 4, we give a summary of models
with extra dimensions. In Chapter 5, we investigate the branching
ratios of lepton flavor violating radion decays $r\rightarrow
e^{\pm},\mu^{\pm}$, $r\rightarrow e^{\pm},\tau^{\pm}$ and
$r\rightarrow\mu^{\pm},\tau^{\pm}$ in the 2HDM, in the framework
of the Randall Sundrum scenario (RS1). Chapter 6 represents our
conclusions. In Appendix A, we present the global and local gauge
invariance. Appendix B is devoted to the detailed calculations of
Einstein equations, that we use in our review. Appendix C
represents the calculation of spin connection.

\newpage
\chapter{THE STANDARD MODEL}\label{SM}
In the second half of the twentieth century physicists made an
impressive contribution to the progresses in particle physics with
complementary theoretical and experimental studies, whereupon the
SM of elementary particles has emerged \cite{WSalam, WSalam2}. The
SM, being a QFT (see for example \cite{kaku}), describes all of
the known elementary particles together with the three out of four
fundamental interactions of nature. According to the SM, the
elementary particles constituting the universe are called as
fermions (i.e., they have spin one-half), namely quarks and
leptons and the fundamental interactions are the electromagnetic
force, the weak force (responsible for radioactive decay) and the
strong force (which holds atomic nuclei together). The SM is based
on the principle of gauge symmetry (see Apendix A), which means
that the properties and interactions of elementary particles are
governed by certain symmetries which are related to the
conservation laws. Therefore, the electromagnetic, weak, and
strong forces are all gauge forces and they are mediated by the
exchange of certain particles, called gauge bosons (i.e., they
have spin one) which are the photon, the $W^{\pm}$ and $Z^{0}$
bosons, and eight gluons, respectively. Several attempts have been
made to fit gravity, the remaining force, into this gauge
framework but these attempts are resulted in failure. However, the
gravity is too weak to change particle physics predictions in the
current experimental energy scales.

With these in mind, it is worthwhile summarizing the Fermi theory
\cite{fermi} which describes the weak interaction phenomenology in
the mid-1950.

\section{The Fermi Theory}

The progress in the field theory of the weak interaction was
rather stagnant for many decades, from Fermi's attempt to describe
the $\beta$ decay
\begin{equation}
n\rightarrow p+e^{-}+\bar{\nu}_{e},
\end{equation}
in 1933, to the advent of the gauge theories in the 1970s. Fermi
expressed this decay mathematically as, at a single point in the
space-time, the quantum mechanical wavefunction of a neutron is
transformed into the wavefunction of a proton, and that the
wavefunction of an incoming neutrino is transformed into that of
an electron. He wrote the phenomenological Lagrangian as
\begin{equation}\label{fermi}
\mathcal{L}_{F}=\frac{G_{F}}{\sqrt{2}}J^{\dagger}_{\alpha,had}(x)J^{\alpha}_{lept}(x)+
h.c.,
\end{equation}
where $G_{F}=(1.16639\pm0.00002)\times10^{-5}$ GeV$^{-2}$ is the
Fermi coupling constant,
$J^{\dagger}_{\alpha,had}(x)=\bar{\psi}_{p}(x)\Gamma_{\alpha}\psi_{n}(x)$
and
$J^{\alpha}_{lept}(x)=\bar{\psi}_{e}(x)\Gamma^{\alpha}\psi_{\nu_{e}}(x)$.
This action, from very start, was known to suffer from a series of
problems. First of all, the $\Gamma_{\alpha}$ matrices, that
contain the essence of the weak interaction, consist of all
possible combinations of the $16$ Dirac matrices. It took many
years to narrow down the choice. In $1958$, Feynman and Gell-Mann
\cite{feynman} with the help of further experimental data proposed
that the correct combination of $\Gamma_{\alpha}$ matrices should
only contain a mixture of vector and axial-vector\footnote{an
axial vector (or a pseudovector ) is a quantity that transforms
like a vector under a proper rotation, but gains an additional
sign flip under an improper rotation (a transformation that can be
expressed as an inversion followed by a proper rotation).} (V-A)
quantities written in the form
$\Gamma_{\alpha}=\gamma_{\alpha}(1-\gamma_{5})$ to incorporate the
parity-violating effects of the weak interaction.

\begin{figure}[h!]
\centering
\includegraphics[height=4cm]{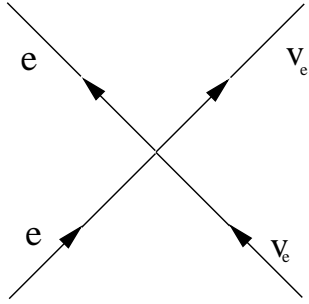}
\caption{The four point $\nu_{e}+e^{-}\rightarrow\nu_{e}+e^{-}$
scattering.\label{fig:fig3}}
\end{figure}

Since the weak force is of extremely short range, Fermi's theory
of point-like interaction yields excellent approximate results at
low energies. However, at high energies the theory suffers from
additional problems. The main problem is the violation of
unitarity. $\nu_{e}+e^{-}\rightarrow\nu_{e}+e^{-}$ scattering is
one of the simplest example of the weak interaction processes. The
Feynman diagram for this scattering in the Fermi's picture of four
point interaction is given in the Fig. \ref{fig:fig3}.
\bigskip
In the center of mass frame (CM), the differential cross section
is found as
\begin{equation}
\frac{d\sigma}{d\Omega}=\frac{G_{F}^{2}k^{2}}{\pi^{2}},
\end{equation}
with CM four momentum $k$, and
\begin{equation}
\sigma=\frac{4G_{F}^{2}k^{2}}{\pi},
\end{equation}
where $k^{2}\gg m_{e}^{2}$. Since the four fermion interaction
takes place at a single point in space-time, the differential
cross section is a pure s-wave. Partial wave unitarity for s-wave
requires that
\begin{equation}
\sigma<\frac{\pi}{2k^{2}}.
\end{equation}
Using the above equation, k is obtained as
\begin{equation}
k^{4}<\frac{\pi^{2}}{8G_{F}^{2}}.
\end{equation}
Then, above a certain energy (i.e., $k>300$ GeV), Fermi theory
violates unitarity. Thus, we can say that it is an effective
theory up to the energies $k<300$ GeV.

\begin{figure}[h!]
\centering
\includegraphics[height=4cm]{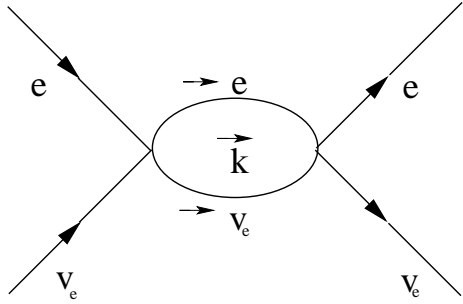}
\caption{One loop correction to the
$\nu_{e}+e^{-}\rightarrow\nu_{e}+e^{-}$ scattering
.\label{fig:fig4}}
\end{figure}

\begin{figure}[h!]
\centering
\includegraphics[height=6cm]{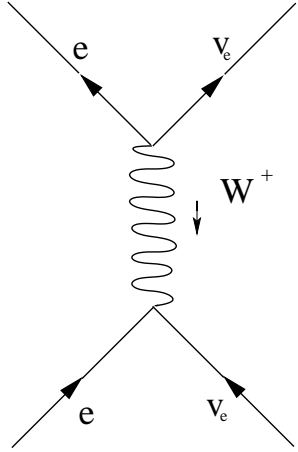}
\caption{The $\nu_{e}+e^{-}\rightarrow\nu_{e}+e^{-}$ scattering
mediated by intermediate massive vector boson
exchange.\label{fig:fig5}}
\end{figure}

Another problem in this theory is the non-renormalizability.
Unfortunately, even in the lowest order approximation in four
fermion interactions, one encounters horrible divergences which
cannot be eliminated by proper renormalization. The Feynman
diagram of the four point interaction including the lowest order
correction for the scattering
$\nu_{e}+e^{-}\rightarrow\nu_{e}+e^{-}$ is shown in the Fig.
\ref{fig:fig4} below which is generated by multiplying the
current-current interaction with itself and the amplitude for this
scattering is proportional to
$\propto\frac{d^{4}k}{k^{2}}=\infty^{2}$.
To eliminate these problems, the idea is that the weak interaction
is mediated by intermediate massive vector boson exchange. The
Feynman diagram for this process is shown in the Fig.
\ref{fig:fig5}.
Therefore, we replace $\mathcal{L}_{F}$ defined in the eq.
\ref{fermi} with
\begin{equation}\label{newlagrangian}
\mathcal{L}_{W}=g_{W}J^{\alpha}(x)W_{\alpha}(x)+h.c.,
\end{equation}
where $W_{\alpha}(x)$ is the weak intermediate vector boson. Now,
in the lowest order diagram the differential cross section of the
$\nu_{e}+e^{-}\rightarrow\nu_{e}+e^{-}$ scattering is found as
\begin{equation}
\frac{d\sigma}{d\Omega}=\frac{2g_{W}^{4}k^{2}}{\pi^{2}(q^{2}-m_{W}^{2})^{^2}},
\end{equation}
for again $k^{2}\geq m_{e}^{2}$. Here, $m_{W}$ and $q$ are the
mass and the momentum transfer vector of the W boson,
respectively. As $q^{2}\rightarrow 0$, the new Lagrangian in the
eq. \ref{newlagrangian} reduces to Fermi Lagrangian given in the
eq. \ref{fermi} provided
\begin{equation}
\frac{g_{W}^{2}}{m_{W}^{2}}=\frac{G_{F}}{\sqrt{2}}.
\end{equation}
However, the interaction is no longer point-like, but mediated by
the force carrier particles. In the scope of the SM, all
fundamental interactions are mediated by the exchange of gauge
bosons. These interactions together with the corresponding gauge
bosons which mediate the forces are listed in the following table
in order of decreasing strength.

\bigskip
\begin{table}[h]
 \caption{The four fundamental forces in nature.}
 \label{forces}
 \begin{center}
 \begin{tabular}{|l|l|l|l|l|}
  \hline
  \textbf{Force}& \textbf{Strength} & \textbf{Range} & \textbf{Theory} & \textbf{Mediator} \\
  \hline
  Strong & $10$ & $<10^{-15}m$ & Chromodynamics & Gluon \\
  \hline
  Electromagnetic & $10^{-2}$ & $\infty$ & Electrodynamics & Photon \\
  \hline
  Weak & $10^{-13}$ & $<10^{-18}m$ & Flavordynamics & $W^{\pm},Z$ \\
  \hline
  Gravitational & $10^{-42}$ & $\infty$ & Geometrodynamics & Graviton \\
  \hline
\end{tabular}
\end{center}
\end{table}
\bigskip

According to the QFT, the short range of the weak force could mean
only one thing: the weak gauge bosons had to have non zero masses.
The mechanism that gives rise to the masses of gauge bosons is
known as the Higgs mechanism \cite{higgs} which relies on the
phenomenon of spontaneous symmetry breaking (SSB) which we will
consider in the following section.

\section{Spontaneously Broken Symmetries}

Symmetry is one of the most important aspects of theoretical
particle physics, since the basis of our current description of
nature originate in symmetries so that every continuous symmetry
leads to a conservation law. A system is said to be symmetric if
it remains invariant after applying a set of transformation rules
that constitute a mathematical group. Symmetries are categorized
into two: spatial symmetry and internal symmetry. In the case of
spatial symmetry physics threats all directions and all positions
as the same, internal symmetries tell us that physical laws act
the same way on distinct, but effectively indistinguishable
objects. The fundamental forces, electromagnetic, weak, and strong
forces all involve internal symmetries. (Gravity is related to the
symmetries of space and time). Exact symmetries are fairly rare in
nature. Thus, the symmetries that the usual 4D theories possess
can be broken explicitly or spontaneously. In the case of SSB of
gauge symmetries, which is one of the crucial ingredients of the
SM, if the broken symmetry is global, the Goldstone theorem
\cite{goldstone} applies, whereas if it is local, then we have
Higgs mechanism \cite{higgs}. In general, the phenomenon of SSB is
simply stated as follows.

\emph{\textquotedblleft A system is said to possess a symmetry
that is spontaneously broken if the ground state of a dynamical
system does not possess the same symmetry properties as the
Lagrangian\textquotedblright}. Here, the ground state -its vacuum
state- is the state in which the field has its lowest possible
energy. Now, we will use a toy model (see \cite{leader} for
details) to explain the Goldstone theorem and the Higgs mechanism.

\subsection{The Goldstone theorem}

\begin{figure}[h!]
\centering
\includegraphics[height=6cm]{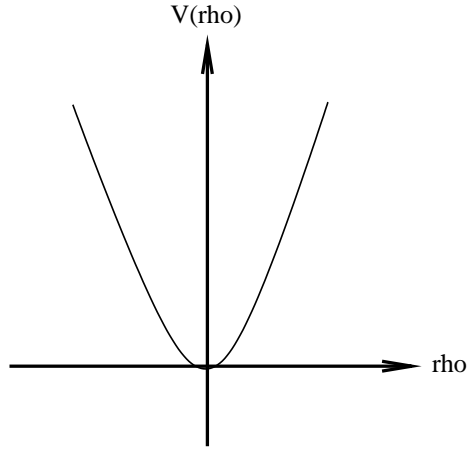}
\caption{The potential function for positive
$\mu^{2}$.\label{fig:fig1}}
\end{figure}

\begin{figure}[h!]
\centering
\includegraphics[height=6cm]{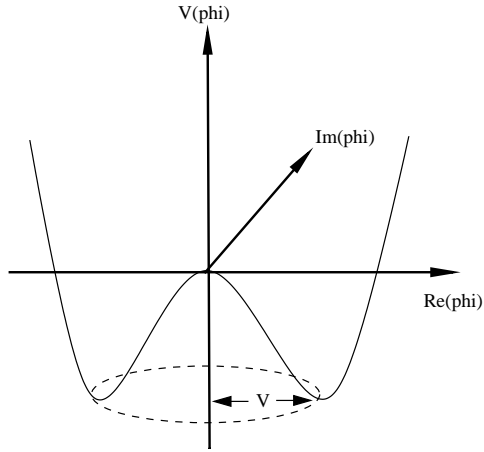}
\caption{The potential function for negative
$\mu^{2}$.\label{fig:fig2}}
\end{figure}

Let us consider the following Lagrangian density which describes a
couple of self interacting complex scalar fields
$\phi(x)=\phi_{1}(x)+i\phi_{2}(x)$ and its complex conjugate
$\phi^{*}(x)=\phi_{1}^{*}(x)-i\phi_{2}^{*}(x)$
\begin{equation}\label{lagrangian}
\mathcal{L}=(\partial_{\mu}\phi)(\partial^{\mu}\phi^{*})
-\mu^{2}\phi\phi^{*}-\lambda(\phi\phi^{*})^{2},
\end{equation}
where $\mu^{2}$ is regarded as the bare mass of the field quanta
and $\lambda$ is the term for self interaction. It is clear that
the Lagrangian remains invariant under the group $U(1)$ of global
gauge transformations (see Appendix A for details). For SSB, we
should check whether the ground state of the system will be
invariant under global gauge transformations or not. For constant
$\phi$ the kinetic term,
$(\partial_{\mu}\phi)(\partial^{\mu}\phi^{*})$ vanishes. Then, the
ground state is obtained when the potential term
$V(\phi,\phi^{*})=\mu^{2}\phi\phi^{*}+\lambda(\phi\phi^{*})^{2}$
corresponds to the minimum. Since the potential term is a function
of $\phi$ and $\phi^{*}$ only in the combination of
$\phi\phi^{*}$, we can make a change of variables so that,
$\rho=\phi\phi^{*}$. Substituting this into $V(\phi,\phi^{*})$ we
get
\begin{equation}
V(\rho)=\mu^{2}\rho+\lambda\rho^{2}.
\end{equation}
The minimum of the potential can be obtained only if $\lambda>0$,
which we take to be so. However, $\mu^2$ can have both positive
and negative values if we do not insist on interpreting $\mu$ as
mass. To find the minimum of the potential, we take the derivative
of $V(\rho)$ with respect to $\rho$ and equate this derivative to
zero such that,
\begin{equation}
\frac{d V(\rho)}{d\rho}=\mu^{2}+2\lambda\rho=0.
\end{equation}
Since
$\rho=\phi\phi^{*}=(\phi_{1}+i\phi_{2})(\phi_{1}^{*}-i\phi_{2}^{*})=(|\phi_{1}|^{2}+|\phi_{2}|^{2})$,
$\rho$ can only take positive values. Therefore, for $\mu^2>0$ a
unique minimum occurs at the origin $\rho=0$, i.e., at $\phi=0$
and the function $V(\rho)$ looks like as in figure \ref{fig:fig1}.
Then, we have a symmetric ground state configuration under the
group $U(1)$ of global gauge transformations for $\mu^2>0$. On the
other hand, for $\mu^2<0$, $\phi=0$ is not a minimum. Instead, the
minimum is at $\rho=-\mu^2/2\lambda$, i.e., at
$|\phi|=\upsilon/\sqrt{2}$ with $\upsilon=\sqrt{-\mu^2/\lambda}$.
Any value of $\phi$ satisfying this relation will give us a true
ground state such that
\begin{equation}
\phi_{vac}=\frac{\upsilon}{\sqrt{2}}e^{i\Lambda},
\end{equation}
where $\Lambda$ is real. Then, we have a continuum degenerate set
of ground states for negative values of $\mu^{2}$. In this case,
the function $V(\phi)$ looks like as in figure \ref{fig:fig2}.
Each will not be symmetric under the global gauge transformation
defined in the eq. \ref{global}. Then, using the definition of SSB
made above one can simply conclude that the symmetry of the
Lagrangian has been spontaneously broken. We are free to choose
any point on the ring of minima since they are equivalent. Let us
choose this point to be on the real axis such that
\begin{equation}
\phi(x)=\frac{1}{\sqrt{2}}[\upsilon+\xi(x)+i\chi(x)],
\end{equation}
where $\xi(x)$, $\chi(x)$ are real fields and $\xi(x)=\chi(x)=0$
in the ground state. Substituting into the eq. \ref{lagrangian}
and ignoring constant terms we get
\begin{equation}
\mathcal{L}=\frac{1}{2}(\partial_{\mu}\xi)^{2}+\frac{1}{2}(\partial^{\mu}\chi)^{2}
-\lambda\upsilon^{2}\xi^{2}-\lambda\upsilon\xi(\xi^{2}+\chi^{2})-\frac{1}{4}\lambda(\xi^{2}+\chi^{2})^{2}.
\end{equation}
Then, we end up with a massless $\chi(x)$ field and a field
$\xi(x)$ with a spontaneously generated mass
\begin{equation}
m_{\xi}(x)=2\lambda\upsilon^{2}.
\end{equation}
Let us now examine a less trivial example \cite{abers} given by
\begin{equation}
\mathcal{L}=\frac{1}{2}\partial_\mu\phi^{i}\partial^\mu\phi^{i}-\frac{1}{2}\mu^2\phi^{i}\phi^{i}-
\frac{1}{4}\lambda(\phi^{i}\phi^{i})^{2},
\end{equation}
where $\phi$ is an n-component real scalar field and $\mathcal{L}$
is invariant under the orthogonal group in n dimensions, $O(n)$.
Again for $\mu^2<0$ we find a whole ring of minima whenever
$\sum_{i}\phi^{i}\phi^{i}=-\mu^2/\lambda$ is satisfied. In this
case, we are free to choose one of the $\phi^{i}$ to be non-zero
in the ground state. Let it be the $n^{th}$ component of $\phi$
such that,
\begin{equation}
\phi_{vac}= \left(
\begin{array}{c}
0\\
0\\
.\\
.\\
.\\
0\\
\upsilon
\end{array}
\right).
\end{equation}
The number of generators that original symmetry group $O(n)$
possesses is $\frac{1}{2}n(n-1)$. There is also a non-trivial
subgroup $O(n-1)$, which has $\frac{1}{2}(n-1)(n-2)$ number of
generators leave the vacuum invariant. Let $L_{ij}$ be the
$\frac{1}{2}n(n-1)$ independent matrices that generates $O(n)$ and
$l_{ij} [l_{ij}=L_{ij}$ for $i,j\neq n]$ be the
$\frac{1}{2}(n-1)(n-2)$ matrices generating $O(n-1)$. There
remains $n-1$ independent matrices which are denoted by
$k_{i}[k_{i}=L_{in}]$ with $1\leq i\leq n-1$. Defining $\eta$ and
$\xi_{i}$ with again $1\leq i\leq n-1$ we get
\begin{equation}
\phi=e^{(i\xi_{i}k_{i}/\upsilon)} \left(
\begin{array}{c}
0\\
0\\
.\\
.\\
.\\
.\\
\upsilon+\eta
\end{array}
\right).
\end{equation}
Note that in general,
\begin{equation}
(L_{ij})_{kl}=-i[\delta_{ik}\delta_{jl}-\delta_{il}\delta_{jk}].
\end{equation}
For j=n we have
\begin{equation}
(L_{in})_{kl}=(k_{i})_{kl}=-i[\delta_{ik}\delta_{nl}-\delta_{il}\delta_{nk}].
\end{equation}
Operating $k_{i}$ on the column vector
$\upsilon_{i}=\upsilon\delta_{in}$ we get
\begin{eqnarray}
(k_{i}\upsilon)_{j}&=& (k_{i})_{jl}\upsilon_{l}\nonumber
\\&=&
-i[\delta_{ij}\delta_{nl}-\delta_{il}\delta_{nj}]\upsilon_{l}\nonumber
\\&=&
-i[\delta_{ij}\upsilon_{n}-\delta_{nj}\upsilon_{i}]\nonumber
\\&=&
-i[\delta_{ij}\upsilon\delta_{nn}-\delta_{nj}\upsilon\delta_{in}]\nonumber
\\&=&
-i\upsilon\delta_{ij}.
\end{eqnarray}
Thus, in the lowest order $\phi_{i}= \xi_{i}(i<n)$ and $\phi_{n}=
\upsilon+\eta$. Then, the Lagrangian density in terms of $\xi_{i}$
and $\eta$ can be presented as
\begin{equation}
\mathcal{L}=\frac{1}{2}[\partial^{\mu}\eta\partial_{\mu}\eta)+\partial^{\mu}\xi_{i}\partial_{\mu}\xi_{i}]-
\frac{1}{2}\mu^{2}(\upsilon+\eta)^{2}-\lambda(\upsilon+\eta)^{4}+...
\end{equation}
Looking at this Lagrangian, we can say that the field $\eta$ has a
positive mass of $-2\mu^{2}$ and the $(n-1)$  $\xi_{i}$ scalar
fields remain massless. These massless bosons are called as
Goldstone bosons. In conclusion, for every broken generator that
leaves the vacuum invariant there exists a massless Goldstone
boson.

\section{The Higgs Mechanism}

Now, we will use the same Lagrangian in the eq. \ref{lagrangian}
but impose invariance under $U(1)$ of local gauge transformations
(see Appendix A for details). To make the Lagrangian invariant
under this transformation we must replace the partial derivative
$\partial_{\mu}$ by the covariant derivative
$D_{\mu}=\partial_{\mu}-ieA_{\mu}$ and add a kinetic term
$-\frac{1}{4}F_{\mu\nu}F^{\mu\nu}$.ç Then, the Lagrangian density
becomes
\begin{equation}\label{lagrangiann2}
\mathcal{L}=-\frac{1}{4}F_{\mu\nu}F^{\mu\nu}+[(\partial^{\mu}
+ieA^{\mu})\phi^{*}(\partial_{\mu}-ieA_{\mu})\phi]
-\mu^{2}\phi^{*}\phi-\lambda(\phi^{*}\phi)^{2},
\end{equation}
where $A_{\mu}$ is the massless gauge field. Under local gauge
transformations we have,
\begin{equation}
\begin{split}
&\phi(x)\rightarrow\phi'(x)=\exp^{-i\theta(x)}\phi(x),\\
&\phi^{*}(x)\rightarrow\phi^{*'}(x)=\exp^{i\theta(x)}\phi(x),\\
&A_{\mu}\rightarrow
A'_{\mu}(x)=A_{\mu}(x)-\frac{1}{e}\partial_{\mu}\theta(x).
\end{split}
\end{equation}
Again we will look at the minimum in the potential. For
$\lambda\geq0$ and $\mu^{2}>0$ we obtain a symmetric ground state
at $\phi= 0$. However, when $\mu^{2}< 0$ there exists again a ring
of degenerate ground states. Proceeding as before we set
\begin{equation}
\phi(x)=\frac{1}{\sqrt{2}}[\upsilon+\xi(x)+i\chi(x)],
\end{equation}
with $\upsilon= \sqrt{-\mu^{2}/\lambda}$ so that
$\phi_{vac}=\upsilon/\sqrt{2}$. Substituting this into
\ref{lagrangiann2}, we obtain the following Lagrangian density
\begin{equation}\label{lagrangiann3}
\begin{split}
\mathcal{L}=&-\frac{1}{4}F_{\mu\nu}F^{\mu\nu}
+\frac{e^{2}\upsilon^{2}}{2}A_{\mu}A^{\mu}+\frac{1}{2}(\partial^{\mu}\xi)^{2}
+\frac{1}{2}(\partial^{\mu}\chi)^{2}\\
&-\frac{1}{2}(2\lambda\upsilon^{2})\xi^{2}-e\upsilon
A_{\mu}\partial^{\mu}\chi+...
\end{split}
\end{equation}
It is surprising that the gauge field $A_{\mu}$ seems to acquire
mass in the quantum picture. The Lagrangian density in the eq.
\ref{lagrangiann3} now seems to describe the interaction of a
massive gauge field $A_{\mu}$ and two scalar fields. To ensure the
gauge invariance, the gauge transformations in terms of $\xi(x)$
and $\chi(x)$ should be in the following form
\begin{equation}
\begin{split}
&\xi(x)\rightarrow\xi'(x)=[\upsilon+\xi(x)]\cos\theta(x)+ \chi(x)\sin\theta(x)-\upsilon,\\
&\chi(x)\rightarrow\chi'(x)=\chi(x)\cos\theta(x)-[\upsilon+\xi(x)]\sin\theta(x),\\
&A_{\mu}\rightarrow\
A'_{\mu}(x)=A_{\mu}(x)-\frac{1}{e}\partial_{\mu}\theta(x).
\end{split}
\end{equation}
We are free to choose $\theta(x)$ to be the phase of $\phi(x)$
since the theory is invariant under any choice of transformation
of this function. Then,
\begin{equation}
\phi'(x)=\exp^{-i\theta(x)}\phi(x)=\frac{1}{\sqrt{2}}[\upsilon+\eta(x)],
\end{equation}
will be real, with $\eta(x)$ is real. Substituting these into the
eq. \ref{lagrangiann2} the Lagrangian density becomes
\begin{equation}
\begin{split}
\mathcal{L}=&-\frac{1}{4}F'_{\mu\nu}F'^{\mu\nu}+
\frac{1}{2}\partial^{\mu}\eta\partial_{\mu}\eta+\frac{1}{2}e^2\upsilon^{2}A'_{\mu}A'^{\mu}\\
&+\frac{1}{2}e^{2}(A'_{\mu})^{2}(2\upsilon\eta+\eta^{2})-\lambda\upsilon^{2}\eta^{2}-
\frac{1}{4}\lambda\eta^{4}...
\end{split}
\end{equation}
with $F'_{\mu\nu}F'^{\mu\nu}=\partial_{\mu}A'_{\nu}-
\partial_{\nu}A'_{\mu}$. By writing the Lagrangian density in this
form, we can say that it describes the interaction of the massive
vector field $A'_{\mu}$ with the massive, real, scalar field
$\eta$. This field is called as Higgs field with a mass of
$2\lambda\upsilon^{2}=-2\mu^{2}$. In this way, all massless
particles completely  disappears. Consequently, in spontaneously
broken symmetries the gauge boson acquires mass due to
disappearing Goldstone boson. Therefore, for each massive gauge
fields we need a complex scalar field, one piece of which
disappears and reappears as the longitudinal mode of the vector
field. Scalar part of this complex field, the so called Higgs
boson, remains.
\section{The Standard Model Lagrangian}
Having discussed the ingredients of the SM, let us turn our
attention to the SM Lagrangian, $\mathcal{L}_{SM}$. As mentioned
above the SM is based on the principle of gauge symmetry. The
overall gauge group of the SM, under which the SM Lagrangian
remains invariant, contains both the Quantum Chromodynamics (QCD)
and the unified EW interaction and is written symbolically as
\begin{equation}
G_{SM}\equiv SU(3)_{C}\otimes SU(2)_{L}\otimes U(1)_{Y}.
\end{equation}
The first group, $SU(3)_{C}$, represents QCD. The subscript $C$
indicates that the gauge bosons of QCD, the eight gluons, couple
only to colored particles, quarks. The remaining part
$SU(2)_{L}\times U(1)_{Y}$ represents the EW interaction, proposed
by Glashow-Weinberg-Salam \cite{WSalam, WSalam2}, with the
subscripts $L$ and $Y$ indicating that the group $SU(2)_{L}$
couples only left handed particles and that the group $U(1)_Y$
couples to weak hypercharged particles where the hypercharge is
obtained using the Gell-Mann-Nishijima relation \cite{gell}
$Q=T_{3}+Y/2$. The EW theory, developed by Glashow-Weinberg-Salam,
predicted the masses of the gauge bosons $W^{\pm}$ and $Z^{0}$ to
be about $80$ GeV and $90$ GeV, respectively. In 1983, physicists
at CERN \cite{cern} led by Carlo Rubia were able to produce and
measure the masses of the $W^{\pm}$ and $Z^{0}$ which were in
complete agreement with the predictions of EW theory. The
discovery of these particles may be considered as the first
experimental evidence of the SSB. In the minimum formulation of
the SM, a complex scalar doublet (the Higgs field) is required,
which is denoted by $\phi$, so that by interacting with the gauge
bosons, it produces the desired breaking
\begin{equation}
G_{SM}\rightarrow SU(3)_{C}\otimes U(1)_{Q},
\end{equation}
where $U(1)_{Q}$ is a subgroup of $SU(2)_{L}\otimes U(1)_{Y}$.
This breaking of symmetry occurs due to the non-zero vacuum
expectation values (VEV) of the scalar field $\phi$ of the form
\begin{equation}
\langle \phi \rangle =\left(
\begin{array}{c}
0 \\
v/\sqrt{2}
\end{array}
\right).
\end{equation}
The reason for why this type of breaking occurs is as follows. The
$W^{\pm}$ and $Z^{0}$ gauge bosons are massive. Therefore,
$SU(2)_{L}\otimes U(1)_{Y}$ can not be a symmetry of the vacuum,
whereas the photon, being massless, reflects that $U(1)_{Q}$ is a
good symmetry of the vacuum.

Now, let us write the most general renormalizable EW SM
Lagrangian. It can be divided into five parts:
\begin{equation}
\mathcal{L}_{SM}=\mathcal{L}_{kinetic}^{f}
+\mathcal{L}_{kinetic}^{H}+\mathcal{L}_{kinetic}^{G}
+\mathcal{L}^{H}_{pot}+\mathcal{L}^{Y}.
\end{equation}

The $\mathcal{L}_{kinetic}^{f}$ term corresponds to the fermionic
sector of the SM Lagrangian. It includes both the left-handed and
right-handed chiralities and can be presented as
\begin{equation}
\mathcal{L}_{kinetic}^{f}=\sum_{\psi_{L}}\bar{\psi}_{L}i\gamma^{\mu}D_{\mu}{\psi_{L}}
+\sum_{\psi_{R}}\bar{\psi}_{R}i\gamma^{\mu}D_{\mu}{\psi_{R}},
\end{equation}
where $\psi_{L}=\frac{1}{2}(1-\gamma_{5})\psi$ stands for the
left-handed  weak isodoublets  and
$\psi_{R}=\frac{1}{2}(1+\gamma_{5})\psi$ for the right-handed weak
isosinglets. The normal derivative, $\partial_{\mu}$, is replaced
by the covariant derivative, $D_{\mu}$:
\begin{equation}\label{patialmu}
D_{\mu}=\partial_{\mu}+igW_{\mu}^{i}\tau_{i}+
i\frac{g'}{2}B_{\mu}Y,
\end{equation}
%
%
%
to preserve the gauge invariance. Here,
$g$ and $g'$ are the coupling constants associated with the groups
$SU(2)_{L}$ and $U(1)_{Y}$ gauge groups, respectively. The
corresponding generators to each gauge group are
$\tau_{i}$ and $Y$ in order. Moreover,
$W_{\mu}^{i}$ are the three weak interaction bosons and $B_{\mu}$
is the single hypercharge boson. Here, the gauge boson fields
$W_{\mu}^{1}$, $W_{\mu}^{2}$, $W_{\mu}^{3}$ couple to weak isospin
and $B_{\mu}$ couple to weak hypercharge.

The second part of the SM Lagrangian is the kinetic term for the
scalar Higgs field, $\phi$ and is responsible for the interaction
of the gauge and Higgs fields. It can be written as
\begin{equation}\label{higgs}
\mathcal{L}_{kinetic}^{H}=(D_{\mu}\phi)^{\dagger}(D_{\mu}\phi),
\end{equation}
with $D_{\mu}$ is defined in the eq. \ref{patialmu}.

The corresponding kinetic term for the gauge fields reads
\begin{equation}
\mathcal{L}_{kinetic}^{G}=-\frac{1}{4}\sum_{i=1}
^{3}F^{\mu\nu}_{i}F_{\mu\nu}^{i}-\frac{1}{4}B^{\mu\nu}B_{\mu\nu},
\end{equation}
where
$F_{\mu\nu}^{i}=\partial_{\mu}W_{\nu}^{i}-\partial_{\nu}W_{\mu}^{i}-g\epsilon^{ijk}W_{\mu}^{j}W_{\nu}^{k}$
is the antisymmetric field strength tensor of the group
$SU(2)_{L}$ with $\epsilon^{ijk}$ being the group structure
constant and
$B_{\mu\nu}=\partial_{\mu}B_{\nu}-\partial_{\nu}B_{\mu}$ is that
of the group $U(1)_{Y}$. After a proper normalization of the gauge
fields, the photon, the neutral weak boson, $Z^{0}$ and the
charged weak boson $W^{\pm}_{\mu}$ fields are obtained as
\begin{equation}
\begin{split}
A_{\mu}&=\sin\theta_{W}W_{\mu}^{3}+\cos\theta_{W}B_{\mu},\\
Z_{\mu}&=\cos\theta_{W}W_{\mu}^{3}-\sin\theta_{W}B_{\mu},\\
W^{\pm}_{\mu}&=\frac{1}{\sqrt{2}}(W_{\mu}^{1}\mp iW^{2}_{\mu}),
\end{split}
\end{equation}
where
\begin{equation}
\sin\theta_{W}=\frac{g'}{\sqrt{g^{2}+g'^{2}}}\qquad;\qquad\cos\theta_{W}=\frac{g}{\sqrt{g^{2}+g'^{2}}},
\end{equation}
with $\theta_{W}$ being the weak mixing angle. Finally, the photon
becomes massless and the mass eigenstates for $W^{\pm}$ and
$Z^{0}$ bosons are obtained as
\begin{equation}
M_{W^{\pm}}=\frac{g\upsilon}{2}\qquad;\qquad
M_{Z^{0}}=\frac{\sqrt{g^{2}+g'^{2}}}{2}.
\end{equation}

The Higgs potential denoted by $\mathcal{L}_{H}^{SM}$ in the SM
Lagrangian reads
\begin{equation}
\mathcal{L}_{pot}^{H}=\mu^{2}(\phi^{\dagger}\phi)+\lambda(\phi^{\dagger}\phi)^{2},
\end{equation}
where $\mu^{2}$ and $\lambda$ are the free parameters. For
$\mu^2<0$ the scalar field $\phi$ develops a non-zero vacuum
expectation value at $|\phi|_{vac}=\upsilon/\sqrt{2}$ with
$\upsilon=\sqrt{-\mu^2/\lambda}$. Thus, the Lagrangian gains a set
of ground states for negative values of $\mu^{2}$. As a result,
symmetry of the Lagrangian is spontaneously broken. In addition,
through the Higgs mechanism, the Higgs mass is yielded to be equal
to $m_{H}=\sqrt{2\lambda\upsilon}$. Notice that, all the terms
explained up to now in the SM Lagrangian are CP invariant.

The final piece of the SM Lagrangian, $\mathcal{L}_{Y}^{SM}$
called the Yukawa Lagrangian describes the interaction among the
fermions and the Higgs field. The general form can be expressed
as\footnote{In the case of massive neutrinos, there is an
additional term in the Lagrangian: $
\mathcal{L'}_{Y}^{SM}=\eta_{ij}^{\nu}{\bar{l}}_{i,L}\tilde{\phi}\nu_{j,R}
+ h.c.$}
\begin{equation}
\mathcal{L}^{Y}=\eta_{ij}^{D}{\bar{Q}}_{i,L}\phi
D_{j,R}+\eta_{ij}^{U}{\bar{Q}}_{i,L}\tilde{\phi}U_{j,R}
+\eta_{ij}^{E}{\bar{l}}_{i,L}\phi E_{j,R}+ h.c.,
\end{equation}
where $\tilde{\phi}=i\tau_{2}\phi^{*}$ and $\eta_{ij}^{U,D,E}$'s
are responsible for the masses of up-down quarks and leptons,
respectively. In addition, $Q_{i,L}$, $U_{j,R}$ and $D_{j,R}$
denote the left handed doublet, right handed up and right handed
down quarks, respectively. Similarly, $l_{i,L}$ represent the left
handed leptons and $E_{i,R}$ the right handed ones. These fermions
are presented in a more elegant way in the following table:

\bigskip
\begin{table}[h]
 \caption{The known fermions.}
 \label{fermions}
\begin{center}
\begin{tabular}{|l|l|l|l|l|}
  \hline
  {\textbf{Generation}}& \multicolumn{2}{|c|}{\textbf{Quarks}} & \multicolumn{2}{|c|}{\textbf{Leptons}}\\
  \hline
  & Charge $2/3$ & Charge $-1/3$ & Charge $-1$ & Charge $0$\\
  \hline
  & Color (R G B) & Color (R G B) & Colorless & Colorless\\
  \hline
  \textbf{First} & u (up) & d (down) & e (electron) & $\nu_{e}$ (electron neutrino)\\
  \textbf{Mass(GeV)} & $0.0015-0.003$ & $0.003-0.007$ & $0.000511$ & $<3\times10^{-9}$\\
  \hline
  \textbf{Second} & c (charmed) & s (strange) & $\mu$ (muon) & $\nu_{\mu}$ (muon neutrino)\\
  \textbf{Mass(GeV)} & $1.25\pm0.09$ & $0.095\pm0.025$ & $0.106$ & $<190\times10^{-6}$\\
  \hline
  \textbf{Third} & t (top) & b (beauty) & $\tau$ (tau) & $\nu_{\tau}$ (tau neutrino)\\
  \textbf{Mass(GeV)} & $174.2\pm3.3$ & $4.2\pm0.07$ & $1.777$ & $<18.2\times10^{-3}$\\
  \hline
\end{tabular}
\end{center}
\end{table}
\bigskip
In this table, quark masses given correspond to the approximate
rest mass energy of quarks confined in hadrons since free quarks
have not been observed yet. For each quark and lepton given in the
table there is a corresponding antiquark and antilepton.


As mentioned before, in the SM, the fundamental fermionic
constituents of the matter are quarks and leptons. All properties
of these particles are summarized in the Table \ref{fermions}.
These particles are placed into SM as left-handed doublets and
right-handed singlets. The left handed doublets for quarks are
given by
\begin{eqnarray}
& & \left( \begin{array}{c}
u \\
d \end{array}\right)_{L}~~~~;~~~~~ \left( \begin{array}{c}
c \\
s \end{array}\right)_{L}~~~~;~~~~~ \left( \begin{array}{c}
t \\
b \end{array}\right)_{L},
\end{eqnarray}
and the right handed singlets for quarks are
\begin{eqnarray}
d_{R}~~~~;~~~~~u_{R}~~~~;~~~~~
s_{R}~~~~;~~~~~c_{R}~~~~;~~~~~b_{R}~~~~;~~~~~t_{R}.
\end{eqnarray}
On the other hand, the lepton doublets are
\begin{eqnarray}
& & \left( \begin{array}{c} \nu_{e} \\
e  \end{array}\right)_{L}~~~~;~~~~~
\left( \begin{array}{c} \nu_{\mu} \\
\mu  \end{array}\right)_{L}~~~~;~~~~~
\left( \begin{array}{c}   \nu_{\tau} \\
\tau \end{array}\right)_{L},
\end{eqnarray}
and the right handed singlets for leptons are
\begin{eqnarray}
e_{R}~~~~;~~~~~ \mu_{R}~~~~;~~~~~ \tau_{R}.
\end{eqnarray}
In the charged weak interactions of leptons, the coupling of
$W^{\pm}$ takes place strictly within a particular generation in
the case of massless neutrinos . In other words, upper members of
left handed lepton doublets couple to the lower members in the
same doublet. That is, only the vertices $e^{-}\nu_{e}W^{-}$,
$\mu^{-}\nu_{\mu}W^{-}$, and $\tau^{-} \nu_{\tau}W^{-}$ appear ,
however, there is no cross generational vertices such as $e^{-}
\nu_{\mu} W^{-}$. The coupling of $W^{\pm}$ to quarks is not quite
so simple since there exist cross generational vertices as well,
such as $\bar{s} u W^{-}$. The idea is that, the quark generations
are rotated for the purposes of weak interactions such that
\begin{eqnarray}
& & \left( \begin{array}{c} u \\
d' \end{array}\right)_{L}~~~~;~~~~~
\left( \begin{array}{c} c \\
s' \end{array}\right)_{L}~~~~;~~~~~
\left( \begin{array}{c} t \\
b' \end{array}\right)_{L},
\end{eqnarray}
where $d'$, $s'$, and $b'$,the linear combinations of the  $d$,
$s$, and $b$, are obtained by using the Cabibbo-Kobayashi-Maskawa
(CKM) mixing matrix $V_{ij}$
\begin{equation}
\left(
\begin{array}{c}
d' \\
s' \\
b'
\end{array}
\right)=\left(
\begin{array}{ccc}
V_{ud} & V_{us} & V_{ub} \\
V_{cd} & V_{cs} & V_{cb} \\
V_{td} & V_{ts} & V_{tb}
\end{array}
\right)\; \left(
\begin{array}{c}
d \\
s \\
b
\end{array}
\right), \label{matrix kin1}
\end{equation}
where the off-diagonal elements of the CKM matrix allow flavor
transitions between different generations. The experimentally
measured values of the matrix elements are \cite{ckm}
\begin{equation}
V_{ij}=\left(
\begin{array}{ccc}
0.97377\pm 0.00027 & 0.2257\pm 0.0021 & 0.00431\pm 0.0003 \\
0.230\pm 0.011 & 0.957\pm0.017\pm0.093 & 0.0416\pm 0.0006 \\
0.0074\pm0.0008 & 0.0406\pm 0.0027 & >0.78
\end{array}
\right). \label{CKM}
\end{equation}

\newpage
\chapter{BEYOND THE STANDARD MODEL}
The SM has been extremely successful in describing the behavior of
all known particles in elementary particle physics up to the EW
energy scale at the order of $10^{3}$ GeV. However, behind this
energy scale, it possesses some conceptual problems which
motivates us to look physics beyond the SM. There are various
alternative extended models proposed for solving these problems as
indicated in the introduction part. MHDM \cite{branco} is one of
them. In this chapter, we will introduce the simplest extension of
the SM, the so called the 2HDM \cite{haber,gunion, Soni}.
\section{The Two Higgs Doublet Model}

Let us first present the motivation for examining the 2HDM;
\begin{itemize}
    \item In the SM, it is assumed that the Higgs sector
    must be minimal having only one physical neutral Higgs scalar.
    However, there is no fundamental reason favoring this
    minimal choice. The 2HDM, being the simple extension of the
    SM, possesses five physical Higgs bosons, namely, a charged pair ($H^{\pm}$),
    two neutral CP even scalars
    ($H^{0}$ and $h^{0}$),
    and a neutral CP odd scalar ($A^{0}$).
    \item
    The ratio between the masses of top and bottom quarks is
    $m_{t}/m_{b}\approx174/5\approx35$. According to the SM,
    both quarks gain mass through interactions with the same
    Higgs doublet. Then, we end up with an unnatural hierarchy between the
    corresponding Yukawa couplings. In the scope of the 2HDM, there
    is a possibility that the bottom receives its mass from one doublet
     (say $\phi_{1}$) and the top from the another one (say $\phi_{2}$).
     Then the hierarchy of their Yukawa couplings would be more natural.
    \item In the framework of the SM the flavor is conserved in
    the lepton sector for massless neutrinos. The LFV interactions,
    carried by the FCNCs, exist
    in the extended SM, the so called $\nu$SM, at least at
    one loop level, which is constructed by taking the neutrinos massive and permitting
    the lepton mixing mechanism \cite{Pontecorvo, illana}. However, even in the $\nu$SM,
    due to the smallness of the neutrino masses,
    the theoretical predictions of the BRs of
    the LFV interactions are too small to reach the experimental
    limits. In addition, in the quark sector
    the FCNCs are prohibited at tree level despite they seem
     not to violate any fundamental law of nature. In that aspect, the 2HDM is an
      elegant framework  to open up the possibility
     of the tree level FCNCs in both lepton and quark sectors which
     is driven by the new scalar Higgs bosons $S$, the CP even scalar
      $h^{0}$, and the CP odd scalar $A^{0}$, and controlled by the Yukawa couplings.
      \item The 2HDM is a minimal extension in that it adds the fewest new arbitrary
      parameters. Instead of one free parameter of the SM,
    this model has six free parameters: the four Higgs masses,
    the ratio of the VEVs, $\tan\beta$,
    and a Higgs mixing angle, $\alpha$. Notice that $v_{1}^{2}+v_{2}^{2}$
    is fixed by the W mass
    $m_W=g^{2}\,\frac{(v_{1}^{2}+v_{2}^{2})}{2}$.
\end{itemize}
Having stated our motivations for an additional scalar doublet,
our next task is to introduce the 2HDM. In the 2HDM, a second
Higgs doublet having the same quantum numbers as the first one is
introduced such that,
\begin{equation}
\Phi _{1}=\left(
\begin{array}{c}
\phi _{1}^{+} \\
\phi _{1}^{0}
\end{array}
\right)\qquad ;\qquad\Phi _{2}=\left(
\begin{array}{c}
\phi _{2}^{+} \\
\phi _{2}^{0}
\end{array}
\right),
\end{equation}
with hypercharges $Y=1$. Parameterizing the doublets in a more
convenient way we can write them in the following form
\begin{equation}
\Phi _{1}=\left(
\begin{array}{c}
\chi^{+} \\
\frac{\upsilon_{1} + H^{0} + i\chi ^{0}}{\sqrt{2}}
\end{array}
\right)\qquad ;\qquad\Phi _{2}=\left(
\begin{array}{c}
H^{+} \\
\frac{\upsilon_{2} + H^{1}+ i H^{2}}{\sqrt{2}}
\end{array}
\right),
\end{equation}
with the VEVs
\begin{equation}
\langle \Phi _{1}\rangle=\frac{1}{\sqrt{2}}\left(
\begin{array}{c}
0\\
v_{1}
\end{array}
\right)\qquad ;\qquad\langle \Phi
_{2}\rangle=\frac{1}{\sqrt{2}}\left(
\begin{array}{c}
0 \\
v_{2}
\end{array}
\right),
\end{equation}
where
$\upsilon=(\upsilon_{1}^{2}+\upsilon_{2}^{2})^{1/2}=(\sqrt{2}G_{F})^{-1/2}=246$
GeV. Here, $H^{0}$ and $H^{1}$ are the CP even, $H^{2}$ is the CP
odd neutral Higgs bosons, and $H^{+}$ is the charged Higgs boson.

In the 2HDM, the Higgs part of the SM Lagrangian should be
extended to include the interaction with the second Higgs doublet.
Then, the kinetic term in eq. \ref{higgs} becomes
\begin{equation}
(D_{\mu}\Phi_{1})^{\dagger}(D^{\mu}\Phi_{1})+(D_{\mu}\Phi_{2})^{\dagger}(D^{\mu}\Phi_{2}),
\end{equation}
and the most general renormalizable CP invariant Higgs potential
potential is written in the form
\begin{equation}
\begin{split}
 V(\Phi_{1}, \Phi_{2})&=\lambda_{1}(\Phi_{1}^{\dagger}\Phi_{1}-\upsilon_{1}^{2})^{2}+\lambda_{2}(\Phi_{2}^{\dagger}\Phi_{2}-\upsilon_{2}^{2})^{2}\\
 &+\lambda_{3}[(\Phi_{1}^{\dagger}\Phi_{1}-\upsilon_{1}^{2})+
 (\Phi_{2}^{\dagger}\Phi_{2}-\upsilon_{2}^{2})]^{2}\\
 &+\lambda_{4}[(\Phi_{1}^{\dagger}\Phi_{1})(\Phi_{2}^{\dagger}\Phi_{2})-(\Phi_{1}^{\dagger}\Phi_{2})(\Phi_{2}^{\dagger}\Phi_{1})]\\
 &+\lambda_{5}[Re(\Phi_{1}^{\dagger}\Phi_{2})-\upsilon_{1}\upsilon_{2}]^{2}\\
 &+\lambda_{6}[Im(\Phi_{1}^{\dagger}\Phi_{2})]^{2},
\end{split}
\end{equation}
where the parameters $\lambda_{i}$ are real.

Then, what remains is the Yukawa piece of the SM Lagrangian in the
presence of the two scalar doublets which is written as follows:
\begin{equation}\label{yukawa}
\begin{split}
{\mathcal{L}}^{Y}_{2HDM}&=\eta_{ij}^{U}{\bar{Q}}_{i,L}\tilde{\Phi}_{1}U_{j,R}+
\eta_{ij}^{D}{\bar{Q}}_{i,L}\Phi_{1}D_{j,R}+
\xi_{ij}^{U}{\bar{Q}}_{i,L}\tilde{\Phi}_{2}U_{j,R}+
\xi_{ij}^{D}{\bar{Q}}_{i,L}\Phi_{2}D_{j,R}\\
&+\eta_{ij}^{E}{\bar{l}}_{i,L}\Phi_{1}E_{j,R}+
\xi_{ij}^{E}{\bar{l}}_{i,L}\Phi_{2}E_{j,R}+ h.c.,
\end{split}
\end{equation}
where $\Phi_{i}$, for $i=1,2$ are the two scalar Higgs doublets,
${\tilde{\Phi}}_{i}=i\sigma_{2}\Phi_{i}$, $\eta_{ij}^{U,D,E}$ and
$\xi_{ij}^{U,D,E}$ are off diagonal $3\times3$ matrices of the
Yukawa couplings where i, j denote family indices (see Chapter
 \ref{SM} for the definitions of terms appearing in the
 Lagrangian).

As stated before, FCNCs with massless neutrino are naturally
suppressed in the tree level in the SM. To avoid FCNCs at tree
level, one can explicitly impose the following \emph{ad hoc}
discrete symmetry sets
\begin{eqnarray}
(I)\qquad\Phi_{1}&\rightarrow&-\Phi_{1},\;\;\;\;\Phi_{2}\rightarrow \Phi_{2},\;\;\;\;D_{j,R}\rightarrow-D_{j,R}, \;\;\;\; U_{j,R}\rightarrow-U_{j,R},\nonumber\\
(II)\qquad\Phi_{1}&\rightarrow&-\Phi_{1},\;\;\;\;\Phi_{2}\rightarrow \Phi_{2},\;\;\;\;D_{j,R}\rightarrow-D_{j,R},\;\;\;\; U_{j,R}\rightarrow+U_{j,R},\nonumber\\
\end{eqnarray}
into the Yukawa Lagrangian, ${\mathcal{L}}^{Y}_{2HDM}$. Imposing
these discrete symmetry sets, the so called model I and model II
are obtained depending on whether the up-type and the down-type
quarks are coupled to the same or two different scalar doublets,
respectively. If no discrete symmetry is implemented into the
${\mathcal{L}}^{Y}_{2HDM}$, both up-type and down-type quarks and
also leptons will have flavor changing (FC) couplings. This type
of 2HDM is called as model III where we should take into account
all the terms in the Yukawa Lagrangian given in eq. \ref{yukawa}.

It is possible to make a rotation of the doublets in such a way
that only one of the doublets acquire VEV so that
\begin{equation}
<\Phi _{1}>=\left(
\begin{array}{c}
0 \\
\frac{\upsilon}{\sqrt{2}}
\end{array}
\right)\qquad ;\qquad<\Phi _{2}>=0.
\end{equation}
The two doublets in this case arise of the form
\begin{equation}
\Phi _{1}=\left(
\begin{array}{c}
\chi^{+} \\
\frac{v + H^{0} + i\chi ^{0}}{\sqrt{2}}
\end{array}
\right)\qquad ;\qquad\Phi _{2}=\left(
\begin{array}{c}
H^{+} \\
\frac{H^{1}+ i H^{2} }{\sqrt{2}}
\end{array}
\right).
\end{equation}
Here
$H^{0}$ and $H^{1}$  are not the neutral mass eigenstates. The
neutral mass eigenstates are obtained from $(H^{0}, H^{1} ,H^{2})$
as follows
\begin{eqnarray}
\overline{H}^0 &=& [(H^0-\upsilon)cos\alpha - H^1sin\alpha],
\nonumber \\
h^0 &=& [-(H^0-\upsilon)sin\alpha + H^1\cos\alpha],
\nonumber \\
A^0 &=& H^2,
\end{eqnarray}
where $\alpha$ is the mixing angle. It is also possible to express
$H^{0}$, $H^{1}$ and $H^{2}$ as functions of mass eigenstates
\begin{eqnarray}
H^0 &=& (\overline{H}^0 cos\alpha - h^0 sin\alpha)+\upsilon,
\nonumber \\
H^1 &=& (h^0 cos\alpha + \overline{H}^0 sin\alpha)
,\nonumber \\
H^2 &=& A^0.
\end{eqnarray}
Choosing $\alpha=0$, $H_{1}$ becomes the well known mass
eigenstate $h_{0}$. As mentioned before, the model III version of
the 2HDM opens up the possibility of FCNCs at the tree level and
the FC part of the Yukawa Lagrangian reads
\begin{equation}
\pounds _{FC}^{III, Y} =\xi
_{ij}^{U}\overline{Q}_{i,L}\widetilde{\Phi }_{2}U_{j,R}+\xi
_{ij}^{D}\overline{Q}_{i,L}\Phi _{2}D_{j,R}+ \xi _{ij}^{E}%
\overline{l}_{i,L}\Phi _{2}E_{j,R}+h.c.
\end{equation}

There is another version of the 2HDM, the so called Model IV such
that $\phi_{1}$ couples and give masses to up-type quarks and
$\phi_{2}$ couples and gives masses to the down-type quarks and
the VEV of the doublets are chosen
\begin{equation}
<\Phi _{1}>=\left(
\begin{array}{c}
0\\
\upsilon_{1}
\end{array}
\right)\qquad ;\qquad<\Phi _{2}>=\left(
\begin{array}{c}
0\\
\upsilon_{2}e^{i\xi}
\end{array}
\right).
\end{equation}
In this case the term containing $\lambda_{6}$ is replaced by
 $\lambda_{6}(Im(\phi_{1}^{+}\phi_{2})-\upsilon_{1}\upsilon_{2}\sin\xi)^{2})$
 and is responsible for the CP violation in the Higgs sector.

\newpage
\chapter{EXTRA DIMENSIONS}\label{extra}
In recent years, models with extra dimensions, not yet experienced
and not yet entirely understood, have been studied extensively in
the literature (see for example \cite{akama}-\cite{RS1}). The
strong motivation to study these scenarios comes from the fact
that they resolve some of the most basic mysteries of our universe
such as the hierarchy problem between the two fundamental energy
scales, the EW scale ($m_{EW}\sim10^3$ GeV) and the Planck scale
($M_{Pl}\sim10^{19}$) GeV, where the strength of gravity becomes
comparable to the one of other gauge interactions. In fact, there
is an important difference between these two energy scales. While
the electroweak interactions have been probed at distances $\sim
m_{EW}^{-1}$, gravitational interactions has not remotely been
probed at distances $\sim M_{Pl}^{-1}$: it has only been
accurately measured in the $1$ cm range. Apart from the hierarchy
problem, the cosmological constant problem (for reviews see
\cite{ccp, witten}), the puzzle of why the vacuum energy is driven
to be a very small number, has also been tried to be solved within
the extra dimensional scenarios. One possible explanation for the
smallness of the cosmological constant problem can be stated as,
if there is a 4D theory with only 4D sources, these will
necessarily lead to an expanding universe. However, if there is 4D
sources in 5D, the effects of brane sources can be balanced by a
5D cosmological constant to get a theory where the effective 4D
cosmological constant would be vanishing. In this way, for an
observer on a brane, the universe will still appear to be static
and flat. The 5D background, however, will be curved since there
exists a bulk cosmological constant in there. Such extra
dimensional theories are called as warped extra dimensions, where
the brane is kept flat and the extra dimensions are curved. This
solution to the cosmological constant problem first pointed out by
Rubakov and Shaposhnikov \cite{rubakov}. Finally, the extra
dimensional scenario, named as the split fermion scenario
\cite{split, Schmaltz, chang}, provides an alternative view for
the fermion mass hierarchies by assuming that the fermions were
located at different points in the extra dimensions with the
exponentially small overlaps of their wavefunctions.

Among the models with extra dimensions, emergence of the ordinary
four dimensional SM as the low energy effective theory of more
fundamental theory lying in higher dimensions found acceptance.
The process of passing from a fundamental theory to the effective
theory includes the compactification of the extra dimension(s).
This compactification leads to the appearance of towers of heavy
KK \cite{Klein} modes of particles in $4D$ effective theory.

\section{Large Extra Dimensions}

In 1998, Nima Arkani-Hamed, Savas Dimopoulos and Gia Dvali
\cite{arkani, arkani1} tried to explain the hierarchy problem
between the scales $m_{EW}$ and $M_{Pl}$ by assuming the existence
of $n$ extra compact spatial dimensions of large size. According
to this model (called as ADD Model), the SM fields are confined to
the 4D brane while  gravity is free to propagate in large extra
dimensions. In other words, the gravitational field has extra
components in $n$ large extra dimensions and this extra components
cause it to be weaker than the other forces at long distances
because it would have been diluted by the large volume of the
extra dimension. In this scenario, since the SM particles are
confined to a 4D brane, everything that does not involve gravity
would look exactly the same as it would without the extra
dimensions, even if the extra dimensions were extremely large.
These extra dimensions, being compact, lead to the appearance of
towers of heavy KK modes. However, since the SM particles, which
are confined to a brane, would not have KK partners. The only
particle in this model that must have the KK partner is the
graviton. However, the graviton's KK partners interact far more
weakly than the SM KK partners (see for example \cite{csaki1} and
the references therein). Therefore, the KK partners of graviton
would be much harder to be observed. One question ADD wanted to
address with their set up is how large the size of the extra
dimensions can be without getting into conflict with observations
made up to date. To answer this question we need to match the 4D
effective theory to the fundamental higher dimensional one. Let us
assume that the higher dimensional action takes the same form with
Einstein-Hilbert action:
\begin{equation}
S_{4+n}\sim\int d^{4+n}x\sqrt{g^{(4+n)}}R^{(4+n)}.
\end{equation}
Here $\sqrt{g^{(n+4)}}$ and $R^{(n+4)}$ are the
metric\footnote{Metric is a quantity that establishes the
measurement scale that determines the physical distances and the
angles. A metric on 3D space can take the form
$ds^{2}=a_{x}dx^{2}+a_{y}dy^{2}+a_{z}dz^{2}$, where x, y, z are
the three coordinates of space, and $a_{x}$, $a_{y}$, $a_{z}$ can
be numbers or functions of x, y, z. If $a_{x}=a_{y}=a_{z}=1$, we
have flat space. More complicated metrics can have cross terms,
such as $dxdy$. In that case, the metric must be described by a
tensor with two indices that is denoted by the coefficients
$a_{ij}$ which is the coefficient in front of $dx_{i}dx_{j}$.} and
curvature scalars in $4+n$ dimensions. The action should be
dimensionless and we need to multiply the higher dimensional
Lagrangian by the appropriate power of the fundamental Planck
scale $(M_{*})$ , to make the action dimensionless. Since terms
$d^{4+n}x$, $\sqrt{g^{(4+n)}}$ and $R^{(4+n)}$ carry dimensions of
$-n-4$, $0$ and $2$, respectively, the power of $M_{*}$ has to be
$n+2$. Thus, we write the higher dimensional action as
\begin{equation}\label{hda}
S_{4+n}=-M_{*}^{n+2}\int d^{4+n}x\sqrt{g^{(4+n)}}R^{(4+n)}.
\end{equation}
Now, we will try to find how the usual 4D action
\begin{equation}\label{4da}
S_{4}=-M_{Pl}^{2}\int d^{4}x\sqrt{g^{(4)}}R^{(4)},
\end{equation}
is contained in the higher dimensional expression. For time being,
it is assumed that the space-time is flat, and the extra
dimensions are compact. So the metric is written as
\begin{equation}
ds^{2}=(\eta_{\mu\nu}+h_{\mu\nu})dx^{\mu}dx^{\nu}-r^{2}d\Omega^{2}_{(n)},
\end{equation}
where $x_{\mu}$ is a four dimensional coordinate with
$\mu=0,1,2,3$, $d\Omega^{2}_{(n)}$ is the line element of the flat
extra dimensional space, $r$ corresponds to the radius of the
extra dimension, $\eta_{\mu\nu}$ is the flat 4D metric and
$h_{\mu\nu}$ is the 4D fluctuation of the metric around its
minimum. Since our goal is to find out how a 4D theory emerges
from a higher dimensional one, we put only 4D fluctuation into the
metric. Then, the expressions for $\sqrt{g^{(n+4)}}$ and
$R^{(n+4)}$ are obtained as
\begin{equation}
\sqrt{g^{(4+n)}}=r^{n}\sqrt{g^{(4)}}\qquad;\qquad R^{(4+n)}=R^{4}.
\end{equation}
Substituting these expressions into the eq. \ref{hda} we get
\begin{equation}\label{hhda}
S_{4+n}=-M_{*}^{n+2}\int d\Omega_{(n)}r^{n}\int
d^{4}x\sqrt{g^{(4)}}R^{(4)},
\end{equation}
where the factor $\int d\Omega_{(n)}r^{n}$ is simply the volume of
the extra dimensional space, $V_{(n)}$. Comparing the eqs.
\ref{4da} and \ref{hhda}, we obtain the matching relation for the
gravitational couplings as
\begin{equation}\label{gracoup}
M^{2}_{Pl}=M_{*}^{n+2}V_{(n)}\sim r^{n}M_{*}^{n+2}.
\end{equation}
Now, we will find the matching relation for gauge couplings. Based
on the assumption that every field propagates in all dimensions we
can write the action as
\begin{equation}\label{agc}
S^{(4+n)}=-\int
d^{4+n}x\frac{1}{4g_{*}^{2}}F_{MN}F^{MN}\sqrt{g^{(4+n)}}.
\end{equation}
Here, $F_{MN}$ with $M,N=0,1,...3+n$ is the higher dimensional
field strength tensor and $g_{*}$ is the fundamental gauge
coupling. Taking the integral over the extra dimension we get
\begin{equation}
S^{(4)}=-\int
d^{4}x\frac{V_{(n)}}{4g_{*}^{2}}F_{\mu\nu}F^{\mu\nu}\sqrt{g^{(4)}}.
\end{equation}
Then, we obtain the matching of gauge couplings as
\begin{equation}\label{gc}
\frac{1}{g_{eff}^{2}}=\frac{V_{n}}{g_{*}^{2}}.
\end{equation}
Notice that, in the eq. \ref{agc}, $d^{4+n}x$, $\sqrt{g^{(4+n)}}$
and $F_{MN}$ carry dimensions $-n-4$, $0$ and $2$, respectively.
Thus, the mass dimension of $g_{*}$ has to be $-n/2$ so that the
action remains dimensionless. Looking at the mass dimensions of
$g_{*}$ and $M_{*}$ we can write
\begin{equation}
g_{*}\sim\frac{1}{M_{*}^{\frac{n}{2}}}.
\end{equation}
Substituting this into the eq. \ref{gc} we get
\begin{equation}\label{geff}
\frac{1}{g_{eff}^{2}}=V_{n}M_{*}^{n}\sim r^{n}M_{*}^{n}.
\end{equation}
Using the eq. \ref{gracoup} we obtain an equation for $M_{*}$ as
\begin{equation}
M_{*}\sim\frac{M_{Pl}^{\frac{2}{n+2}}}{r^{\frac{n}{n+2}}}.
\end{equation}
If we substitute this into the eq. \ref{geff} we get
\begin{equation}
\frac{1}{g_{eff}^{2}}\sim\frac{M_{Pl}^{\frac{2n}{n+2}}}{r^{-\frac{2n}{n+2}}}.
\end{equation}
Then, r becomes
\begin{equation}\label{radius}
r\sim\frac{1}{M_{Pl}}g_{eff}^{\frac{n+2}{n}}.
\end{equation}
Since $r\sim 1/M_{Pl}$ there would be no hope of finding out about
the existence of these tiny extra dimensions in the near future.

Up to now, we have assumed that every field propagate in all
dimensions. Now, we will deal with the restrictions on the size of
extra dimensions when the SM particles are localized to 4D brane
while the unobserved fields such as gravity are to propagate in
extra dimensions. However, in distances as small as the size of
extra dimensions, it is impossible to test gravity. Because, in
such short distances electromagnetic and weak forces become more
dominate than the gravitational force and the gravitational
interactions have been tested at the distances of the order of
millimeter. Therefore, the real bound on the size of extra
dimensions becomes
\begin{equation}
r\leq 0.1 mm,
\end{equation}
if only gravity propagates in the extra dimension. From the
relation $M_{Pl}^{2}\sim M_{*}^{n+2}r^{n}$ we know that if we
increase the value of the radius $r$, the fundamental Planck scale
$M_{*}$ decreases. If $M_{*}< 1 TeV$, we would have observed some
effect of quantum gravity in the collider experiments. Thus, one
has to impose the lowest possible value to be $M_{*}\sim 1 TeV$.
Therefore, one can say that being equal to the fundamental Planck
scale, $m_{EW}$ is the only fundamental short distance scale in
nature and $M_{Pl}$ is valid in a 4D scenario is an effective
scale derived from $m_{EW}$. Such models are called as Large extra
dimensions, proposed by Arkani- Hamed, Dimopoulos and Dvali.

Let us check, how large a radius one would need for the lowest
possible value of $M_{*}$, using the eq. \ref{gracoup} the value
of radius would be
\begin{equation}
\frac{1}{r}=M_{*}(\frac{M_{*}}{M_{Pl}})^{\frac{2}{n}}=(1TeV)10^{-\frac{32}{n}}.
\end{equation}
Using %
\begin{equation}
1GeV^{-1}=2.10^{-14}cm,
\end{equation}
we get
\begin{equation}
r\sim 2.10^{-17}10^{-\frac{32}{n}}cm.
\end{equation}
For $n=1$, the value of $r=2.10^{15}$ cm. Since this value is
larger then the astronomical unit of $1.5\times10^{13}$ cm, we can
conclude that there cannot be one flat large extra dimension. If
there are two extra dimensions, $r\sim 2$ mm. This is just a
borderline for the currently probed gravitational experiments. For
$n>2$, $r<10^{-6}$ cm. This value is so small to be measured in
the near future. Then, the hierarchy problem between the
fundamental Planck scale and the scale of weak interactions would
have been solved so that gravity would be weaker than the other
forces at long distances because it would have been diluted by the
large volume of the extra dimension.

\section{Universal and Non-Universal Extra Dimensions}

As mentioned above, the compactification of the extra dimensions
to a circle $S_1$ with a small radius $R$ makes them imperceptibly
small. But can an extra dimensional universe hide its nature so
completely that none of its features distinguishes it from a 4D
world? That would be hard to believe. If there are extra
dimensions, fingerprints of them sure to exists. Such fingerprints
are called as KK \cite{Klein} particles. These new particles
originate in extra dimensions, but appear to us as extra particles
in our 4D space-time. In other words, they are manifestations of
particles, which are in higher dimensions, in 4D such that every
particle that travel in higher dimensional space is replaced by KK
particles in our 4D space. A universe with extra dimensions
contains both familiar particles and their KK relatives that carry
extra dimensional momentum. However, a 4D dimensional space-time
does not include information about higher dimensional position or
momentum. This extra dimensional momentum would be seen in our 4D
world as mass. Thus, KK particles should be like the ones we know
(having the same charge), but heavier. If the universe contains
additional dimensions, these heavier KK particles will be the
first real evidence of them.

The wavefunction of a KK particle is written as Fourier
decomposition of the higher dimensional wavefunction. To be
concrete let us imagine a space with only one additional spatial
dimension which is compactified on $S^{1}/Z_{2}$ orbifold $(Z_{2}:
y\rightarrow -y)$. Then, the KK decompositions of the five
dimensional (5D) fields can be written as follows,
\begin{equation}
\begin{split}
\phi(x,y)&= \frac{1}{\sqrt{\pi R}}\phi^{(0)}(x)+ \frac{\sqrt{2}}{\sqrt{\pi R}}\sum_{n=1}^{\infty}\phi^{(n)}(x)\cos(\frac{ny}{R}),\\
A_{\mu}(x,y)&= \frac{1}{\sqrt{\pi R}}A_{\mu}^{(0)}(x)+ \frac{\sqrt{2}}{\sqrt{\pi R}}\sum_{n=1}^{\infty}A_{\mu}^{(n)}(x)\cos(\frac{ny}{R}),\\
A_{5}(x,y)&= \frac{\sqrt{2}}{\sqrt{\pi R}}\sum_{n=1}^{\infty}A_{5}^{(n)}(x)\sin(\frac{ny}{R}),\\
Q(x,y)&= \frac{1}{\sqrt{\pi R}}Q_{L}^{(0)}(x)+
\frac{\sqrt{2}}{\sqrt{\pi
R}}\sum_{n=1}^{\infty}[Q_{L}^{(n)}(x)\cos(\frac{ny}{R})+
Q_{R}^{(n)}(x)\sin(\frac{ny}{R})],\\
U(x,y)&= \frac{1}{\sqrt{\pi R}}U_{R}^{(0)}(x)+\frac{\sqrt{2}}{\sqrt{\pi R}}\sum_{n=1}^{\infty}[U_{R}^{(n)}(x)\cos(\frac{ny}{R})+U_{L}^{(n)}(x)\sin(\frac{ny}{R})],\\
D(x,y)&= \frac{1}{\sqrt{\pi
R}}D_{R}^{(0)}(x)+\frac{\sqrt{2}}{\sqrt{\pi
R}}\sum_{n=1}^{\infty}[D_{R}^{(n)}(x)\cos(\frac{ny}{R})+
U_{L}^{(n)}(x)\sin(\frac{ny}{R})].
\end{split}
\end{equation}
Fields even under the $Z_{2}$ symmetry will have zero modes and
they correspond to the SM particles in our usual 4D world whereas
fields odd under $Z_{2}$ symmetry will only have KK modes and will
be absent in the low energy spectrum so that, in the four
dimensional Lagrangian, we get rid of the the zero modes of wrong
chirality $(i.e., Q_{R}, U_{L}$, and $D_{L})$ and the fifth
component of the gauge field, $A_{5}$. If we could measure and
study their properties, they would tell us everything about the
higher dimensional space.

The elecroweak, $SU(2)_{L}\times U(1)_{Y}$, part of the SM
Lagrangian in 5D can be written as follows
\begin{equation}\label{5DL}
\mathcal{L}= \int^{\pi R}_{0}dy
(\mathcal{L}^{f}_{kinetic}+\mathcal{L}^{H}+\mathcal{L}^{G}_{kinetic}+\mathcal{L}^{Y}),
\end{equation}
where $y\equiv x^{4}$ is the coordinate along the extra dimension.
The fermionic piece of the 5D Lagrangian is defined as
\begin{equation}
\mathcal{L}_{kinetic}^{f}= \bar{Q}(i\Gamma^{M}D_{M})Q+
\bar{U}(i\Gamma^{M}D_{M})U+ \bar{D}(i\Gamma^{M}D_{M})D,
\end{equation}
where, $M,N = 0, 1, 2, 3, 4$ corresponds to the five-dimensional
Lorentz indices. The fermionic fields Q, U and D denotes five
dimensional generic quark doublet, up type quark singlet, and down
type quark singlet, respectively. Unlike in the SM, these
fermionic fields have both chiralities and are all vector type.
The covariant derivative, $D_{M}$ is defined as
$D_{M}=\partial_{M}-i\tilde{g}W_{M}^{a}T^{a}-i\tilde{g}'B_{M}Y$
with $\tilde{g}$ is the five dimensional gauge coupling constant
of the group $SU(2)_{L}$ and $\tilde{g}'$ is that of $U(1)_{Y}$.
Here, $T^{a}$ and $Y$ are the corresponding generators. In
addition, $\Gamma_{M}$ are the five dimensional gamma matrices
with $\Gamma_{\mu}= \gamma_{\mu}$ and $\Gamma_{4}= i\gamma_{5}$.
The Higgs piece of the 5D Lagrangian is
\begin{equation}
\mathcal{L}^{H}=(D_{M}\phi)^{\dagger}(D^{M}\phi)-V(\phi),
\end{equation}
and the gauge piece is
\begin{equation}
\mathcal{L}_{kinetic}^{G}=
-\frac{1}{4}\sum_{i=1}^{3}F^{MN}_{i}F_{MN}^{i}-\frac{1}{4}F^{MN}F_{MN}.
\end{equation}
The field strength tensors associated with the $SU(2)_{L}$ and
$U(1)_{Y}$ gauge group can be expressed as
\begin{equation}
F_{MN}^{i}=\partial_{M}W_{N}^{i}-\partial_{N}W_{M}^{i}+
g\epsilon^{ijk}W_{M}^{j}W_{N}^{k},
\end{equation}
and
\begin{equation}
F_{MN}=\partial_{M}B_{N}- \partial_{N}B_{M},
\end{equation}
respectively. Finally the Yukawa piece reads
\begin{equation}
\mathcal{L}^{Y}=\bar{Q}\tilde{Y}_{u}\phi^{c}U+\bar{Q}\tilde{Y}_{d}\phi
D.
\end{equation}
In the above equation, the fields $\phi$ and $\phi^{c}=
i\tau^2\phi^{\ast}$ stand for the standard Higgs doublet and its
charge conjugated field as each refers to each in order. Finally,
$\tilde{Y}_{u}$ and $\tilde{Y}_{d}$ correspond the Yukawa matrices
in the five dimensional theory which are responsible for mixing
different generations. For simplicity lepton or gluon indices are
not included.

Substituting the KK decompositions of the five dimensional fields
given above into the eq. \ref{5DL} and integrating over the extra
dimension $y$, we obtain the effective four dimensional
Lagrangian. One can realize simply that the KK excitations receive
mass not only due to the vacuum expectation value of the zero-mode
Higgs but also from the kinetic energy term in the five
dimensional Lagrangian. The mass of the $n^{th}$ KK particle is
given by $m^{KK}_{n}= \sqrt{m_{0}^{2}+m_{n}^{2}}$. Here $m_{0}$
corresponds to the zero mode mass and $m_{n}=n/R$.

Depending on the underlying fundamental theory, extra dimensions
may or may not be accessible to all fields in the model. According
to this accessibility, extra dimensions can be grouped into two,
including "universal extra dimensions" (see for example (UED)
\cite{ued, ued1, ued2} and the references therein) and
"non-universal extra dimensions" (NUED) (see for example
\cite{nued, nued1} and the references therein), respectively. In a
theory with universal extra dimensions, all fields in the model
feel the extra dimensions. Conservation of extra dimensional
momentum leads to the key future of such theories that KK number
at each elementary interaction vertex is conserved. As a result of
this feature, production of an isolated KK particle at colliders
is forbidden. Instead they are produced in pairs. This, in turn,
implies that there is no tree-level contribution to weak decays of
quarks and leptons. They enter into the calculations only through
loop corrections. However, in a theory with non-universal extra
dimensions, some of the SM fields are confined to a 4D brane and
the others live in the bulk. In this case, the Lagrangian contains
localizing delta function which permits KK number violating
couplings. Then, the tree-level interactions of KK modes with the
ordinary particles can exist.

\section{The Randal-Sundrum Model}

The large extra dimensions which are discussed in the previous
section took the advantage of the fact that branes could trap
particles and force but neglected the energy that the branes
themselves could carry. However, according to Einstein's theory of
general relativity, gravitational field is induced by means of
energy, which means that when branes carry energy, they should
curve space and time. Lisa Randall and Raman Sundrum \cite{RS,
RS1} tried to explain how space-time would be curved in the
presence of two 4D energetic branes that bounded the extra
dimension of space where the bulk\footnote{Full higher-dimensional
space.} geometry is anti-de Sitter\footnote{Space-time with
constant negative curvature.}, by solving Einstein's gravity
equations based on the assumption that both the bulk and and the
branes have energy. In this space-time the 4D branes and any
single slice along the fifth dimension are completely flat. But
the 5D space-time under consideration is nonetheless curved. The
technical term for this type of geometry is \textquoteleft
warped\textquoteright. This section focusses on a warped
five-dimensional world that provides an alternative approach to
explain the huge discrepancy between $m_{EW}$ and $M_{Pl}$ without
the need for a large extra dimension. In this scenario (called as
Randall-Sundrum (RS1) Model), the geometry contains two 4D flat
branes, the Planck brane where the gravity is localized and the
TeV brane where all SM particles are confined, that bound a fifth
dimension which is compactified to $S^{1}/Z_{2}$ orbifold. The two
4D flat branes with opposite tensions, which reside at the
orbifold fixed points together with a finely tuned non-vanishing
5D cosmological constant $\Lambda$, serve as sources for 5D
gravity. Since the two branes are completely flat, the induced
metric at every point along the extra dimension has to be the
ordinary flat 4D Minkowski metric, and the components of the 5D
metric depend only on the fifth coordinate, $y$. Thus, the most
general space-time metric satisfying these properties is given by
\begin{equation}\label{ds2}
ds^{2}=e^{-A(y)}\eta_{\mu\nu}dx^{\mu}dx^{\nu}-dy^{2},
\end{equation}
where $e^{-A(y)}$ is called as the warp-factor which determines
the amount of curvature along the extra dimension. It is also
possible to write this metric in a conformally flat form where
there is an overall factor. To go into the conformally flat frame,
all we need to do is to make a coordinate transformation of the
form $z=z(y)$ such that $dy$ and $dz$ are related by
\begin{equation}\label{eAzdz}
e^{-A(z)/2}dz=dy.
\end{equation}
Then, the metric in the eq. \ref{ds2} becomes
\begin{equation}
ds^{2}=e^{-A(z)}(\eta_{\mu\nu}dx^{\mu}dx^{\nu}-dz^{2}),
\end{equation}
and we get
\begin{equation}
g_{MN}=e^{-A(z)}\tilde{g}_{MN} \label{gMN1},
\end{equation}
where $\tilde{g}_{MN}$ is the flat metric,
$\tilde{g}_{MN}=\eta_{MN}$.
\subsection{The Einstein tensor and the brane tensions}
The starting point is the action (see \cite{dirac} for example)
\begin{equation}
S=\int d^5x\sqrt{g}(M_{*}^{3}R) \label{varSfree},
\end{equation}
where $M_{*}$ is the 5D Planck scale, $R=g^{MN}R_{MN}$ and
$R_{MN}$ reads
\begin{equation}
R_{MN}=\Gamma_{MK,N}^{K}-\Gamma_{MN,K}^{K}
-\Gamma_{MN}^{K}\Gamma_{KL}^{L} +\Gamma_{ML}^{K}\Gamma_{NK}^{L},
\end{equation}
with the connections
\begin{equation}
\Gamma_{MN}^{K}=g^{KL}\Gamma_{LMN}=\frac{1}{2}g^{KL}(g_{LM,N}+g_{LN,M}-g_{MN,L}).
\end{equation}
After the variation of this action with respect to $g_{MN}$ (see
\cite{csaki} and \cite{dirac} for details) we get
\begin{equation}
\delta S=\int
d^5x[\delta\sqrt{g}M_{*}^{3}R+\sqrt{g}M_{*}^{3}\delta
g^{MN}R_{MN}],
\end{equation}
where
\begin{equation}
\delta\sqrt{g}=\frac{1}{2}\delta g_{MN}g^{MN}\sqrt{g}\qquad
;\qquad\delta g^{MN}=-g^{MK}g^{NL}\delta g_{KL}.
\end{equation}
Finally we obtain
\begin{equation}\label{varI}
\delta S= -M_{*}^{3}\,\int
d^{5}x(R^{MN}-\frac{1}{2}g^{MN}R)\sqrt{g}\,\delta g_{MN},
\end{equation}
with the Einstein tensor $G_{MN}=R_{MN}-\frac{1}{2}g_{MN}R$. Now,
we will calculate this tensor for the special metric given in eq.
\ref{gMN1}. From now on, since $\tilde{g}_{MN}=\eta_{MN}$, all
covariant derivatives $\tilde{\nabla}_{M}$ (which are with respect
to the metric $\tilde{g}$) will be replaced by the normal
derivative $\partial_{M}$.  Let us calculate $R_{MN}$ term by
term. The first term is:
\begin{equation}
\begin{split}
\Gamma_{MK,N}^{K}&=\{g^{KS}\Gamma_{SMK}\}_{,N}\\
&=\frac{1}{2}\{g^{KS}[g_{SM,K}+g_{SK,M}-g_{MK,S}]\}_{,N}\\
&=\frac{1}{2}\{e^{A}\eta^{KS}[(e^{-A}\eta_{SM})_{,K}+(e^{-A}\eta_{SK})_{,M}-(e^{-A}\eta_{MK})_{,S}]\}_{,N}\\
&=\frac{1}{2}\{\eta^{KS}[-\eta_{SM}\partial_{K}A-\eta_{SK}\partial_{M}A+\eta_{MK}\partial_{S}A]\}_{,N}\\
&=\frac{1}{2}\{\eta^{KS}[-\eta_{SM}\partial_{N}\partial_{K}A-\eta_{SK}\partial_{N}\partial_{M}A+\eta_{MK}\partial_{N}\partial_{S}A]\}\\
&=-\frac{1}{2}\eta^{K}_{M}\partial_{N}\partial_{K}A-\frac{1}{2}\eta_{S}^{S}\partial_{N}\partial_{M}A+\frac{1}{2}\eta_{M}^{S}\partial_{N}\partial_{S}A,
\end{split}
\end{equation}
where
\begin{equation}
\eta_{M}^{N}=\delta_{MN}\qquad \eta_{M}^{M}=d,
\end{equation}
with the space-time dimension $d$. Substituting these equations
into $\Gamma_{MK,N}^{K}$ we obtain the first term in $R_{MN}$ as
\begin{equation}
\Gamma_{MK,N}^{K}=-\frac{d}{2}\partial_{N}\partial_{M}A.
\end{equation}
By using the same procedure we get the other terms as:
\begin{equation}
\begin{split}
\Gamma_{MN,K}^{K}&=\{g^{KS}\Gamma_{SMN}\}_{,K}\\
&=\frac{1}{2}\{g^{KS}[g_{SM,N}+g_{SN,M}-g_{MN,S}]\}_{,K}\\
&=\frac{1}{2}\{e^{A}\eta^{KS}[(e^{-A}\eta_{SM})_{,N}+(e^{-A}\eta_{SN})_{,M}-(e^{-A}\eta_{MN})_{,S}]\}_{,K}\\
&=\frac{1}{2}\{\eta^{KS}[-\eta_{SM}\partial_{N}A-\eta_{SN}\partial_{M}A+\eta_{MN}\partial_{S}A]\}_{,K}\\
&=\frac{1}{2}\{\eta^{KS}[-\eta_{SM}\partial_{K}\partial_{N}A-\eta_{SN}\partial_{K}\partial_{M}A+\eta_{MN}\partial_{K}\partial_{S}A]\}\\
&=-\frac{1}{2}\eta^{K}_{M}\partial_{K}\partial_{N}A-\frac{1}{2}\eta_{N}^{K}\partial_{K}\partial_{M}A+\frac{1}{2}\eta^{KS}\eta_{MN}\partial_{K}\partial_{S}A\\
&=-\partial_{M}\partial_{N}A+\frac{1}{2}\eta_{MN}\partial^2A,
\end{split}
\end{equation}
\begin{equation}
\begin{split}
\Gamma_{MN}^{K}\Gamma_{KL}^{L}&=g^{KS}\Gamma_{SMN}g^{LP}\Gamma_{PKL}\\
&=\frac{1}{2}g^{KS}[g_{SM,N}+g_{SN,M}-g_{MN,S}]\frac{1}{2}g^{LP}[g_{PK,L}+g_{PL,K}-g_{KL,P}]\\
&=\frac{1}{4}e^{A}\eta^{KS}[(e^{-A}\eta_{SM})_{,N}+(e^{-A}\eta_{SN})_{,M}-(e^{-A}\eta_{MN})_{,S}]\\
&\times e^{A}\eta^{LP}[(e^{-A}\eta_{PK})_{,L}+(e^{-A}\eta_{PL})_{,K}-(e^{-A}\eta_{KL})_{,P}]\\
&=\frac{1}{4}\eta^{KS}[-\eta_{SM}\partial_{N}A-\eta_{SN}\partial_{M}A+\eta_{MN}\partial_{S}A]\\
&\times\eta^{LP}[-\eta_{PK}\partial_{L}A-\eta_{PL}\partial_{K}A+\eta_{KL}\partial_{P}A]\\
&=\frac{1}{4}[-\eta^{K}_{M}\partial_{N}A-\eta_{N}^{K}\partial_{M}A+\eta^{KS}\eta_{MN}\partial_{S}A]\\
&\times[-\eta^{L}_{K}\partial_{L}A-\eta_{L}^{L}\partial_{K}A+\eta_{K}^{P}\partial_{P}A]\\
&=\frac{1}{4}[-\eta^{K}_{M}\partial_{N}A-\eta_{N}^{K}\partial_{M}A+\eta_{MN}\partial^{K}A][-d\partial_{K}A]\\
&=\frac{d}{2}\partial_{M}A\partial_{N}A-\frac{d}{4}\eta_{MN}(\partial
A)^{2},
\end{split}
\end{equation}
\begin{equation}
\begin{split}
\Gamma_{ML}^{K}\Gamma_{NK}^{L}&=g^{KS}\Gamma_{SML}g^{LP}\Gamma_{PNK}\\
&=\frac{1}{2}g^{KS}[g_{SM,L}+g_{SL,M}-g_{ML,S}]\frac{1}{2}g^{LP}[g_{PN,K}+g_{PK,N}-g_{NK,P}]\\
&=\frac{1}{4}e^{A}\eta^{KS}[(e^{-A}\eta_{SM})_{,L}+(e^{-A}\eta_{SL})_{,M}-(e^{-A}\eta_{ML})_{,S}]\\
&\times e^{A}\eta^{LP}[(e^{-A}\eta_{PN})_{,K}+(e^{-A}\eta_{PK})_{,N}-(e^{-A}\eta_{NK})_{,P}]\\
&=\frac{1}{4}\eta^{KS}[-\eta_{SM}\partial_{L}A-\eta_{SL}\partial_{M}A+\eta_{ML}\partial_{S}A]\\
&\times\eta^{LP}[-\eta_{PN}\partial_{K}A-\eta_{PK}\partial_{N}A-\eta_{NK}\partial_{P}A]\\
&=\frac{1}{4}[-\eta^{K}_{M}\partial_{L}A-\eta_{L}^{K}\partial_{M}A+\eta^{KS}\eta_{ML}\partial_{S}A]\\
&\times[-\eta^{L}_{N}\partial_{K}A-\eta_{K}^{L}\partial_{N}A+\eta^{LP}\eta_{NK}\partial_{P}A]\\
&=\frac{1}{4}[-\eta^{K}_{M}\partial_{L}A-\eta_{L}^{K}\partial_{M}A+\eta_{ML}\partial^{K}A]\\
&\times[-\eta^{L}_{N}\partial_{K}A-\eta_{K}^{L}\partial_{N}A+\eta_{NK}\partial_{L}A]\\
&=\frac{1}{4}(2+d)\partial_{M}A\partial_{N}A-\frac{1}{2}\eta_{MN}(\partial
A)^{2}.
\end{split}
\end{equation}
Finally $R_{MN}$ becomes
\begin{equation}\label{RMN}
\begin{split}
R_{MN}&=\frac{2-d}{2}\partial_{M}\partial_{N}A+\frac{2-d}{4}\partial_{M}A\partial_{N}A\\
&+\frac{d-2}{4}\eta_{MN}(\partial
A)^{2}-\frac{1}{2}\eta_{MN}\partial^{2}A.
\end{split}
\end{equation}
Now, we will find $R$ using $R=g^{MN}R_{MN}$ as follows:
\begin{equation}\label{R}
\begin{split}
R&=e^{A}\eta^{MN}R_{MN}\\
&=e^{A}[\frac{2-d}{2}\partial^{2}A+\frac{2-d}{4}(\partial A)^{2}+\frac{d(d-2)}{4}(\partial A)^{2}-\frac{d}{2}\partial^{2}A]\\
&=e^{A}[(1-d)\partial^{2}A+\frac{(2-d)(1-d)}{4}(\partial A)^{2}].
\end{split}
\end{equation}
Substituting the eqs. \ref{RMN} and \ref{R} into the Einstein
tensor we get
\begin{equation}
\begin{split}
G_{MN}&=\frac{2-d}{2}\{\frac{1}{2}\partial_{M}A\partial_{N}A+\partial_{M}\partial_{N}A\\
&-\eta_{MN}[\partial^{K}\partial_{K}A-\frac{d-3}{4}\partial^{K}A\partial_{K}A]\}.
\end{split}
\end{equation}
Here $d=5$ stands for the number of dimensions. Using this
expression we can evaluate the non-vanishing terms $G_{55}$ and
$G_{\mu\nu}$. Let us start with $G_{55}$:
\begin{equation}
G_{55}=-\frac{3}{2}\{\frac{1}{2}A'^{2}+A''-\eta_{55}[-A''+\frac{1}{2}A'^{2}]\},
\end{equation}
where $A'=\partial_{5}A$ and $\eta_{55}=-1$ and we get
\begin{equation}\label{g55}
G_{55}=-\frac{3}{2}A'^{2}.
\end{equation}
$G_{\mu\nu}$ can be obtained in the same way:
\begin{equation}
G_{\mu\nu}=-\frac{3}{2}\{\frac{1}{2}\partial_{\mu}A\partial_{\nu}A
+\partial_{\mu}\partial_{\nu}A-\eta_{\mu\nu}[\partial^{K}\partial_{K}
A-\frac{1}{2}\partial^{K}A\partial_{K}A]\}.
\end{equation}
Since $A=A(z)$, the terms including 4D differentiation vanish and
we obtain,
\begin{equation}\label{gmunu}
G_{\mu\nu}=\frac{3}{2}\eta_{\mu\nu}(-A''+\frac{1}{2}A'^2).
\end{equation}
Now, we consider the 5D Einstein action for gravity with a bulk
cosmological constant $\Lambda$:
\begin{equation}
S=\int d^5x\sqrt{g}(M_{*}^{3}R-\Lambda) \label{varS}.
\end{equation}
%
Taking the variation of the action with respect to metric
\begin{equation}
\delta S=\int
d^5x[\delta\sqrt{g}M_{*}^{3}R+\sqrt{g}M_{*}^{3}\delta
g^{MN}R_{MN}-\delta\sqrt{g}\Lambda],
\end{equation}
we get  (see eq. \ref{varI} for the variation of the first term in
eq. \ref{varS})
\begin{equation}
\delta S=\int d^5x[-M_{*}^{3}G^{MN}-\frac{1}{2}\Lambda
g^{MN}]\sqrt{g}\delta g_{MN}.
\end{equation}
Then, we can simply write the Einstein tensor as
\begin{equation}
G_{MN}=-\frac{1}{2M_{*}}\Lambda g_{MN}.
\end{equation}
The 55 component of Einstein tensor will then be:
\begin{equation}
\frac{3}{2}A'^{2}=\frac{1}{2M_{*}}\Lambda
g_{55}=-\frac{1}{2M_{*}}\Lambda e^{-A(z)}.
\end{equation}
and $A'$ reads,
\begin{equation}
A'=\sqrt{\frac{-\Lambda }{3M_{*}^{3}}}e^{-A(z)/2} \label{Aprime}.
\end{equation}
There exists a solution if and only if $\Lambda<0$. This means
that the important case for us will be considering anti-de Sitter
spaces, that is the spaces with negative cosmological constant.
Now, let us take $f=e^{-A(z)/2}$ and, therefore,
$f'=\frac{1}{2}A'(z)f$. Substituting them into the eq.
\ref{Aprime}  we obtain
\begin{equation}
-\frac{f'}{f^{2}}=\frac{1}{2}\sqrt{\frac{-\Lambda}{3M_{*}^{3}}}.
\end{equation}
Solving this differential equation we get a relation for $f$ such
that,
\begin{equation}
f=e^{-A(z)}=\frac{1}{(kz+c_{0})^{2}},
\end{equation}
where $k^{2}=\frac{-\Lambda}{12M_{*}^{3}}$. If we choose
$e^{-A(0)}=1$,
\begin{equation}
e^{-A(0)}=\frac{1}{c_{0}^{2}}=1,
\end{equation}
we get $c_{0}=1$ and, finally, we obtain
\begin{equation}
e^{-A(z)}=\frac{1}{(kz+1)^{2}}.
\end{equation}
This solution must be symmetric under $z\rightarrow-z$ reflection
since we are on a $S^{1}/Z_{2}$ orbifold and, therefore, we take
\begin{equation}\label{euAz}
e^{-A(z)}=\frac{1}{(k|z|+1)^{2}}.
\end{equation}
The RS metric is then obtained as
\begin{equation}
ds^{2}=\frac{1}{(k|z|+1)^{2}}(\eta_{\mu\nu}dx^{\mu}dx^{\nu}-dz^{2}).
\end{equation}
Using the eq. \ref{eAzdz}, we obtain $A(z)$ as follows:
\begin{equation}
\frac{dz}{k|z|+1}=dy,
\end{equation}
and solving for $y$ we get
\begin{equation}
y=\frac{\ln(k|z|+1)}{kz/|z|}+c_{1}.
\end{equation}
If we choose $y=0$ to correspond to $z=0$ the constant $c_{1}$
becomes $c_{1}=0$. Then, we have
\begin{equation}
\frac{1}{(k|z|+1)^{2}}=e^{-2k|y|},
\end{equation}
and the final form of the RS metric in $y$ coordinates becomes
\begin{equation}
ds^{2}=e^{-2k|y|}\eta_{\mu\nu}dx^{\mu}dx^{\nu}-dy^{2},
\end{equation}
where $y$ corresponds to the physical distance along the extra
dimension, since in that metric there is no warp factor in front
of $dy^{2}$ term and the Planck (TeV) brane is located at $y=0$
($y=r_{0}$).

At this stage, we would like to check whether the $4D$ components
of the Einstein tensor given in eq. \ref{gmunu} are also satisfied
or not. Using the eq. \ref{euAz}, $A(z)$ is obtained as
\begin{equation}
A(z)=\ln(k|z|+1)^{2}.
\end{equation}
The derivative of $A(z)$ with respect to $z$ gives
\begin{equation}
A'(z)=\frac{2(k|z|+1)kz/|z|}{(k|z|+1)^{2}}=\frac{2k\varepsilon(z)}{(k|z|+1)},
\end{equation}
where $\frac{z}{|z|}=\varepsilon(z)=(\theta(z)-\theta(-z))$ and
one more derivative of $A(z)$ with respect to $z$ reads
\begin{equation}
A''(z)=-\frac{2k^{2}}{(k|z|+1)^{2}}+\frac{4k}{k|z|+1}(\delta(z)-\delta(z-z_{1})).
\end{equation}
Substituting $A'$ and $A''$ into eq. \ref{gmunu} we get
\begin{equation}
G_{\mu\nu}=-\frac{3}{2}\eta_{\mu\nu}\Big\{\frac{4k^{2}}{(k|z|+1)^{2}}-\frac{4k[\delta(z)-\delta(z-z_{1})]}{k|z|+1}\Big\}.
\end{equation}
Here the first term is the contribution of the bulk cosmological
constant into the energy momentum tensor. The remaining delta
functions should be compensated by the additional sources onto the
branes. To do this we need to find the enenrgy- momentum tensor of
a brane tension term $V$ using the action
\begin{equation}
S=\int d^{4}xV\sqrt{g^{ind}}=\int
d^{5}xV\frac{\sqrt{g}}{\sqrt{g^{55}}}\delta(y),
\end{equation}
for a flat brane at $y=0$. Taking the variation of this action
with respect the metric $g_{\mu\nu}$ we get
\begin{equation}
\delta S=\int d^{4}xV\frac{1}{2}g_{\mu\nu}\delta
g^{\mu\nu}\sqrt{g_{ind}},
\end{equation}
and by using the energy-momentum tensor for a brane
\begin{equation}
T_{\mu\nu}=\frac{1}{\sqrt{g}}\frac{\delta S}{\delta g^{\mu\nu}},
\end{equation}
we obtain $T_{\mu\nu}$ as
\begin{equation}
T_{\mu\nu}=\frac{1}{2\sqrt{g^{55}}}g_{\mu\nu}V\delta(y).
\end{equation}
%
%
Therefore, at the branes we need to have two brane tensions to
compensate the delta functions. Thus, we need the equality
\begin{equation}
-\frac{3}{2}\eta_{\mu\nu}\Big[-\frac{4k(\delta(z)-\delta(z-z_{1}))}{k|z|+1}\Big]=
\frac{\eta_{\mu\nu}}{2M_{*}^{3}}\Big[\frac{V_{0}\delta(z)+V_{1}\delta(z-z_{1})}{k|z|+1}\Big].
\end{equation}
Then, one can simply conclude that, the brane tensions at the two
fixed points will have opposite signs and they are given by
\begin{equation}
V_{0}=-V_{1}=12kM_{*}^{3}.
\end{equation}
Finally, using the expression $k^{2}=\frac{-\Lambda}{12M_{*}^{3}}$
the bulk cosmological constant is obtained as
\begin{equation}
\Lambda=-\frac{V_{0}^{2}}{12M_{*}^{3}}\qquad;\qquad V_{1}=-V_{0}.
\end{equation}
\subsection{The Radion}
There is a potential gap in this scenario that needs to be filled.
In the Randall-Sundrum scenario, it is assumed that the brane
dynamics would naturally lead to branes to be located at a modest
distance apart but it is not explicitly explained that how the
distance between the two branes is established since their
solution is valid for any choice of $r_{0}$. If the distance
between the two branes remains undetermined, when the energy or
the temperature of the universe evolves, the branes will have the
potential to move toward or against to each other. If the brane
separation could change, the universe would not evolve in the way
it is supposed to in 4D and thus the warped 5D universe would not
agree with the cosmological observations. Goldberger and Wise (GW)
\cite{goldberger} did the important research that closed this gap
in the theory by fixing\footnote{In the negative tension brane
where we live all mass scales are exponentially suppressed,
$e^{-kr_{0}}M_{Pl}\sim 1$ TeV. Therefore, $kr_{0}\sim ln
10^{16}\sim 30$ since the Planck scale $M_{Pl}=10^{16}$ TeV.}
$r_{0}\sim 30/k$ without introducing any large finetuning. They
suggested that, in addition to the graviton, there is a massive
particle that lives in the 5D bulk for which the equilibrium
configuration for the field and the branes would involve a modest
brane separation. Denoting the scalar field in the bulk by $\phi$,
the action under consideration will be (see for example
\cite{csaki1} and \cite{csaki})
\begin{equation}\label{action}
\begin{split}
S=&\int d^5x\sqrt{g}M_{*}^{3}R+\int
d^5x\sqrt{g}\frac{1}{2}[(\nabla\phi)^{2}-V(\phi)]-\int
d^4x\sqrt{g_{4}}\lambda_{P}(\phi)\\
&-\int d^4x\sqrt{g_{4}}\lambda_{T}(\phi),
\end{split}
\end{equation}
where the first term is the usual 5D Einstein-Hilbert action and
the second term is the bulk action for the scalar field, while the
next two terms, with $\sqrt{g_{4}}$ being the induced metric on
the branes, are the brane induced potentials for the scalar field
on the Planck and on the TeV branes. We will look for an ansatz of
the background metric again of generic form as in the RS case such
that,
\begin{equation}
ds^{2}=e^{-2A(y)}\eta_{\mu\nu}dx^{\mu}dx^{\nu}-dy^{2}.
\end{equation}
The Einstein equations will be exactly the same as we have derived
for the RS case, except the energy momentum tensor that is derived
from the action of the scalar field. Let us take the variation of
this action with respect to metric term by term. The variation of
the first term is given in eq. \ref{varI}.
%
The metric variation for the second term
\begin{equation}
\int d^{5}x\sqrt{g}[\frac{1}{2}\nabla\phi\nabla\phi-V(\phi)]=\int
d^{5}x\sqrt{g}[\frac{1}{2}g^{MN}\partial_{M}\phi\partial_{N}\phi-V(\phi)],
\end{equation}
reads
\begin{equation}
\begin{split}
\delta\int d^{5}x\sqrt{g}([\frac{1}{2}\nabla\phi\nabla\phi-V(\phi)]&=\int d^{5}x\delta\sqrt{g}[\frac{1}{2}(\partial\phi)^{2}-V(\phi)]\\
&+\int d^{5}x\sqrt{g}\frac{1}{2}\delta g^{MN}\partial_{M}\phi\partial_{N}\phi\\
&=\int d^{5}x(\frac{1}{2}\sqrt{g}g^{KL}\delta g_{KL})[\frac{1}{2}(\partial\phi)^2-V(\phi)]\\
&-\int d^{5}x\sqrt{g}\frac{1}{2}(g^{MK}g^{NL}\delta
g_{KL})\partial_{M}\phi\partial_{N}\phi.
\end{split}
\end{equation}
By making simplification, we obtain the variation as,
\begin{equation}
\begin{split}
\delta\int d^{5}x\sqrt{g}[\frac{1}{2}\nabla\phi\nabla\phi-V(\phi)]=\int d^{5}x\{&\frac{1}{2}g^{KL}[\frac{1}{2}(\partial\phi)^2-V(\phi)]\\
-&\frac{1}{2}\partial^{K}\phi\partial^{L}\phi\}\sqrt{g}\delta
g_{KL}.
\end{split}
\end{equation}
Finally, we will look at the metric variation of the last two
terms which can be written in a more compact form as
\begin{equation}
\int d^{4}x\sqrt{g_{4}}\lambda_{P}(\phi)+\int
d^{4}x\sqrt{g_{4}}\lambda_{T}(\phi)=\int
d^{5}x\frac{\sqrt{g}}{\sqrt{g_{55}}}\sum_{i}\lambda_{i}(\phi)\delta(y-y_{i}),
\end{equation}
where $i$ denote TeV and Planck branes. If we do the metric
variation to the combination of third and the fourth terms we get
\begin{equation}
\begin{split}
\int d^{4}x\sqrt{g_{4}}\lambda_{P}(\phi)+\int
d^{4}x\sqrt{g_{4}}\lambda_{T}(\phi)&=\int
d^5x\delta\sqrt{g}\frac{1}{\sqrt{g_{55}}}\sum_{i}\lambda_{i}(\phi)\delta(y-y_{i})\\
&=\int
d^5x(\frac{1}{2}\delta g_{KL}g^{KL}\sqrt{g})\frac{1}{\sqrt{g_{55}}}\sum_{i}\lambda_{i}(\phi)\delta(y-y_{i})\\
&=\int
d^5x[\frac{1}{2}g^{K}_{\mu}g^{L}_{\nu}g^{\mu\nu}\sum_{i}\lambda_{i}(\phi)\delta(y-y_{i})]\sqrt{g}\delta
g_{KL}.
\end{split}
\end{equation}
Replacing $K$ and $L$ with $M$ and $N$, respectively, we obtain
$T^{MN}$ as
%
%
%
\begin{equation}
\begin{split}
T^{MN}&=\frac{1}{2}g^{MN}[\frac{1}{2}(g_{RS}\partial^{R}\phi\partial^{S}\phi)-V(\phi)]\\
&-\frac{1}{2}\partial^{M}\phi\partial^{N}\phi-\frac{1}{2}
g^{K}_{\mu}g^{L}_{\nu}g^{\mu\nu}\sum_{i}\lambda_{i}(\phi)\delta(y-y_{i}).
\end{split}
\end{equation}
By considering the equality (see Appendix B for its derivation)
\begin{equation}
R_{MN}=\kappa^{2}\tilde{T}_{MN},
\end{equation}
where
\begin{equation}
\tilde{T}_{MN}=T_{MN}-\frac{1}{3}g_{MN}T,
\end{equation}
we get
\begin{equation}\label{firsteinsteinequation}
4A'^{2}-A''=-\frac{2\kappa^{2}}{3}V(\phi_{0})-\frac{\kappa^{2}}{3}
\sum_{i}\lambda_{i}(\phi_{0})\delta(y-y_{i}).
\end{equation}
On the other hand, the $55$ component of the Einstein equation is
obtained as
\begin{equation}
G_{55}=\frac{\kappa^{2}}{2}\phi_{0}'^{2} -\kappa^{2}V(\phi_{0}),
\end{equation}
and  we get
\begin{equation}\label{secondeinsteinequation}
A'^{2}=\frac{\kappa^{2}}{12}\phi_{0}'^{2}-\frac{\kappa^{2}}{6}V(\phi_{0}).
\end{equation}
In these equations $\phi_{0}$ denotes the solution of the scalar
field, which is assumed to be only a function of y:
$\phi=\phi_{0}(y)$. In addition to these two equations, the bulk
scalar equation of motion is found by using
\begin{equation}
\partial_{M}\frac{\partial\mathcal{L}}{\partial(\partial_{M}\phi)}-\frac{\partial\mathcal{L}}{\partial\phi}=0,
\end{equation}
as
\begin{equation}
\partial_{M}(\sqrt{g}g^{MN}\partial_{N}\phi)=-\frac{\partial
V}{\partial\phi}\sqrt{g}.
\end{equation}
We can write the term
$\partial_{M}(\sqrt{g}g^{MN}\partial_{N}\phi)$ also in the form
\begin{equation}
\begin{split}
\partial_{M}(\sqrt{g}g^{MN}\partial_{N}\phi)&=\partial_{\mu}(\sqrt{g}g^{\mu\nu}\partial_{\nu}\phi)+\partial_{5}(\sqrt{g}g^{55}\partial_{5}\phi)\\
&=\partial_{\mu}(e^{-4A}e^{2A}\partial_{\nu}\phi)+\partial_{5}(e^{-4A}(-1)\partial_{5}\phi)\\
&=e^{-2A}\partial_{\mu}\partial_{\nu}\phi-2e^{-2A}\partial_{\mu}A\partial_{\nu}\phi-e^{-4A}\partial_{5}\partial_{5}\phi+4e^{-4A}\partial_{5}A\partial_{5}\phi.
\end{split}
\end{equation}
Since $\phi=\phi_{0}(y)$ and $A=A(y)$ we get
\begin{equation}
-e^{-4A}\phi_0''+4e^{-4A}A'\phi_0'=-\frac{\partial
V}{\partial\phi}e^{-4A}.
\end{equation}
Including the brane tensions we can write
\begin{equation}
\phi''_{0}-4A'\phi'_{0}=\frac{\partial
V(\phi_{0})}{\partial\phi}+\sum_{i}\frac{\partial\lambda_{i}(\phi_{0})}{\partial\phi}\delta(y-y_{i}).
\end{equation}
The metric itself have to be continuous. However, there is no
requirement that the derivative of the metric to be continuous. In
the first Einstein equation given in \ref{firsteinsteinequation}
there exists the explicit delta function term
$-\frac{\kappa^{2}}{3}\sum_{i=P,T}\lambda_{i}(\phi_{0})\delta(y-y_{i})$
at the branes. It seems not the be balanced by anything else
unless there is a jump in the derivative $A'$ at the branes. If
the derivative jumps from $A'(0-\epsilon)$ to $A'(0+\epsilon)$,
this implies that locally $A'$ contains a term of the form
\begin{equation}
A'(y=0)\sim [A'(0+\epsilon)-A'(0-\epsilon)]\varepsilon(y),
\end{equation}
where $\varepsilon(y)$ is the unitstep function. Taking one more
derivative with respect to y we get
\begin{equation}
A''(y=0)\sim [A'(0+\epsilon)-A'(0-\epsilon)]\delta(y).
\end{equation}
Thus, the delta function is proportional to the jump of the
derivative of $A'$. In the same way, the delta function in the
bulk scalar equation of motion will be proportional to the jump of
the derivative $\phi'_{0}$. Therefore, the boundary conditions,
(or jump equations) will be given by
\begin{equation}\label{jumps}
\begin{split}
[A']_{i}&=\frac{\kappa^{2}}{3}\lambda_{i}(\phi_{0}),\\
[\phi'_{0}]_{i}&=\frac{\partial\lambda_{i}(\phi_{0})}{\partial\phi}.
\end{split}
\end{equation}
These are coupled second order differential equations. So, they
are quite hard to solve. Only for specific potentials solution can
be simplified. Defining the function $W(\phi)$ such that
\begin{equation}
\begin{split}
A'&\equiv\frac{\kappa^{2}}{6}W(\phi_{0}),\\
\phi'_{0}&\equiv\frac{1}{2}\frac{\partial W}{\partial\phi},
\end{split}
\end{equation}
and substituting these equations into the 55 component of Einstein
equation we get
\begin{equation}
V(\phi)=\frac{1}{8}(\frac{\partial
W}{\partial{\phi}})^{2}-\frac{\kappa^{2}}{6}W(\phi)^{2}.
\end{equation}
This is called as the consistency equation since when we plug in
the expressions for $A'$ and $\phi'_{0}$ into the Einstein and
scalar equations, we will find that all equations are satisfied.
Then, the jump equations are given by
\begin{equation}
\begin{split}
\frac{1}{2}[W(\phi_{0})]_{i}&=\lambda_{i}(\phi_{0}),\\
\frac{1}{2}[\frac{\partial
W}{\partial\phi}]_{i}&=\frac{\partial\lambda_{i}(\phi_{0})}{\partial\phi}.
\end{split}
\end{equation}
If W were given, the coupled second order differential equations
would be reduced to first order equations that are easy to solve.
We can simply start with a superpotential that will produce $V$
with the required properties. In our case, we would like the bulk
potential to include cosmological constant term (independent of
$\phi$) and mass term (quadratic in $\phi$) in its simplest form.
So we choose
\begin{equation}
W(\phi)=\frac{6k}{\kappa^{2}}-u\phi^{2}.
\end{equation}
where the first term is just one needs for cosmological constant
and the second one is for the mass term. Then,
\begin{equation}
\phi'=\frac{1}{2}\frac{\partial W}{\partial\phi}=-u\phi.
\end{equation}
Substituting $\phi=Ce^{my}$ we get $m=-u$. If we use the boundary
condition that at $y=0$, $\phi=\phi_{P}$, we find that
$C=\phi_{P}$. Then we have
\begin{equation}
\phi=\phi_{P}e^{-uy}.
\end{equation}
From this value of the scalar field at the TeV brane at $r$ it is
determined to be
\begin{equation}
\phi_{T}=\phi_{P}e^{-ur}.
\end{equation}
This means that the radius is no longer arbitrary but given by,
\begin{equation}
r=\frac{1}{u}ln\frac{\phi_{P}}{\phi_{T}}.
\end{equation}
This is the GW mechanism. The background metric will then be
obtained from the equation
\begin{equation}
A'=\frac{\kappa^{2}}{6}W(\phi_{0})=k-\frac{u\kappa^{2}}{6}\phi_{P}^{2}e^{-2uy},
\end{equation}
given by the solution
\begin{equation}\label{Aofy}
A(y)=ky+\frac{\kappa^{2}\phi_{P}^{2}}{12}e^{-2uy},
\end{equation}
where the first term is the usual RS warp factor, while the second
term is the backreaction of the metric to the non-vanishing scalar
field in the bulk. Since we want to generate the right hierarchy
between the Planck and Weak scales we need to ensure that
$kr\sim30$ from which we get
\begin{equation}
\frac{k}{u}\ln(\frac{\phi_{P}}{\phi_{T}})\sim30,
\end{equation}
which is the ratio that will set the hierarchy in the RS model.
Since $\phi_{P}/\phi_{T}$ is constant, so $u$ is kept constant.

\subsection{The coupled field equations and the radion mass}
Once we have established the mechanism for the stabilization of
radion, we realize that the radion is no longer massless. Then,
the question is what will be the value of the radion mass in the
GW stabilized RS model? For this we consider spin-0 fluctuations
of the coupled gravity-scalar system over the background. This can
be parameterized in the following way
\begin{equation}\label{ansatzfluctuation}
\begin{split}
&ds^{2}= e^{-2A(y)-2F(x,y)}\eta_{\mu\nu}dx^{\mu}dx^{\nu}-(1+
G(x,y))^{2}dy^{2},\\
&\phi(x,y)=\phi_{0}(y)+\varphi(x,y).
\end{split}
\end{equation}
It looks like as if there would be three different fluctuations,
namely $F$, $G$, $\varphi$. We will use this ansatz to linearize
Einstein and scalar field equations. Then, some coupled equations
for F, G, and $\varphi$ will be obtained. The linearized Einstein
eq. \ref{RMNtildeT} (see Appendix B for details) equations read
\begin{equation}
\delta R_{MN}=\kappa^{2}\delta\tilde{T}_{MN},
\end{equation}
and by using this equation together with the ansatz given in eq.
\ref{ansatzfluctuation} where
\begin{equation}
\begin{split}
&g_{\mu\nu}=e^{-2A(y)-2F(x,y)}\eta_{\mu\nu},\\
&g^{\mu\nu}=e^{2A(y)+2F(x,y)}\eta^{\mu\nu},\\
&g_{55}=-(1+G(x,y))^{2},\\
&g^{55}=-(1+G(x,y))^{-2},
\end{split}
\end{equation}
the linearized form of the $R_{\mu\nu}$ is obtained as (see
Appendix B for the detailed calculations)
\begin{eqnarray}
\delta R_{\mu\nu}&=&-2\partial_{\mu}\partial_{\nu}F+
\partial_{\mu}\partial_{\nu}G-\eta_{\mu\nu}\Box
F+ e^{-2A}\eta_{\mu\nu}\nonumber\\
&\times&[F''-8A'F'-A'G'-2FA''+8FA'^{2}-2GA''+8GA'^{2}].
\end{eqnarray}
Inspecting the $\delta R_{\mu\nu}$ equation, we realize the
$\partial_{\mu}\partial_{\nu}$ term must vanish since to linear
order in perturbations all the terms in $\tilde{T}_{\mu\nu}$ (see
eq. \ref{deltaTmunuson}) are proportional to $\eta_{\mu\nu}$.
Then, one can immediately conclude that $G=2F$ since
\begin{equation}
\delta R_{\mu\nu}= ......-
2\partial_{\mu}\partial_{\nu}F+\partial_{\mu}\partial_{\nu}G+.......
\end{equation}
Substituting this into the equation for $\delta R_{\mu\nu}$ we
get,
\begin{equation}
\delta R_{\mu\nu}=-\eta_{\mu\nu}\Box F+
\eta_{\mu\nu}e^{-2A}[F''-10A'F'-6FA''+24A'^{2}F].
\end{equation}
Similarly, $\delta R_{\mu 5}$ and $\delta R_{55}$ is obtained as
\begin{equation}
\delta R_{\mu 5}=-3\partial_{\mu}F'+6A'\partial_{\mu}F,
\end{equation}
and
\begin{equation}
\delta R_{55}=-4F''-2e^{2A}\Box F+16A'F'.
\end{equation}
(see Appendix B for details). Now, we will present the linearized
source terms $\delta\tilde{T}_{\mu\nu}$, $\delta\tilde{T}_{\mu5}$,
and $\delta\tilde{T}_{55}$ (see Appendix B for their derivations):
\begin{equation}
\begin{split}
\delta\tilde{T}_{\mu\nu}=\;\;&\frac{1}{3}e^{-2A}\eta_{\mu\nu}
[\varphi V'(\phi_{0})-2V(\phi_{0})F]\\
&+\frac{1}{6}e^{-2A}\eta_{\mu\nu}\sum_{i}[\frac{\partial
\lambda_{i}(\phi_{0})}{\partial\phi}\varphi-4
\lambda_{i}(\phi_{0})F]\delta(y-y_{i}),\\
\delta\tilde{T}_{\mu5}=\;\;&-\frac{1}{2}\phi_{0}'\partial_{\mu}\varphi,\\
\delta\tilde{T}_{55}=\;\;&-\frac{4}{3}V(\phi_{0})F-\frac{1}{3}\varphi
V'(\phi_{0})-\varphi'\phi_{0}'\\
&-\frac{2}{3}\sum_{i}[\frac{\partial
\lambda_{i}(\phi_{0})}{\partial\phi}\varphi+2
\lambda_{i}(\phi_{0})F]\delta(y-y_{i}).
\end{split}
\end{equation}
Finally, we present the linearized scalar field equation for
completeness:
\begin{equation}\label{linearscalarfield}
\begin{split}
e^{2A}\Box\varphi-\varphi''+4A'\varphi'+\frac{\partial^{2}V(\phi_{0})}{\partial\phi^{2}}\varphi=
&-\sum_{i}(\frac{\partial^{2}\lambda_{i}(\phi_{0})}{\partial\phi^{2}}\varphi+
2\frac{\partial\lambda_{i}(\phi_{0})}{\partial\phi}F)\delta(y-y_{i})\\
&-6\phi'_{0}F'-4\frac{\partial V}{\partial\phi}F.
\end{split}
\end{equation}
Using the equation $\delta
R_{\mu5}=\kappa^{2}\delta\tilde{T}_{\mu5}$ we get
\begin{equation}
3(\partial_{\mu}F'-2A'\partial_{\mu}F)=
\kappa^{2}\phi_{0}'\partial_{\mu}\varphi.
\end{equation}
This can be integrated immediately to obtain
\begin{equation}\label{Phi0Phi}
\phi_{0}'\varphi=\frac{3}{\kappa^{2}}(F'-2A'F)+C(y),
\end{equation}
where $C(y)$ is the integration constant. We set this constant as
zero since we require that the fluctuations $F$ and $\varphi$ are
also localized in x. Let us find the boundary conditions for $F$
and $\varphi$ on the two branes. Now, we use the equation $\delta
R_{55}=\kappa^{2}\delta\tilde{T}_{55}$ which leads to,
\begin{equation}
\begin{split}
2e^{2A}\Box
F+4F''-16A'F'&=\kappa^{2}\{2\phi_{0}'\varphi'+\frac{2}{3}V'(\phi_{0})\varphi
+\frac{8}{3}V(\phi_{0})F\\
&+\frac{4}{3}\sum_{i}[\frac{\partial\lambda_{i}(\phi_{0})}{\partial\phi}
+2\lambda_{i}(\phi_{0})F]\delta(y-y_{i})\},
\end{split}
\end{equation}
and, taking in to account the continuity of the metric but not its
derivative, we get the equation
\begin{equation}
[F']=\frac{2\kappa^{2}}{3}\lambda_{i}(\phi_{0})F+\frac{\kappa^{2}}{3}
\frac{\partial\lambda_{i}(\phi_{0})}{\partial\phi}\varphi.
\end{equation}
Here the delta function is proportional to the jump of the
derivative of $F'$,
\begin{equation}
F''(y=0)\sim [F'(0+\epsilon)-F'(0-\epsilon)]\delta(y),
\end{equation}
with
\begin{equation}
F'(y=0)\sim [F'(0+\epsilon)-F'(0-\epsilon)]\varepsilon(y),
\end{equation}
where $\varepsilon(y)$ is a unitstep function. In the same manner,
by using linearized scalar field equation, eq.
\ref{linearscalarfield}, we can take
\begin{equation}
[\varphi']|_{i}=\frac{\partial^{2}\lambda_{i}(\phi_{0})}
{\partial\phi^{2}}\varphi+2\frac{\partial\lambda_{i}}{\partial\phi}F.
\end{equation}
Using the jump equations, eqs. \ref{jumps} and \ref{Phi0Phi}, with
$C(y)=0$ we get
\begin{equation}
\begin{split}
[F']&=\frac{\kappa^{2}}{3}[\phi_{0}']\varphi+2[A']F\\
&=\frac{2\kappa^{2}}{3}\lambda_{i}(\phi_{0})F+
\frac{\kappa^{2}}{3}\frac{\partial\lambda_{i}(\phi_{0})}{\partial\phi}\varphi.
\end{split}
\end{equation}
Thus, it provides no new constraints. Then, only the second
boundary condition must be taken into account. For a convenient
limit
$\frac{\partial^{2}\lambda_{i}(\phi_{0})}{\partial\phi^{2}}\gg1$,
the second boundary condition is simply $\varphi|_{i}=0$. Then, in
this limit the first boundary condition is just
\begin{equation}
(F'-2A'F)|_{i}=0.
\end{equation}
Now a single equation for F is obtained as follows. Considering
the combination $e^{2A}\delta R_{\mu\nu}+\eta_{\mu\nu}\delta
R_{55}$ in the bulk we get
\begin{equation}
e^{2A}\delta R_{\mu\nu}+\eta_{\mu\nu}\delta
R_{55}=3\eta_{\mu\nu}[e^{2A}\Box F+F''-2A'F'+2F(A''-4A'^2)],
\end{equation}
where $A''-4A'^2=\frac{2\kappa^{2}}{3}V(\phi_{0})$. Here, we do
not take into account the $\delta$ terms since we work in the
bulk. The similar combination for the source terms reads
\begin{equation}
e^{2A}\kappa^{2}\delta\tilde{T}_{\mu\nu}+\eta_{\mu\nu}\kappa^{2}
\delta\tilde{T}_{55}=4\kappa^{2}\eta_{\mu\nu}V(\phi_{0})F+2\kappa^{2}
\eta_{\mu\nu}\phi_{0}'\varphi',
\end{equation}
and equating them to get
\begin{equation}\label{e2AboxF}
e^{2A}\Box F+F''-2A'F'=\frac{2\kappa^{2}}{3}\phi_{0}'\varphi'.
\end{equation}
Using eq. \ref{Phi0Phi}, $\varphi'$ can be obtained in terms of F
as
\begin{equation}
\varphi=\frac{3}{\kappa^{2}}\frac{F'-2A'F}{\phi_{0}'}.
\end{equation}
If we take one more derivative with respect to y we get
\begin{equation}
\varphi'=\frac{3}{\kappa^{2}}\frac{(F''-2A'F'-2A''F)\phi_{0}'-(F'-2A'F)
\phi_{0}''}{\phi_{0}'^{2}},
\end{equation}
and by making some arrangements in this equation we obtain,
\begin{equation}
\frac{2\kappa^{2}}{3}\phi_{0}'\varphi'=2(F''-2A'F'-2A''F)-2(F'-2A'F)
\frac{\phi_{0}''}{\phi_{0}'}.
\end{equation}
If we substitute into the eq. \ref{e2AboxF} we get
\begin{equation}\label{radionequation}
F''-2A'F'-4A''F-2\frac{\phi_{0}''}{\phi_{0}'}F'+4A'\frac{\phi_{0}''}{\phi_{0}'}=
e^{2A}\Box F,
\end{equation}
to be solved in the bulk. It is important to note that each
eigenmode $\Box F_{n}=-m_{n}^{2}F_{n}$ to this equation has two
integration constants and one mass eigenvalue. The first constant
corresponds to the overall normalization. while the remaining one
is fixed by the boundary condition on the Planck brane, and the
mass is determined by the boundary condition on the TeV brane.

Now we are ready to calculate the radion mass. In the following we
will show how backreaction generates a non-vanishing mass for the
radion field by using eq. \ref{radionequation}. Substituting
$\phi_{0}(y)=\phi_{P}e^{-uy}$ into the above equation we get
\begin{equation}\label{Fdoubleprime}
F''-2A'F'-4A''F+2uF'-4uA'F+m^{2}e^{2A}F=0,
\end{equation}
where $A(y)$ is given in eq. \ref{Aofy}. The backreaction will be
treated as perturbation such that
\begin{equation}
A(y)=k|y|+\frac{l^{2}}{6}e^{-2u|y|},
\end{equation}
where $l=\kappa\phi_{0}/\sqrt{2}$. Then, we will look the solution
in terms of the perturbative series in $l$. The solution is
written as follows
\begin{equation}
F_{0}=e^{2k|y|}(1+l^{2}f_{0}(y))\qquad;\qquad
m_{r}^{2}=l^{2}\tilde{m}^{2}.
\end{equation}
By substituting the solution into the eq. \ref{Fdoubleprime} and
keeping only the leading terms in $l^{2}$ we get
\begin{equation}\label{f0doubleprime}
f_{0}''+2(k+u)f_{0}'=-\tilde{m}^{2}e^{2k|y|}-\frac{4}{3}(k-u)u
e^{-2u|y|}.
\end{equation}
By solving this equation we get
\begin{equation}\label{f0prime}
f_{0}'(y)=Ce^{-2(k+u)|y|}-\frac{\tilde{m}^{2}}{2(2k+u)}e^{2k|y|}-2\frac{(k-u)u}{3k}e^{-2u|y|}.
\end{equation}
If we make the same substitution in the boundary condition
$F'-2A'F=0$ we get
\begin{equation}
l^{2}f_{0}'+\frac{2}{3}ul^{2}e^{-2u|y|}+\frac{4}{3}ul^{4}
f_{0}e^{-2u|y|}=0,
\end{equation}
and by keeping only the leading terms in $l^{2}$ we obtain
\begin{equation}
f_{0}'+\frac{2}{3}ue^{-2u|y|}=0.
\end{equation}
at the boundary, where $|y|=r_0$. If we substitute this into the
eq. \ref{f0doubleprime}, $\tilde{m}$ is obtained as
\begin{equation}
\tilde{m}^{2}=\frac{4}{3}\frac{2k+u}{k}u^{2}e^{-2(u+k)r_0},
\end{equation}
and the radion mass reads
\begin{equation}
m^{2}_{radion}=\frac{4l^{2}(2k+u)u^{2}}{3k}e^{-2(u+k)r_0}.
\end{equation}
\subsection{Coupling to SM fields and the normalized radion field}
In the previous section, including the backreaction, a mass scale
$\mathcal{O}(TeV^{2})$ is obtained for the radion mass. Then, the
wavefunction can be written as
\begin{equation}
F_{0}(x,y)=e^{2k|y|}(1+l^{2}f_{0}(y))R(x),
\end{equation}
where $f_{0}(y)$ obtained using the integral in the eq.
\ref{f0prime}. Based on the assumption $l^{2}\ll 1$, we see that
the backreaction induces a small correction to the unperturbed
wavefunction. So for purposes of determining the coupling of the
radion to the TeV brane, we can use the unperturbed wavefunction,
namely $F(x,y)=e^{2k|y|}R(x)$. Let us try to find the coefficient
of $(\partial F)^{2}$ term in the Lagrangian $\sqrt{g}R$ with
$\sqrt{g}=e^{-4A-4F}(1+2F)$, so that we are able to write the
normalized wave function $R(x)$. $R$ can be splited as
$R=g^{\mu\nu}R_{\mu\nu}+g^{55}R_{55}$. Then, using the
$R_{\mu\nu}$  (see Appendix B for details) read
\begin{eqnarray}
R_{\mu\nu}&=&-2\partial_{\mu}\partial_{\nu}F+2\frac{\partial_{\mu}
\partial_{\nu}F}{1+2F}-\eta_{\mu\nu}\Box
F+\eta_{\mu\nu}\frac{e^{-2A-2F}}{(1+2F)^{2}}\nonumber\\
&\times&[A''+F''-4(A'+F')^{2}-\frac{2(A'+F')F'}{(1+2F)}]\nonumber\\
&+&4\frac{\partial_{\mu}F\partial_{\nu}F}{1+2F}-2\eta_{\mu\nu}\frac{(\partial
F)^{2}}{1+2F}-2\partial_{\mu}F\partial_{\nu}F+2\eta_{\mu\nu}(\partial
F)^{2}.
\end{eqnarray}
and multiplying by $\sqrt{g}g^{\mu\nu}$ from left, for the first
term in R, including $(\partial F)^{2}$, we get
\begin{equation}
\begin{split}
\sqrt{g}g^{\mu\nu}R_{\mu\nu}&\cong e^{-2A}(1-2F)(-4(1+2F)F\Box
F-4(1+2F)\Box F+
6(1+2F)(\partial F)^{2}\\
&-4(\partial F)^{2}...),
\end{split}
\end{equation}
where $F\Box
F=F\partial_{\mu}\partial^{\mu}F=\partial_{\mu}(F\partial^{\mu}F)-(\partial
F)^{2}$. Substituting this quality into the above equation we
obtain,
\begin{equation}
\begin{split}
g^{\mu\nu}R_{\mu\nu}&\cong
e^{-2A}(1-2F)(-4[\partial_{\mu}(F\partial^{\mu}F)
-(\partial F)^{2}](1+2F)\\
&-4(\partial_{\mu}(F\partial^{\mu}F)-(\partial
F)^{2}) F-8[\partial_{\mu}(F\partial^{\mu}F)-(\partial F)^{2}]\\
&-2(\partial F)^{2}+...).
\end{split}
\end{equation}
Similarly, using the equation for $R_{55}$ (see Appendix B for
details) below
\begin{equation}
\begin{split}
R_{55}&=-4(A''+F'')+4e^{2A+2F}(1+2F)(\partial F)^{2}\\
&-2e^{2A+2F}\Box F(1+2F)+\frac{8(A'+F')F'}{1+2F}+4(A'+F')^{2},
\end{split}
\end{equation}
the second term in $R$ is obtained as
\begin{equation}
\begin{split}
\sqrt{g}g^{55}R_{55}&=\frac{-e^{-4A-4F}(1+2F)}{(1+2F)^{2}}R_{55}\\
&=e^{-4A-4F}(1+2F)\Big(\frac{4(A''+F'')}{(1+2F)^{2}}-\frac{4e^{2A+2F}(\partial
F)^{2}}{1+2F}+\frac{2e^{2A+2F}\Box F}{1+2F}\\
&-\frac{8(A'+F')F'}{(1+2F)^{3}}+\frac{4(A'+F')^{2}}{(1+2F)^{2}}\Big)\\
&\cong e^{-4A}(1-2F)\Big(4(A''+F'')(1-4F)-4e^{2A}(1-4F^2)(\partial F)^{2}\\
&+2e^{2A}(1-4F^2)(\partial_{\mu}(F\partial^{\mu}F)-(\partial
F)^{2}) F-8(A'+F')F'(1-6F)\\
&-4(A'+F')^{2}(1-4F)\Big).
\end{split}
\end{equation}
So, one can simply find the coefficient of $(\partial F)^{2}$ in
$\sqrt{g}$R in linear order as $e^{-2A}6$. Then, a straightforward
calculation gives
\begin{equation}
-M^{3}_{*}\int dy\sqrt{g}R\supset6M^{3}_{*}(\partial R)^{2}\int
e^{-2A}e^{4k|y|}=\frac{6M^{3}_{*}}{k}(e^{2kr_{0}}-1)(\partial
R)^{2}.
\end{equation}
Therefore, the normalized radion $r(x)$ is
$R(x)=r(x)e^{-kr{0}}/\sqrt{6}M_{Pl}$, which is obtained by using
$M^{3}_{*}/k=M_{Pl}^{2}/2$.

\newpage
\chapter{LEPTON FLAVOR VIOLATING RADION DECAYS IN THE
RANDALL-SUNDRUM SCENARIO} The hierarchy problem between weak and
Planck scales could be explained by introducing the extra
dimensions. One of the possibility is to pull down the Planck
scale to TeV range by considering the compactified extra
dimensions of large size \cite{arkani, arkani1}. The assumption
that the extra dimensions are at the order of submilimeter
distance, for two extra dimensions, the hierarchy problem in the
fundamental scales could be solved and the true scale of quantum
gravity would be no more the Planck scale but it is of the order
of EW scale. This is the case that the gravity spreads over all
the volume including the extra dimensions, however, the matter
fields are restricted in four dimensions, so called 4D brane.
Another possibility, which is based on the non-factorizable
geometry, is introduced by Randall and Sundrum \cite{RS, RS1} and,
in this scenario, the extra dimension is compactified to $S^1/Z_2$
orbifold with two 4D brane boundaries. Here, the gravity is
localized in one of the boundary, so called the Planck brane,
which is away from another boundary, the TeV brane where we live.
The size of extra dimension is related to the vacuum expectation
of a scalar field and its fluctuation over the expectation value
is called the radion field (see section 2 for details). The radion
in the RS1 model has been studied in several works in the
literature \cite{goldberger}-\cite{mahanta} (see \cite{csaki1} for
extensive discussion).

In the present work, \cite{Makale}  we study the possible LFV
decays of the radion field $r$. The LFV interactions exist at
least in one loop level in the extended SM, so called $\nu$SM,
which is constructed by taking neutrinos massive and by permitting
the lepton mixing mechanism \cite{Pontecorvo}. Their negligibly
small Brs stimulate one to go beyond and they are worthwhile to
examine since they open a window to test new models and to ensure
considerable information about the restrictions of the free
parameters, with the help of the possible accurate measurements.
The LFV interactions are carried by the FCNCs and in the SM with
extended Higgs sector (the multi Higgs doublet model) they can
exist at tree level. Among multi Higgs doublet models, the 2HDM is
a candidate for the lepton flavor violation. In this model, the
lepton flavor violation is driven by the new scalar Higgs bosons
$S$, scalar $h^0$ and pseudo scalar $A^0$, and it is controlled by
the Yukawa couplings appearing in lepton-lepton-S vertices.

Here, we predict the BRs of the LFV $r$ decays in the 2HDM, in the
framework of the RS1 scenario. We observe that the BRs of the
processes we study are at most of the order of $10^{-8}$, for the
small values of radion mass $m_r$ and their sensitivities to $m_r$
decrease with the increasing values of $m_r$. Among the LFV decays
we study, the $r\rightarrow \tau^{\pm}\, \mu^{\pm}$ decay would be
the most suitable one to measure its BR.
\section{The LFV RS1 model radion decay in the 2HDM}
The RS1 model is an interesting candidate in order to explain the
well known hierarchy problem. It is formulated as two 4D surfaces
(branes) in 5D world in which the extra dimension is compactified
into $S^1/Z_2$ orbifold. In this model, the SM fields are assumed
to live on one of the brane, so called the TeV brane. On the other
hand, the gravity peaks near the other brane, so called the Planck
brane and extends into the bulk with varying strength. Here, 5D
cosmological constant is non vanishing and both branes have equal
and opposite tensions so that the low energy effective theory has
flat 4D spacetime. The metric of such 5D world reads
\begin{eqnarray}
ds^2=e^{-2\,A(y)}\,\eta_{\mu\nu}\,dx^\mu\,dx^\nu-dy^2\, ,
\label{metric1}
\end{eqnarray}
where $A(y)=k\,|y|$, k is the bulk curvature constant, y is the
extra dimension parametrized as $y=R\,\theta$. The exponential
factor $e^{-k\,L}$ with $L=R\,\pi$, is the warp factor which
causes that all the mass terms are rescaled in the TeV brane. With
a rough estimate  $L\sim 30/k$, all mass terms are brought down to
the TeV scale. The size $L$ of extra dimension is related to the
vacuum expectation of the field $L(x)$ and its fluctuation over
the expectation value is called the radion field $r$. In order to
avoid the violation of equivalence principle, $L(x)$ should
acquire a mass and, to stabilize $r$, a mechanism was proposed by
Goldberger and Wise \cite{goldberger}, by introducing a potential
for $L(x)$. Finally the metric in 5D is defined as
\cite{charmousis}.
\begin{eqnarray}
ds^2=e^{-2\,A(y)-2\,F(x)}\,\eta_{\mu\nu}\,dx^\mu\,dx^\nu- (1+2\,
F(x))\,dy^2\, , \label{metric2}
\end{eqnarray}
where the radial fluctuations are carried by the scalar field
$F(x)$,
\begin{eqnarray}
F(x)=\frac{1}{\sqrt{6}\,M_{Pl}\,e^{-k\,L}}\, r(x)\, . \label{Fx}
\end{eqnarray}
Here the field $r(x)$ is the normalized radion field (see
\cite{csaki}). At the orbifold point $\theta=\pi$ (TeV brane) the
induced metric reads,
\begin{eqnarray}
g^{ind}_{\mu\nu}=e^{-2\,A(L)-2\frac{\gamma}{v}\,r(x)}\,\eta_{\mu\nu}
\, . \label{metricind}
\end{eqnarray}
Here the parameter $\gamma$ reads
$\gamma=\frac{v}{\sqrt{6}\,\Lambda}$ with
$\Lambda=M_{Pl}\,e^{-k\,L}$ and $v$ is the vacuum expectation
value of the SM Higgs boson. The radion is the additional degree
of freedom of the 4D effective theory and we study the possible
LFV decays of this field.

\begin{figure}
\begin{tabular}{p{6cm} p{5cm}} \vskip -7.9truein
\parbox[b]{16cm}{\epsffile{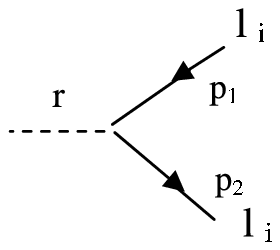}} &
\raisebox{5.ex}{$\frac{-i\,\gamma}{v}\,\left[\frac{3}{2}\,
(p_1\!\!\!\!\!/+p_2\!\!\!\!\!/\,)-4\,m_{l_i}\right]$}
\\ \textbf{(a)} \\ \\ \vskip -7.0truein
\parbox[b]{6cm}{\epsffile{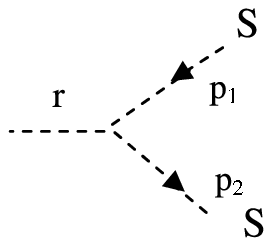}} & \\  \vskip 0.2truein \hspace{6.3cm}
\raisebox{5.ex} {$\frac{-2\,i\,\gamma}{v}\, (p_1.p_2-m_S^2)$}
\\ \textbf{(b)}\\ \\ \vskip -7.1truein
\parbox[b]{6cm}{\epsffile{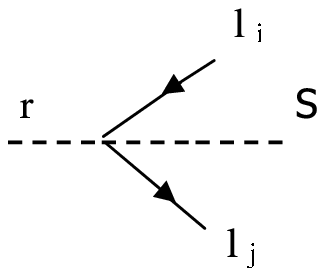}} & \vspace{0.7cm}\\ \vskip -0.4truein
\hspace{4.7cm}
\raisebox{5.ex}{$(S=h^0)\,\,\,\,\,\,\,\frac{4\,i\gamma}{2\sqrt{2}\,v}
\left[(\xi_{ij}^{E}+
\xi_{ji}^{E*})+(\xi_{ij}^{E}-\xi_{ji}^{E*})\gamma_5\right]$}
\vspace{0.7cm}\\ \vskip -0.5truein \hspace{4.7cm}
\raisebox{5.ex}{$(S=A^0)\,\,\,\,\,\,\,\frac{-4\,\gamma}{2\sqrt{2}\,v}\left[
(\xi_{ij}^{E}-\xi_{ji}^{E*})+(\xi_{ij}^{E}+\xi_{ji}^{E*})\gamma_5\right]$}
\textbf{(c)}\\ \\ \\ \vskip -7.3truein
\parbox[b]{6cm}{\epsffile{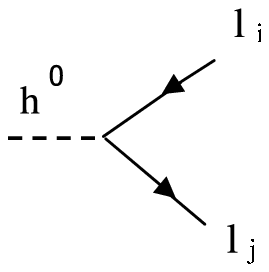}} & \vspace{0.5cm}\\ \hspace{6.3cm}
\raisebox{5.ex}{$\frac{-i}{2\sqrt{2}}\left[(\xi_{ij}^{E}+
\xi_{ji}^{E*})+(\xi_{ij}^{E}-\xi_{ji}^{E*})\gamma_5\right]$}
\vspace{-1.5cm} \textbf{(d)} \\ \\
\\ \\ \\ \\
\vskip -7.7truein
\parbox[b]{6cm}{\epsffile{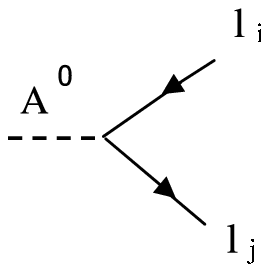}} & \vspace{-1.5cm}\\ \vskip -0.4truein
\hspace{6.3cm}
\raisebox{5.ex}{$\frac{1}{2\sqrt{2}}\left[(\xi_{ij}^{E}-
\xi_{ji}^{E*})+(\xi_{ij}^{E}+\xi_{ji}^{E*})\gamma_5\right]$}  \textbf{(e)}\\ \\
\\ \\ \\ \\
\end{tabular}
%
\label{figvert1} \caption{The vertices used in the present work.}
\end{figure}

The FCNCs at tree level can exist in the 2HDM and they induce the
FV interactions with large BRs.  The FV $r$ decays, $r\rightarrow
l_1^- l_2^+$, can exist at least in one loop level in the
framework of the 2HDM. The part of action which carries the
interaction, responsible for the LFV processes reads
\begin{eqnarray}
{\cal{S}}_{Y}&=& \int d^4x \sqrt{-g^{ind}}\Bigg(
\eta^{E}_{ij} \bar{l}_{i L} \phi_{1} E_{j R}+ \xi^{E}_{ij}
\bar{l}_{i L} \phi_{2} E_{j R} + h.c. \Bigg) \,\,\, ,
\label{lagrangian1}
\end{eqnarray}
where $L$ and $R$ denote chiral projections $L(R)=1/2(1\mp
\gamma_5)$, $\phi_{i}$ for $i=1,2$, are two scalar doublets, $l_{i
L}$ ($E_{j R}$) are lepton doublets (singlets), $\xi^{E}_{ij}$
\footnote{In the following, we replace $\xi^{E}$ with
$\xi^{E}_{N}$ where "N" denotes the word "neutral".} and
$\eta^{E}_{ij}$, with family indices $i,j$ , are the Yukawa
couplings and $\xi^{E}_{ij}$ induce the FV interactions in the
leptonic sector. Here $g^{ind}$ is the determinant of the induced
metric on the TeV brane where the 2HDM particles live. Here, we
assume that the Higgs doublet $\phi_1$ has a non-zero vacuum
expectation value to ensure the ordinary masses of the gauge
fields and the fermions, however, the second doublet has no vacuum
expectation value, namely, we choose the doublets $\phi_{1}$ and
$\phi_{2}$ and their vacuum expectation values as
\begin{eqnarray}
\phi_{1}=\frac{1}{\sqrt{2}}\left[\left(\begin{array}{c c}
0\\v+H^{0}\end{array}\right)\; + \left(\begin{array}{c c} \sqrt{2}
\chi^{+}\\ i \chi^{0}\end{array}\right) \right]\, ;
\phi_{2}=\frac{1}{\sqrt{2}}\left(\begin{array}{c c} \sqrt{2}
H^{+}\\ H_1+i H_2 \end{array}\right) \,\, , \label{choice}
\end{eqnarray}
and
\begin{eqnarray}
<\phi_{1}>=\frac{1}{\sqrt{2}}\left(\begin{array}{c c}
0\\v\end{array}\right) \,  \, ; <\phi_{2}>=0 \,\, .\label{choice2}
\end{eqnarray}
This choice ensures that the mixing between neutral scalar Higgs
bosons is switched off and it would be possible to separate the
particle spectrum so that the SM particles are collected in the
first doublet and the new particles in the second one
\footnote{Here $H_1$ ($H_2$) is the well known mass eigenstate
$h^0$ ($A^0$).}. The action in eq. (\ref{lagrangian1}) is
responsible for the tree level $S-l_1-l_2$ ($l_1$ and $l_2$ are
different flavors of charged leptons, $S$ denotes the neutral new
Higgs boson, $S=h^0,A^0$) interaction (see Fig. \ref{figvert1}-d,
e) and the four point $r-S-l_1-l_2$ interaction (see Fig.
\ref{figvert1}-c) where $r$ is the radion field. The latter
interaction is coming from the determinant factor
$\sqrt{-g^{ind}}=e^{-4\,A(L)-4\frac{\gamma}{v}\,r(x)}$. Notice
that the term $e^{-4\,A(L)}$ in $\sqrt{-g^{ind}}$ is embedded into
the redefinitions of the fields on the TeV brane, namely, they are
warped as $S\rightarrow e^{A(L)}\,S_{warp}$, $l\rightarrow
e^{3\,A(L)/2}\,l_{warp}$ and in the following we use warped fields
without the $warp$ index.

On the other hand, the part of new scalar action
\begin{eqnarray}
{\cal{S}}_{2}&=& \int d^4x \sqrt{-g^{ind}} \Bigg( g^{ind\,\,
\mu\nu}\, (D_\mu\,\phi_2)^\dagger \, D_\nu\, \phi_2-m_S^2\,
\phi_2^\dagger \,\phi_2\Bigg)  \label{lagrangian2}
\end{eqnarray}
leads to
\begin{eqnarray}
{\cal{S'}}_{2}\!\!\!\!&=&\!\! \!\!\frac{1}{2}\,\int d^4x \,\Bigg\{
e^{-2\frac{\gamma}{v}\, r}\eta^{\mu\nu}\,
\Big(\partial_\mu\,h^0\partial_\nu\,h^0+\partial_\mu\,A^0\partial_\nu\,A^0
\Big)\nonumber\\
&-&e^{-4\frac{\gamma}{v}\,
r}\,(m_{h^0}^2\,h^0\,h^0+m_{A^0}^2\,A^0\,A^0)\Bigg\} \, ,
\label{lagrangian2a}
\end{eqnarray}
which carries the $S-S-r$ interaction\footnote{In general, there
is no symmetry which forbids the curvature-scalar interaction,
\begin{eqnarray}
{\cal{S}}_{\xi}=\int d^4x
\,\sqrt{-g^{ind}}\,\xi\,{\cal{R}}\,H^\dagger\,H \, ,
\label{curvscalarmix}
\end{eqnarray}
where $\xi$ is a restricted positive parameter and $H$ is the
Higgs scalar field \cite{csaki1, csaki, Kingman}. This interaction
results in the radion-(SM or new) Higgs mixing which can bring a
sizeable contribution to the physical quantities studied. Here, we
assume that there is no mixing between first and second doublet
and only the first Higgs doublet has vacuum expectation value.
Therefore, we choose that there exists a mixing between the radion
and the SM Higgs field, but not between the radion and the new
Higgs fields. This is the case that the lepton flavor violation is
not affected by the mixing since the SM Higgs field is not
responsible for the FCNC current at tree level.} (see Fig.
\ref{figvert1} 1-b).

Finally, the interaction of leptons with the radion field is
carried by the action (see \cite{das})
\begin{eqnarray}
{\cal{S}}_{3}&=& \int d^4x \sqrt{-g^{ind}}\, \Bigg( g^{ind\,\,
\mu\nu}\, \bar{l}\,\gamma_\mu\,i\,D_\nu\,l-m_l\,\bar{l}\,l\Bigg)
\, , \label{lagrangian3}
\end{eqnarray}
where
\begin{eqnarray}
D_\mu \,l=\partial_\mu\,l+\frac{1}{2}\,w_\mu^{ab}\,\Sigma_{ab}\,l
\, , \label{Dmuf}
\end{eqnarray}
with $\Sigma_{ab}=\frac{1}{4}[\gamma_a,\gamma_b]$. Here
$w_\mu^{ab}$ is the spin connection and, by using the vierbein
fields $e^a_\mu$, it can be calculated (linear in $r$) as
\begin{eqnarray}
w_\mu^{ab}=-\frac{\gamma}{v}\partial_\nu\,r\,(e^{\nu
b}\,e^a_\mu-e^{\nu a}\,e_\mu^b )\, , \label{vierbein}
\end{eqnarray}
(see Appendix C for details). Notice that the vierbein fields are
the square root of the metric and they satisfy the relation
\begin{eqnarray}
e^\mu_a\,e^{a \nu}=g^{ind\,\, \mu\nu}\, . \label{metrvierbein}
\end{eqnarray}
Using eqs. (\ref{lagrangian3})-(\ref{metrvierbein}), one gets the
part of the action which describes the tree level $l-l-r$
interaction  (see Fig. \ref{figvert1}-a) as
\begin{eqnarray}
{\cal{S'}}_{3}&=&\int d^4x \,\Bigg\{ -3\frac{\gamma}{v}\, r\,
\bar{l}\,i\,\partial\!\!\!/ l-3\frac{\gamma}{2\,v}\,
\bar{l}\,i\,\partial\!\!\! / r\, l+ 4\,\frac{\gamma}{v}\, m_l\,r\,
\bar{l}\,l \Bigg\} \, . \label{lagrangian3a}
\end{eqnarray}

\begin{figure}[htb]
\vskip -7.0truein \centering \epsfxsize=9.5in
\leavevmode\epsffile{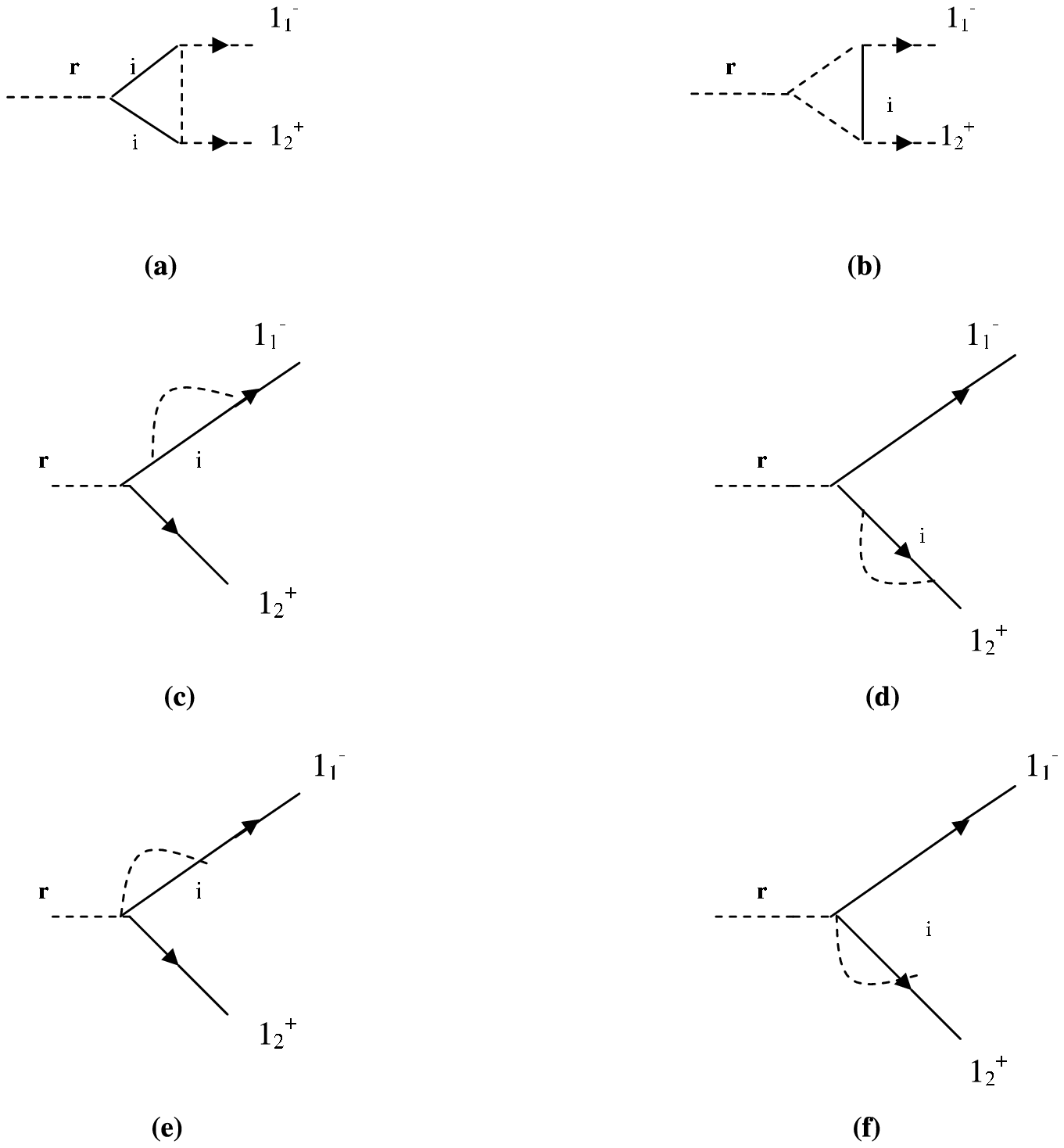} \vskip 3.5truein \caption{One
loop diagrams contribute to $r\rightarrow l_1^-\,l_2^+$ decay due
to the neutral Higgs bosons $h_0$ and $A_0$ in the 2HDM. $i$
represents the internal lepton, $l_1^-$ ($l_2^+$) outgoing lepton
(anti lepton), internal dashed line the $h_0$ and $A_0$ fields.}
\label{figselfvert}
\end{figure}

Now, we are ready to calculate the matrix element for the LFV
radion decay. The decay of the radion to leptons with different
flavors exits at least in one loop order, with the help of
internal new Higgs bosons $S=h^0, A^0$. The possible vertex and
self energy diagrams are presented in Fig. \ref{figselfvert}.
After addition of all these diagrams, the divergences which occur
in the loop integrals are eliminated and the matrix element square
for this decay is obtained as
\begin{eqnarray}
|M|^2= 2\Big( m_r^2 -(m_{l_1^-}+m_{l_2^+})^2\Big)\,|A|^2 \, ,
\label{Matrx2}
\end{eqnarray}
where
\begin{eqnarray}
A= f^{self}_{h^0}+f^{self}_{A^0}+f^{vert}_{h^0}+f^{vert}_{A^0}+
f^{vert}_{h^0 h^0}+f^{vert}_{A^0 A^0} \, , \label{fh0A0}
\end{eqnarray}
and their explicit expressions are given by
\begin{eqnarray}
f^{self}_{h^0}&=& \frac{\gamma}{128\,v\,\pi^2\,(w'_h-w_h) }
\int_0^1\,dx\, m_{h^0}\,
 \Bigg \{\Big( \eta_i^V\,(x-1)\,w_h-\eta_i^+\,z_{ih} \Big)\, \nonumber\\
 &\times& \Big(
3\,w'_h-5\,w_h \Big)\,\,ln\,\frac{L^{self}_{1,
h^0}\,m^2_{h^0}}{\mu^2}\nonumber \\
&+& \Big( \eta_i^V\,(x-1)\,w'_h-\eta_i^+\,z_{ih} \Big)\, \Big(
5\,w'_h-3\,w_h \Big)\,\,ln\,\frac{L^{self}_{2,
h^0}\,m^2_{h^0}}{\mu^2}\Bigg \} \, , \nonumber \\
f^{self}_{A^0}&=& \frac{\gamma}{128\,v\,\pi^2\,(w'_A-w_A) }
\int_0^1\,dx\ ,m_{A^0}\,
 \Bigg \{\Big( \eta_i^V\,(x-1)\,w_A+\eta_i^+\,z_{iA} \Big)\, \nonumber\\
 &\times& \Big(
3\,w'_A-5\,w_A \Big)\,\,ln\,\frac{L^{self}_{1,
A^0}\,m^2_{A^0}}{\mu^2}\nonumber \\
&+& \Big( \eta_i^V\,(x-1)\,w'_A+\eta_i^+\,z_{iA} \Big)\, \Big(
5\,w'_A-3\,w_A \Big)\,\,ln\,\frac{L^{self}_{2,
A^0}\,m^2_{A^0}}{\mu^2}\Bigg \} \, , \nonumber \\
f^{vert}_{h^0}&=& \frac{\gamma}{128\,v\,\pi^2} \int_0^1\,dx\,
\int_0^{1-x} \, dy \, \Bigg\{
\frac{m_{h^0}}{L^{ver}_{h^0}}\,\Bigg[ \eta_i^V
\Bigg( 3\, z_{rh}^2\,\Big(y\,(1-y)\,w'_h+x^2\,(4\,y-1)\,w_h \nonumber \\
&+& x\,((1-3\,y)\,w_h+y\,(4\,y-3)\,w'_h)\Big)
+ 5\, z_i^2\, \Big( (2\,y-1)\,w'_h+(2\,x-1)\,w_h \Big)\nonumber
\\&-&
3\, (x+y-1)\,\Big(x\,(4\,x-3)\,w_h^3+y\,(4\,y-3)\,w_h^{'3}\Big) \nonumber \\
&-& 3\, w_h\, w'_h \,(x+y-1)\,\Big(
 (1-y+x\,(4\,y-2))\,w_h+  (1-2\,y+x\,(4\,y-1))\,w'_h \Big)\nonumber \\
&+& 3\, (x+y-1)\,\Big((2\,x-1)\,w_h+(2\,y-1)\,w'_h \Big)
 \Bigg)\nonumber \\
&+& \eta_i^+ \,z_{ih}\,\Bigg(
(x+y-1)\,\Big(-4+2\,w'_h\,w_h+w_h^{'2}\,(8\,y-3)+w_h^{2}\,(8\,x-3)\,\Big)
\nonumber \\
&-& \Big(8\,z_{ih}^2+z_{rh}^2\,((8\,y-3)\,x-3\,y)\Big)\,\Bigg)
\Bigg]
\nonumber \\
&-& m_{h^0}\,ln\,\frac{L^{ver}_{h^0}\,m^2_{h^0}}{\mu^2}\,\Bigg(
9\, \eta_i^V \Big(w'_h\,(2\,y-1)+w_h\,(2\,x-1)\Big)- 8\,\eta_i^+
\,z_{ih} \Bigg)
\Bigg \}\, , \nonumber
\end{eqnarray}
\begin{eqnarray}
f^{vert}_{A^0}&=& \frac{\gamma}{128\,v\,\pi^2} \int_0^1\,dx\,
\int_0^{1-x} \, dy \, \Bigg\{
\frac{m_{A^0}}{L^{ver}_{A^0}}\,\Bigg[ \eta_i^V
\Bigg( 3\, z_{rA}^2\,\Big(y\,(1-y)\,w'_A+x^2\,(4\,y-1)\,w_A \nonumber \\
&+& x\,((1-3\,y)\,w_A+y\,(4\,y-3)\,w'_A)\Big)
+ 5\, z_i^2\, \Big( (2\,y-1)\,w'_A+(2\,x-1)\,w_A \Big)\nonumber
\\&-&
3\, (x+y-1)\,\Big(x\,(4\,x-3)\,w_A^3+y\,(4\,y-3)\,w_A^{'3}\Big) \nonumber \\
&-& 3\, w_A\, w'_A \,(x+y-1)\,\Big(
 (1-y+x\,(4\,y-2))\,w_A+  (1-2\,y+x\,(4\,y-1))\,w'_A \Big)\nonumber \\
&+& 3\, (x+y-1)\,\Big((2\,x-1)\,w_A+(2\,y-1)\,w'_A \Big)
 \Bigg)\nonumber \\
&+& \eta_i^+ \,z_{iA}\,\Bigg(
(x+y-1)\,\Big(-4+2\,w'_A\,w_A+w_A^{'2}\,(8\,y-3)+w_A^{2}\,(8\,x-3)\,\Big)
\nonumber \\
&-& \Big(8\,z_{iA}^2+z_{rA}^2\,((8\,y-3)\,x-3\,y)\Big)\,\Bigg)
\Bigg]
\nonumber \\
&-& m_{A^0}\,ln\,\frac{L^{ver}_{A^0}\,m^2_{A^0}}{\mu^2}\,\Bigg(
9\, \eta_i^V \Big(w'_A\,(2\,y-1)+w_A\,(2\,x-1)\Big)+ 8\,\eta_i^+
\,z_{iA} \Bigg)
\Bigg \}\, , \nonumber \\
f^{vert}_{h^0\,h^0}&=& \frac{\gamma}{64\,v\,\pi^2} \int_0^1\,dx\,
\int_0^{1-x} \, dy \, \Bigg \{
\frac{m_{h^0}}{L^{ver}_{h^0\,h^0}}\,\Bigg[ \eta_i^V
\Bigg( z_{rh}^2\,\Big( y-1+x\,(1-4\,y)\Big)\,(x\,w_h+y \,w'_h) \nonumber \\
&+&
y\,(x+y-1)\,w'_h\,\Big( (4\,x-1)\,w_h^2+(4\,y-1)\,w_h^{'2} \Big) \nonumber \\
&+& w_h^3\,x\,(x+y-1)\,(4\,x-1)+
(x+y-1)\,\Big(2\,y\,w'_h+x\,w_h\,(2+w_h^{'2}\,(4\,y-1))\,\Big)
 \Bigg) \nonumber \\
&+& \eta_i^+ \Bigg(
(x+y-1)\,z_{ih}\,\Big((4\,y-1)\,w_h^{'2}+(4\,x-1)\,w_h^2+2\Big)\nonumber\\&-&
z_{ih}\,z_{rh}^2\,\Big((4\,y-1)\,x-y+1\Big)\Bigg)\,\nonumber
\Bigg] \nonumber \\
&-&
m_{h^0}\,ln\,\frac{L^{ver}_{h^0\,h^0}\,m^2_{h^0}}{\mu^2}\,\Bigg(
\, \eta_i^V \Big(w'_h\,(1-6\,y)+w_h\,(1-6\,x)\Big)- 4\,\eta_i^+
\,z_{ih} \Bigg)
\Bigg \}\, , \nonumber
\end{eqnarray}
\begin{eqnarray}
f^{vert}_{A^0\,A^0}&=& \frac{\gamma}{64\,v\,\pi^2} \int_0^1\,dx\,
\int_0^{1-x} \, dy \, \Bigg \{
\frac{m_{A^0}}{L^{ver}_{A^0\,A^0}}\,\Bigg[ \eta_i^V
\Bigg( z_{rA}^2\,\Big( y-1+x\,(1-4\,y)\Big)\,(x\,w_A+y \,w'_A) \nonumber \\
&+&
y\,(x+y-1)\,w'_A\,\Big( (4\,x-1)\,w_A^2+(4\,y-1)\,w_A^{'2} \Big) \nonumber \\
&+& w_A^3\,x\,(x+y-1)\,(4\,x-1)+
(x+y-1)\,\Big(2\,y\,w'_A+x\,w_A\,(2+w_A^{'2}\,(4\,y-1))\,\Big)
 \Bigg) \nonumber \\
&+& \eta_i^+ \Bigg(
(x+y-1)\,z_{iA}\,\Big((4\,y-1)\,w_A^{'2}+(4\,x-1)\,w_A^2+2\Big)\nonumber\\
&-& z_{iA}\,z_{rA}^2\,\Big((4\,y-1)\,x-y+1\Big)\Bigg)\,\nonumber
\Bigg] \nonumber \\
&-&
m_{A^0}\,ln\,\frac{L^{ver}_{A^0\,A^0}\,m^2_{A^0}}{\mu^2}\,\Bigg(
\, \eta_i^V \Big(w'_A\,(1-6\,y)+w_A\,(1-6\,x)\Big)+ 4\,\eta_i^+
\,z_{iA} \Bigg)
\Bigg \}\, , \label{fVAME}
\end{eqnarray}
where
\begin{equation}
\begin{split}
&L^{self}_{1, h^0 (A^0)}=1+x^2\,w_{h(A)}^{2}+x\,(z^2_{ih(iA)}-
w_{h(A)}^{2}-1)\, , \\
&L^{self}_{2, h^0 (A^0)}=1+x^2\,w_{h(A)}^{'2}+x\,(z^2_{ih
(iA)}-w_{h(A)}^{'2}-1)
\, , \\
&L^{ver}_{h^0
(A^0)}=x^2\,w_{h(A)}^{2}+(y-1)\,(w_{h(A)}^{'2}\,y-1)+x\,(y\,w_{h(A)}^{'2}+
(y-1)\,w_{h(A)}^{2}-y\,z^2_{rh (rA)}-1)
 \, , \\
&L^{ver}_{h^0\,h^0
(A^0\,A^0)}\!\!\!\!=\!\!\!\!x^2\,w_{h(A)}^{2}+(1+w_{h(A)}^{'2}\,(y-1))\,y+
x\,(1+w_{h(A)}^{2}\,(y-1)+w_{h(A)}^{'2}\,y-z^2_{rh (rA)}\,y) \, ,
 \label{Lh0A0}
 \end{split}
\end{equation}
with the parameters $w_{h(A)}=\frac{m_{l_1^-}}{m_{h^0 (A^0)}}$,
$w'_{h(A)}=\frac{m_{l_2^+}}{m_{h^0 (A^0)}}$, $z_{rh
(rA)}=\frac{m_r}{m_{h^0(A^0)}}$, $z_{ih
(iA)}=\frac{m_i}{m_{h^0(A^0)}}$ and
\begin{eqnarray}
\eta_i^V&=&\xi^{E}_{N,l_li}\xi^{E\,*}_{N,il_2}+
\xi^{E\,*}_{N,il_1} \xi^{E}_{N,l_2 i} \nonumber \, , \\
\eta_i^+&=&\xi^{E\,*}_{N,il_1}\xi^{E\,*}_{N,il_2}+
\xi^{E}_{N,l_1i} \xi^{E}_{N,l_2 i} \, . \label{etaVA}
\end{eqnarray}
In eq. (\ref{etaVA}), the flavor changing couplings $\xi^{E}_{N,
l_ji}$ represent the effective interaction between the internal
lepton $i$, ($i=e,\mu,\tau$) and the outgoing $j=1\,(j=2)$ lepton
(anti lepton). Here, we choose the couplings $\xi^{E}_{N, l_ji}$
real.

Finally, the BR for $r\rightarrow l_1^-\,l_2^+$ can be obtained by
using the matrix element square as
\begin{eqnarray}
BR (r\rightarrow l_1^- \,l_2^+)=\frac{1}{16\,\pi\,m_r}\,
\frac{|M|^2}{\Gamma_r}\, , \label{BR1}
\end{eqnarray}
where $\Gamma_r$ is the total decay width of radion $r$. In our
numerical analysis,  we consider the BR due to the production of
sum of charged states, namely
\begin{eqnarray}
BR (r\rightarrow l_1^{\pm}\,l_2^{\pm})= \frac{\Gamma(r\rightarrow
(\bar{l}_1\,l_2+\bar{l}_2\,l_1))}{\Gamma_r} \, .\label{BR2}
\end{eqnarray}
\section{Numerical Analysis and Discussion}
In four dimensions, the higher dimensional gravity is observed as
it has new states with spin 2,1 and 0, so called, the graviton,
the gravivector, the graviscalar. These states interact with the
particles in the underlying theory. In the RS1 model with one
extra dimension, the spin 0 gravity particle radion $r$ interacts
with the particles of the theory (2HDM in our case) on the TeV
brane and this interaction occurs over the trace of the
energy-momentum tensor $T^\mu_\mu$ with the strength
$1/\Lambda_r$,
\begin{eqnarray}
{\cal{L}}_{int}=\frac{r}{\Lambda_r}\,T^\mu_\mu \, ,\label{Radint}
\end{eqnarray}
where $\Lambda_r$  is at the order of TeV. The radion interacts
with gluon ($g$) pair or photon ($\gamma$) pair in one loop order
from the trace anomaly. For the radion mass $m_r\le 150\,GeV$, the
decay width is dominated by $r\rightarrow gg$.  For the masses
which are beyond the WW and ZZ thresholds, the main decay mode is
$r\rightarrow WW$. In the present work, we study the possible LFV
decays of the RS1 radion in the 2HDM and estimate the BRs of these
decays for different values of radion masses. We take the total
decay width $\Gamma_r$ of the radion by considering the dominant
decays $r\rightarrow gg\, (\gamma\gamma,  ff, W^+W^-,ZZ, SS)$
where $S$ are the neutral Higgs particles (see \cite{cheung} for
the explicit expressions of these decay widths). Here, we include
the possible processes in the $\Gamma_r$ according to the mass of
the radion.

The flavor violating  $r$ decays $r\rightarrow l_1^- l_2^+$ can
exist at least in one loop level, in the framework of the 2HDM and
the flavor violation is carried by the Yukawa couplings
$\bar{\xi}^{E}_{N,ij}$\footnote{The dimensionfull Yukawa couplings
$\bar{\xi}^{E}_{N,ij}$ are defined as
$\xi^{E}_{N,ij}=\sqrt{\frac{4\,G_F}{\sqrt{2}}}\,
\bar{\xi}^{E}_{N,ij}$.}. In the version of 2HDM where the FCNC are
permitted, these couplings are free parameters which should be
restricted by using the present and forthcoming experiments. At
first, we assume that these couplings are symmetric with respect
to the flavor indices $i$ and $j$. Furthermore, we take that the
couplings which contain at least one $\tau$ index are dominant and
we choose a broad range for these couplings, by respecting the
upper limit prediction of $\bar{\xi}^{E}_{N,\tau \mu}$ (see
\cite{iltan} and references therein) which is obtained by using
the experimental uncertainty, $10^{-9}$, in the measurement of the
muon anomalous magnetic moment and by assuming that the new
physics effects can not exceed this uncertainty. For the coupling
$\bar{\xi}^{E}_{N,\tau e}$, the restriction is estimated by using
this upper limit and the experimental upper bound of BR of
$\mu\rightarrow e \gamma$ decay, BR $\leq 1.2\times 10^{-11}$
\cite{brooks}. Finally, this coupling is taken in the range
$10^{-3}-10^{-1}\, GeV$ (see \cite{iltan1}). For the Yukawa
coupling $\bar{\xi}^{E}_{N,\tau \tau}$, we have no explicit
restriction region and we use the numerical values which are
greater than $\bar{\xi}^{E}_{N,\tau \mu}$.
%
%
Throughout our calculations we use the input values given in Table
(\ref{input}).
\begin{table}[h]
\caption{The values of the input parameters used in the numerical
          calculations.}
        \begin{center}
        \begin{tabular}{|l|l|}
        \hline
        \multicolumn{1}{|c|}{Parameter} &
                \multicolumn{1}{|c|}{Value}     \\
        \hline \hline
        $m_{\mu}$                   & $0.106$ (GeV) \\
        $m_{\tau}$                  & $1.78$ (GeV) \\
        $m_{h^0}$           & $100$   (GeV)  \\
        $m_{A^0}$           & $200$   (GeV)  \\
        $G_F$             & $1.16637 10^{-5} (GeV^{-2})$  \\
        \hline
        \end{tabular}
        \end{center}
\label{input}
\end{table}

\begin{figure}[htb]
\vskip -3.0truein
\centering\epsfxsize=5.8in\epsffile{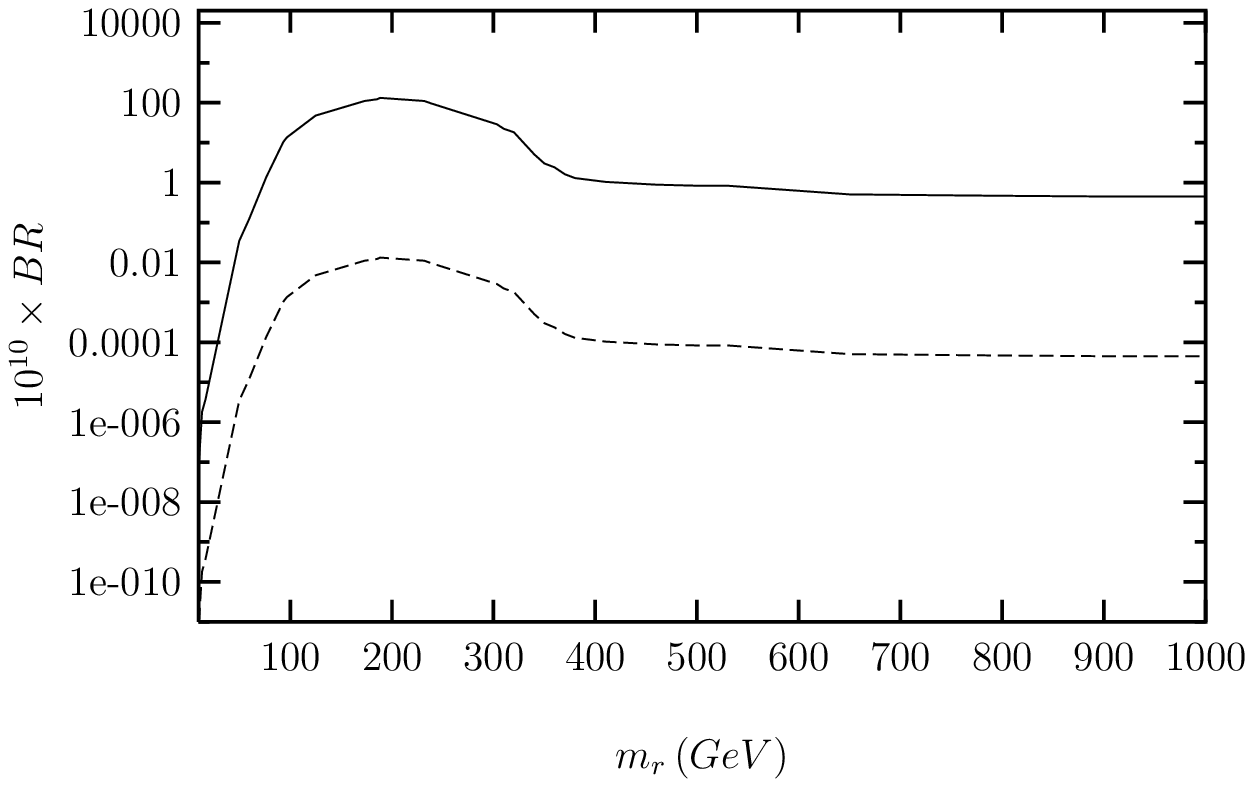} \vskip
-3.0truein \caption{ $m_r$ dependence of the BR $(r\rightarrow
\tau^{\pm}\, \mu^{\pm})$. The solid-dashed lines represent the
BR$(r\rightarrow \tau^{\pm}\, \mu^{\pm})$ for
$\bar{\xi}^{E}_{N,\tau \tau}=100\,GeV$, $\bar{\xi}^{E}_{N,\tau
\mu}=10\,GeV$- $\bar{\xi}^{E}_{N,\tau \tau}=10\,GeV$,
$\bar{\xi}^{E}_{N,\tau \mu}=1\,GeV$.} \label{RtotaumumR}
\end{figure}

In Fig.\ref{RtotaumumR} we present  $m_r$ dependence of the BR
$(r\rightarrow \tau^{\pm}\, \mu^{\pm})$. The solid-dashed lines
represent the BR $(r\rightarrow \tau^{\pm}\, \mu^{\pm})$ for
$\bar{\xi}^{E}_{N,\tau \tau}=100\,GeV$, $\bar{\xi}^{E}_{N,\tau
\mu}=10\,GeV$- $\bar{\xi}^{E}_{N,\tau \tau}=10\,GeV$,
$\bar{\xi}^{E}_{N,\tau \mu}=1\,GeV$. It is observed that the BR
$(r\rightarrow \tau^{\pm}\, \mu^{\pm})$ is of the order of the
magnitude of $10^{-8}$ for the large values of the couplings and
the radion mass values $\sim 200\,GeV$. For the heavy masses of
the radion the BR is stabilized to the values of the order of
$10^{-9}$.

\begin{figure}[htb]
\vskip -3.0truein \centering \epsfxsize=5.8in
\leavevmode\epsffile{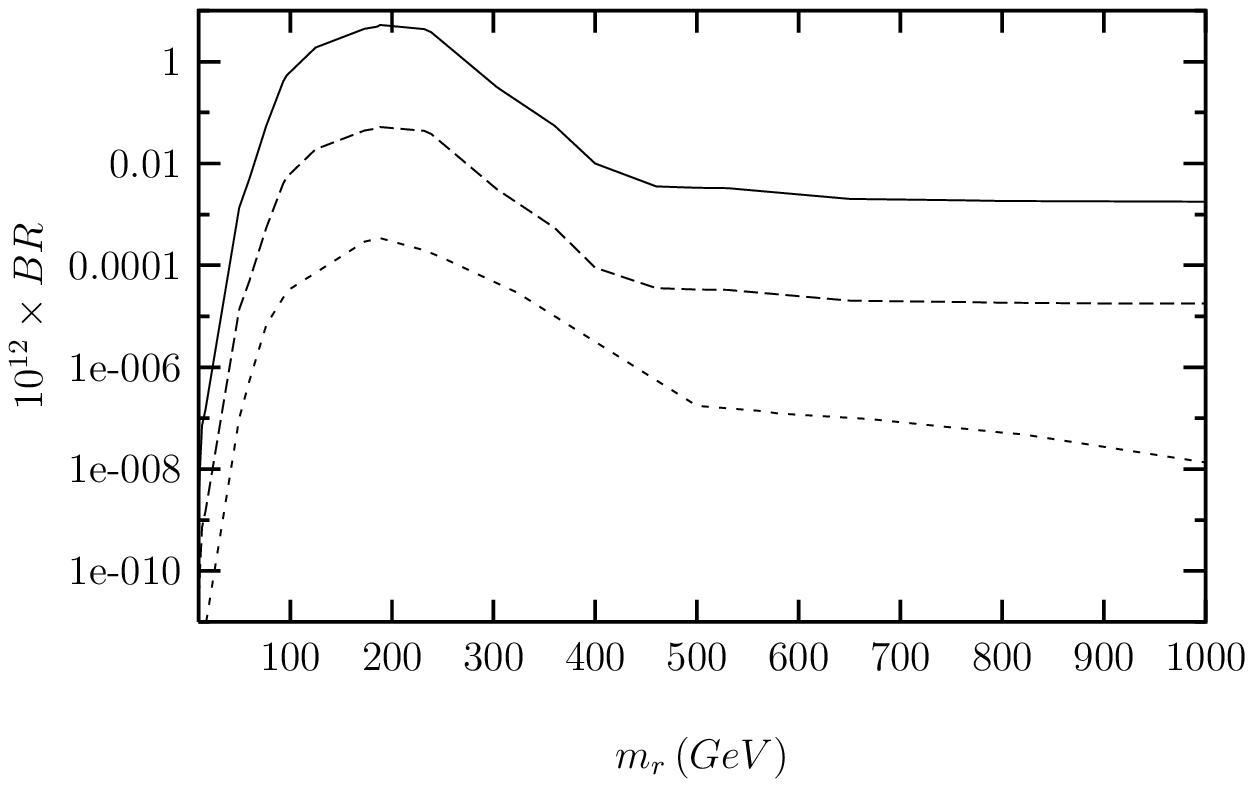} \vskip -3.0truein \caption{
$m_r$ dependence of the BR $(r\rightarrow \l_1^{\pm}\,
l_2^{\pm})$. The solid-dashed lines represent the BR$(r\rightarrow
\tau^{\pm}\, e^{\pm})$ for $\bar{\xi}^{E}_{N,\tau \tau}=100\,GeV$,
$\bar{\xi}^{E}_{N,\tau e}=0.1\,GeV$- $\bar{\xi}^{E}_{N,\tau
\tau}=10\,GeV$, $\bar{\xi}^{E}_{N,\tau e}=0.1\,GeV$. The small
dashed line represents the BR $(r\rightarrow \mu^{\pm}\, e^{\pm})$
for $\bar{\xi}^{E}_{N,\tau \mu}=1\,GeV$, $\bar{\xi}^{E}_{N,\tau
e}=0.1\,GeV$.} \label{RtomuetauemR}
\end{figure}

Fig.\ref{RtomuetauemR} is devoted to $m_r$ dependence of the BR
$(r\rightarrow \tau^{\pm}\, e^{\pm})$ and BR $(r\rightarrow
\mu^{\pm}\, e^{\pm})$. The solid-dashed lines represent the BR
$(r\rightarrow \tau^{\pm}\, e^{\pm})$ for $\bar{\xi}^{E}_{N,\tau
\tau}=100\,GeV$, $\bar{\xi}^{E}_{N,\tau e}=0.1\,GeV$-
$\bar{\xi}^{E}_{N,\tau \tau}=10\,GeV$, $\bar{\xi}^{E}_{N,\tau
e}=0.1\,GeV$. The small dashed line represents the BR
$(r\rightarrow \mu^{\pm}\, e^{\pm})$ for $\bar{\xi}^{E}_{N,\tau
\mu}=1\,GeV$, $\bar{\xi}^{E}_{N,\tau e}=0.1\,GeV$. This figure
shows that the BR $(r\rightarrow \tau^{\pm}\, \mu^{\pm})$ is of
the order of the magnitude of $10^{-12}$ for the large values of
the couplings and the radion mass values $\sim 200\,GeV$. For the
heavy masses of the radion, this BR reaches to the values less
than $10^{-14}$. The BR $(r\rightarrow \mu^{\pm}\, e^{\pm})$ is of
the order of $10^{-15}$ for $m_r\sim 200\,GeV$ and for the
intermediate values of Yukawa couplings. These BRs, especially BR
($r\rightarrow \mu^{\pm}\, e^{\pm}$), are negligibly small.

Now, we present the Yukawa coupling dependencies of the BRs of the
decays under consideration, for different radion masses .

\begin{figure}[htb]
\vskip -3.0truein \centering \epsfxsize=5.8in
\leavevmode\epsffile{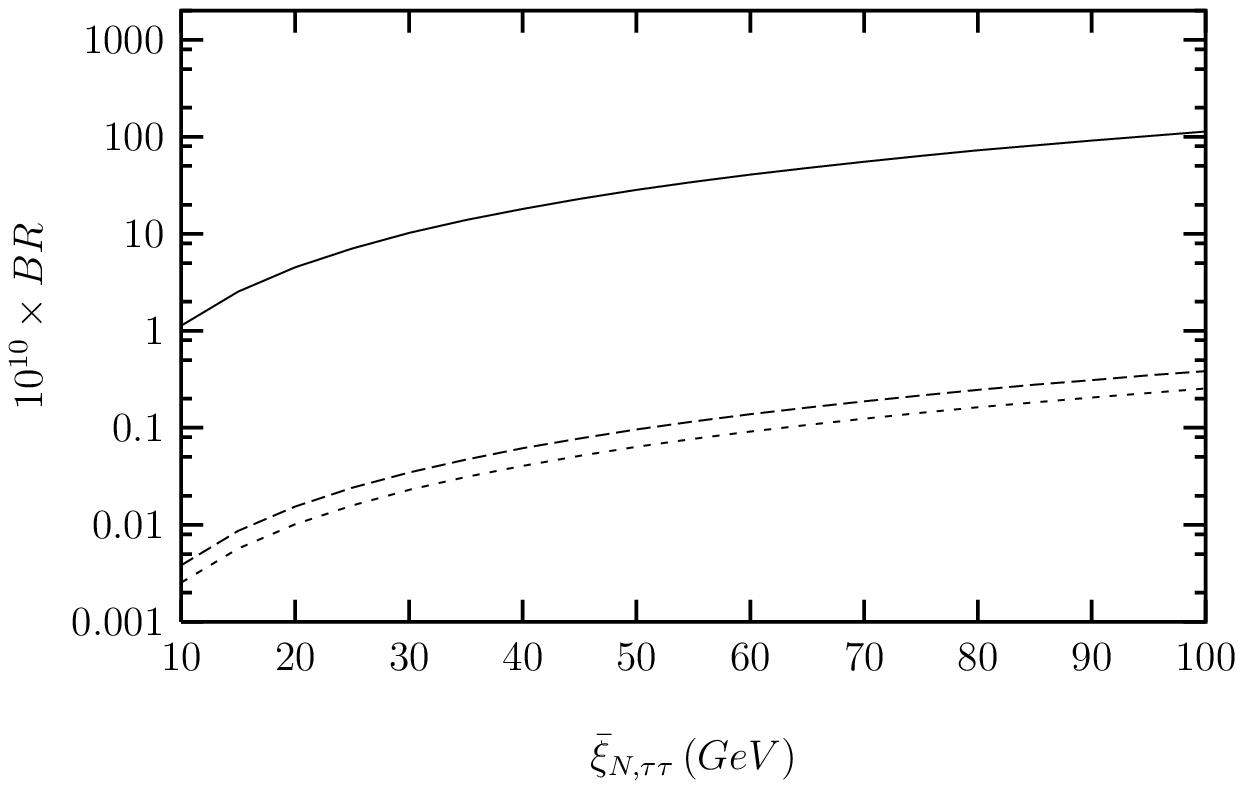} \vskip -3.0truein
\caption{$\bar{\xi}^{E}_{N,\tau \tau}$ dependence of the
BR$(r\rightarrow \tau^{\pm}\, \mu^{\pm})$ for
$\bar{\xi}^{E}_{N,\tau \mu}=10 \,GeV$. The solid-dashed-small
dashed lines represent the BR for the radion masses
$m_r=200-500-1000\,GeV$. } \label{Rtotaumuxi}
\end{figure}

Fig.\ref{Rtotaumuxi} represents  the $\bar{\xi}^{E}_{N,\tau \tau}$
dependence of the BR $(r\rightarrow \tau^{\pm}\, \mu^{\pm})$ for
$\bar{\xi}^{E}_{N,\tau \mu}=10 \,GeV$. The solid-dashed-small
dashed lines represent the BR for the radion masses
$m_r=200-500-1000\, GeV$. This figure shows that the BR is
sensitive to the radion mass and, obviously, it is enhanced two
orders of magnitude in the range $10\,GeV\leq
\bar{\xi}^{E}_{N,\tau \tau}\leq 100\,GeV$.

\begin{figure}[htb]
\vskip -3.0truein \centering \epsfxsize=5.8in
\leavevmode\epsffile{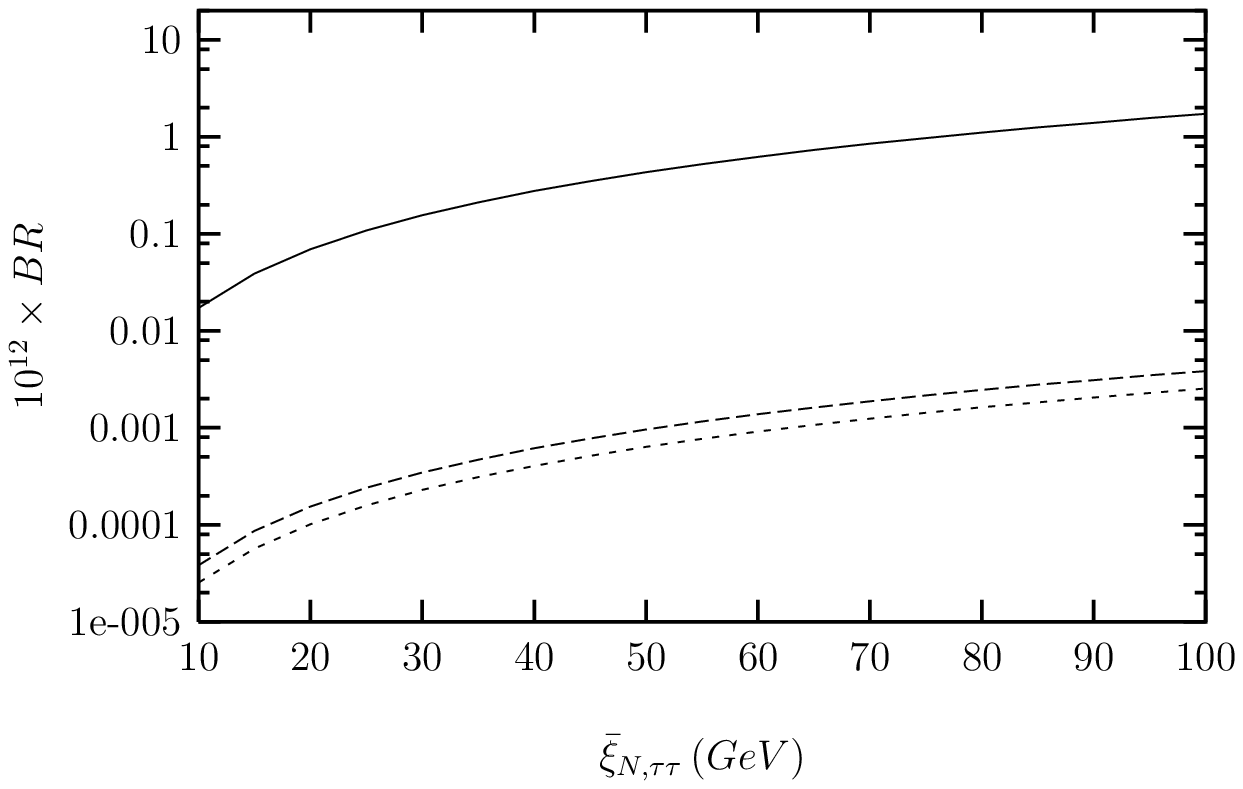} \vskip -3.0truein
\caption{$\bar{\xi}^{E}_{N,\tau \tau}$ dependence of the
BR$(r\rightarrow \tau^{\pm}\, e^{\pm})$ for $\bar{\xi}^{E}_{N,\tau
e}=0.1 \,GeV$. The solid-dashed-small dashed lines represent the
BR for the radion masses $m_r=200-500-1000\,GeV$.}
\label{Rtotauexi}
\end{figure}

In Fig.\ref{Rtotauexi}, we present the $\bar{\xi}^{E}_{N,\tau
\tau}$ dependence of the BR $(r\rightarrow \tau^{\pm}\, e^{\pm})$
for $\bar{\xi}^{E}_{N,\tau e}=0.1 \,GeV$. The solid-dashed-small
dashed lines represent the BR for the radion masses
$m_r=200-500-1000\,GeV$. Similar to the $r\rightarrow \tau^{\pm}\,
\mu^{\pm}$ decay, the BR is strongly sensitive to the radion mass.

As a summary, the LFV decays of the radion in the RS1 model
strongly depend on the radion mass and the Yukawa couplings. The
BR for $r\rightarrow \tau^{\pm}\, \mu^{\pm}$ decay is of the order
of $10^{-8}$ for the small values of radion mass $m_r$ and it
decreases with the increasing values of $m_r$. On the other hand,
the BRs for $r\rightarrow \tau^{\pm}\, e^{\pm}$ $(r\rightarrow
\mu^{\pm}\, e^{\pm})$ decays are of the order of $10^{-12}$
($10^{-15}$) for the small values of $m_r$. These results show
that, among these processes, the LFV $r\rightarrow \tau^{\pm}\,
\mu^{\pm}$ decay would be the most appropriate one to measure its
BR.
With the possible production of the radion (the most probable
production is due to the gluon fusion, $gg\rightarrow r$
\cite{cheung}), hopefully, the future experimental results of this
decay would  be useful in order to test the possible signals
coming from the extra dimensions and new physics which results in
flavor violation. 
\section{The vertices appearing in the present work}
In this section we present the vertices appearing in our
calculations. Here $S$ denotes the new neutral Higgs bosons $h^0$
and $A^0$.

\newpage
\chapter{CONCLUSION}
Over the last thirty years or so, the SM has been subjected to
many diverse experiments and most of these experiments have been
found to be consistent with the predictions of the SM except the
recently performed ones on the FCNCs including LFV interactions.
In the framework of the SM, LFV interactions exist in the $\nu$SM
by taking the neutrinos massive and permitting the lepton mixing
mechanism \cite{Pontecorvo, illana} in order to accommodate the
present data on neutrino masses and mixings. However, the neutrino
masses are so small that, the predicted BRs of the LFV
interactions are too small to explain the experimental data
obtained. Therefore, it is clear that there must be a theory which
goes beyond the SM, but which reproduces the results of the SM
where the SM has been shown experimentally to be correct. The LFV
interactions have still been probed at many experiments such as
the MEG experiment \cite{MEG}, the MEGA experiment \cite{MEGA}, in
the Belle detector at the KEKB \cite{KEKB}, and BABAR detector at
the PEP-II \cite{PEP1, PEP2}. In addition, at CERN there is a
particle accelerator, which is currently under construction,
namely the Large Hadron Collider (LHC), is expected to operate in
May $2008$ \cite{LHC}. The collider tunnel in LHC consists of two
pipes and each pipe contains a proton beam having an energy of
$7.0$ TeV which travels in opposite directions around the ring.
This means that in total the collision energy will be 14 TeV.
Furthermore, the International Linear Collider (ILC), which is
also under construction, is planned to be completed in the late
2010s \cite{LHC}. The initial collision energy is planned to be
$500$ GeV and this collision will be  between electron and
positron beams. It is expected that the physics beyond the SM will
be detected at the LHC and ILC.

The simplest extension of the SM, the so called the 2HDM,
possesses five physical Higgs bosons, namely, a charged pair
($H^{\pm}$), two neutral CP even scalars ($H^{0}$ and $h^{0}$),
and a neutral CP odd scalar ($A^{0}$). Observation of these
particles in the experiments would become a clear signature for
the physics beyond the SM.

Beside this phenomenological hint, the incompatibility of the
predicted BRs of the LFV interactions with the experimental data
obtained, the SM also possesses some conceptual problems which
motivate us to look physics beyond, such as the hierarchy problem
between EW and Planck scales. This problem could be achieved by
introducing the extra dimensions. One possibility is proposed by
Arkani-Hamed, Savas Dimopoulos and Gia Dvali \cite{arkani,
arkani1} with n compact extra spatial dimensions of large size. In
this model, gravity spreads over all the volume including the
extra dimensions while the matter fields are restricted in 4D
brane. Another possibility is introduced by Randall and Sundrum
\cite{RS, RS1}. In this scenario, the extra dimension is
compactified to $S^{1}/Z_{2}$ orbifold with two 4D brane
boundaries. Even if the compactified extra dimensions are so small
that we perceive the universe as 4D, if there are extra
dimensions, fingerprints of them sure to exist and such
fingerprints are the particles called KK particles which are the
additional ingredients of the universe with extra dimensions. The
masses of KK particles are determined by the higher dimensional
geometry. In the large extra dimensions, for example, the masses
of the KK particles are proportional to the inverse size of the
extra dimension (for two extra dimensions, the size of the extra
dimension is $\sim 1$ mm). This tells us that, the current and
future accelerators should be able to discover them. Since in
large extra dimensions, the only particle that can travel along
the extra dimensions is called as graviton and thus, it is the
only particle that has KK partners. However, the KK partners of
the graviton interacts as weakly as the graviton itself.
Therefore, KK partners of graviton would be hard to observe. In
warped extra dimensions, however, we cannot take just the the
inverse size of the extra dimension as the masses of KK particles
which gives us the Planck scale mass and we know that on the TeV
brane, nothing much heavier than a TeV can exist. If one calculate
the masses of KK particles taking into account the warped
space-time, the KK particles of graviton turn out to have masses
of about a TeV and these KK particles, not as in the case of large
extra dimensions, will interact sixteen orders of magnitude larger
than the graviton itself in this geometry. This is good news,
since depending on the ultimate energy reach in LHC, there is a
probability of finding the KK partners of the graviton. Apart from
the graviton, in the Randall Sundrum scenario, there exists an
additional scalar field, that lives in the 5D bulk, such that the
size of extra dimension is proportional to its vacuum expectation
value and its fluctuation over this expectation value is called as
the radion field. In order to avoid the conflict with the
equivalence principle, the introduced field should be massive and
to stabilize the distance between the branes, a potential for this
field is proposed by Goldberger and Wise \cite{goldberger}. The
radion decays are interesting since a considerable information
about the scenario under consideration (the RS1 scenario) can be
obtained with the help of accurate measurements.

In the present work we study the possible LFV decays of the radion
field $r$ and predict the BRs of the LFV r decays $r\rightarrow
l^{-}_{1}l^{+}_{2}$ in the RS1 model. We observe that the BRs of
the processes we study are at most of the order of $10^{-8}$, for
the small values of radion mass $m_r$ and their sensitivities to
$m_r$ decrease with the increasing values of $m_r$. On the other
hand, the BRs for $r\rightarrow \tau^{\pm}\, e^{\pm}$
$(r\rightarrow \mu^{\pm}\, e^{\pm})$ decays are of the order of
$10^{-12}$ ($10^{-15}$) for the small values of $m_r$. These
results show that, among these processes, the LFV $r\rightarrow
\tau^{\pm}\, \mu^{\pm}$ decay would be the most appropriate one to
measure its BR.

\newpage
\addtocontents{toc}{\protect\contentsline {part}{}{}}
\addcontentsline{toc}{chapter}{\bibname}
\bibliographystyle{amsplain}

\begin{thebibliography}{99}
\singlespacing

\bibitem{WSalam} S. L. Glashow, \emph{Nucl. Phys.} \textbf{22} (1961);
S. Weinberg, \emph{Phys. Rev. Lett.} \textbf{19} (1967) 1264; A.
Salam, \emph{in Elementary Particle Theory} (Nobel Symposium No.
8), edited by N. Svartholm, (Almquist and Wiksell, Stockholm,
1968).
%
\bibitem{WSalam2} S. L. Glashow, I. Iliopoulos, and L. Maiani, \emph{Phys. Rev.} \textbf{D2} (1970) 1285;
S. L. Glashow, \emph{Rev. Mod. Phys.} \textbf{53} (1980) 539; S.
Weinberg, \emph{Rev. Mod. Phys.} \textbf{52} (1980) 515; A. Salam,
\emph{Rev. Mod. Phys.} \textbf{52} (1980) 525.
%
\bibitem{GWeinberg} T. P. Cheng and L. F. Li, \emph{Gauge Theory of Elementary Particle
Physics} (Claredon Press, Oxford, 1984).
%
\bibitem{leader} E. Predazzi and E. Leader, \emph{An Introduction to Gauge Theories and
Modern Particle Physics} (Cambridge University Press, Cambridge,
1996).
%
\bibitem{abers} E. S. Abers and B. W. Lee, \emph{Phys. Rep.}
\textbf{9} (1973) 1.
%
\bibitem{branco} S. Weinberg, \emph{Physical Review Letters}, \textbf{37} (1976)
11; G. C. Branco, A. J. Buras, and J. M. Gerard, \emph{Nucl.
Phys.} \textbf{B259} (1985) 306; P. Krawczyk and S. Pokorski,
\emph{Phys. Rev. Lett.} \textbf{60} (1988) 182; J. Kalinowski and
S. Pokorski, \emph{Phys. Lett.} \textbf{B259} (1989) 116.
%
\bibitem{haber} H. E. Haber, G. L. Kane, \emph{Phys. Rep.} \textbf{C 117} (1985)
75; J. Gunion, H. Haber, G. Kane, and S. Dawson, \emph{The Higgs
Hunter's Guide} (Addison-Wesley, New York, 1990).
%
\bibitem{gunion} V. Barger, R. J. N. Phillips, and D.P. Roy, \emph{Phys. Lett.} \textbf{B324} (1994)
236; J. F. Gunion and S. Geer, \emph{hep-ph/9310333} (1993); J. F.
Gunion, \emph{Phys. Lett.} \textbf{B322} (1994) 125; D. J. Miller,
S. Moretti, D. P. Roy, and W. J. Stirling, \emph{Phys. Rev.}
\textbf{D61} (2000) 055011; S. Moretti and D. J. Roy, \emph{Phys.
Lett.} \textbf{B470} (1999) 209, K. Odagiri, \emph{hep-ph/9901432}
(1999); S. Raychaudhuri, D. P. Roy, \emph{Phys. Rev.} \textbf{D53}
(1996) 4902.
%
\bibitem{Soni} D. Atwood, L. Reina, and A. Soni, \emph{Phys.Rev.} \textbf{D55} (1997)
3156.
\bibitem{sohnius} M. F. Sohnius, \emph{Phys. Rep.} \textbf{2} (1985)
39.
\bibitem{barbieri} R. Barbieri and L. J. Hall, \emph{Phys. Lett.} \textbf{B338} (1994) 212.
\bibitem{barbieri2} R. Barbieri, L. J. Hall, and A. Strumia, \emph{Nucl. Phys.} \textbf{B445} (1995) 219.
\bibitem{barbieri3} R. Barbieri, L. J. Hall, and A. Strumia, \emph{Nucl. Phys.} \textbf{B449} (1995) 437.
\bibitem{pati} J. C. Pati and A. Salam \emph{Phys. Rev.} \textbf{D10} (1974) 275;
R. N. Mohapatra and J. C. Pati, \emph{Phys. Rev.} \textbf{D11}
(1975) 2558; G. Senjanovic, R. N. Mohapatra, \emph{Phys. Rev.}
\textbf{D12} (1975) 1502.
\bibitem{ghosal} A. Ghosal, Y. Koide, and H. Fusaoka, \emph{Phys.Rev.} \textbf{D64} (2001)
053012.
\bibitem{minkowski} P. Minkowski, \emph{Phys. Lett.} \textbf{B67} (1977) 421; R. N. Mohapatra and G.
Senjanovic, \emph{Phys. Rev. Lett.} \textbf{B94} (1980) 912.
\bibitem{yue} C. Yue, W. Wang, and F. Zhang, \emph{J. Phys.} \textbf{G30} (2004)
1065.
%
%
\bibitem{akama} K. Akama, \emph{Lect. Notes Phys.} \textbf{176} (1982) 267, \emph{
hep-th/0001113.}; K. Akama, \emph{Prog. Theor. Phys.} \textbf{60}
(1978) 1900; \textbf{78} (1987) 184; \textbf{79} (1988) 1299;
\textbf{80} (1988) 935; K. Akama and T. Hattori, \emph{Mod. Phys.
Lett.} \textbf{A15} (2000) 2017.
%
\bibitem{antoni} I. Antoniadis, \emph{Phys. Lett.} \textbf{B246} (1990) 377.
%
\bibitem{arkani} N. Arkani-Hamed, S. Dimopoulos, and G. R. Dvali, \emph{Phys. Lett.} \textbf{B429}
(1998) 263, \emph{hep- ph/9803315}; I. Antoniadis, et al.,
\emph{Phys. Lett.} \textbf{B436} (1998) 257,
\emph{hep-ph/9804398}; I. Antoniadis, S. Dimopoulos, and G. Dvali,
\emph{Nucl. Phys.} \textbf{B516} (1998) 70; I. Antoniadis, C.
Bachas, D. Lewellen, and T. Tomaras, \emph{Phys. Lett.} \textbf{B
207}, (1988) 441.
\bibitem{arkani1} N. Arkani-Hamed, S. Dimopoulos, and G. R. Dvali, \emph{Phys. Rev.} \textbf{D59} (1999) 086004,\emph{ hep-
ph/9807344}.
%
\bibitem{ued} T. Appelquist, H.-C. Cheng, and B. A. Dobrescu, \emph{Phys. Rev.} \textbf{D64} (2001) 035002; T.
Appelquist and H. Yee, \emph{Phy. Rev.} \textbf{D67} (2003)
055002.
\bibitem{ued1} J. Papavassiliou and A. Santamaria, \emph{Phys. Rev.} \textbf{D63} (2001)
016002.
\bibitem{ued2} D. Chakraverty, K. Huitu, and A. Kundu, \emph{Phys.Lett.} \textbf{B558} (2003) 173; A. J. Buras, M.
Spranger, and A. Weiler, \emph{Nucl.Phys.} \textbf{B660} (2003)
225.
\bibitem{nued} K. R. Dienes, E. Dudas, and T. Gherghetta,\emph{ Nucl. Phys.} \textbf{B557} (1999) 25; Q. H. Shrihari, S.
Gopalakrishna, and C. P. Yuan, \emph{Phys. Rev.} \textbf{D69}
(2004) 115003.
\bibitem{nued1} C. S. Lam, \emph{hep-ph/0302227} (2003); C. A. Scrucca, M. Serona, and L. Silvestrini, \emph{Nucl.Phys.}
\textbf{B669} (2003) 128; M. Gozdz and W. A. Kaminsk, \emph{Phys.
Rev.} \textbf{D68} (2003) 057901; C. Biggio, et.al,
\emph{Nucl.Phys.} \textbf{B677} (2004) 451; M. Carena, et.al,
\emph{Phys. Rev.} \textbf{D68} (2003) 035010; A. J. Buras, et.
al., \emph{Nucl.Phys.} \textbf{B678} (2004) 455; T. G. Rizzo,
\emph{JHEP} \textbf{0308} (2003) 051; A. J. Buras,
\emph{hep-ph/0307202} (2003); S. Matsuda and S. Seki,
\emph{hep-ph/0307361} (2003); R. N. Mohapatra, \emph{Phys. 10
Rev.} \textbf{D68} (2003) 116001; B. Lillie, \emph{JHEP}
\textbf{0312} (2003) 030; A.A Arkhipov, \emph{hep-ph/0309327}
(2003).
\bibitem{split} N. Arkani-Hamed and M. Schmaltz, \emph{Phys. Rev.} \textbf{D61} (2000) 033005; N. Arkani-Hamed,
Y. Grossman, and M. Schmaltz, \emph{Phys. Rev.} \textbf{D61}
(2000) 115004.
\bibitem{Schmaltz} E. A. Mirabelli, Schmaltz, \emph{Phys. Rev.} \textbf{D61} (2000) 113011.
\bibitem{chang} W. F. Chang, I. L. Ho, and J. N. Ng, \emph{Phys. Rev.} \textbf{D66} (2002) 076004.
\bibitem{RS} L. Randall and R.Sundrum, \emph{Phys. Rev. Lett.} \textbf{83} (1999) 3370, \emph{hep-ph/9905221}.
\bibitem{RS1} L. Randall and R.Sundrum, \emph{Phys. Rev. Lett.} \textbf{83} (1999) 4690, \emph{hep-th/9906064}.
\bibitem{Klein} T. Kaluza, \emph{Sitzungober. Preuss. Akad. Wiss. Berlin} (1921) 966; O.
Klein, \emph{Z. Phys.} \textbf{37} (1926) 895.


\bibitem{kaku} M. Kaku, \emph{Quantum field theory : a modern
introduction} (Oxford University Press, New York 1993).
\bibitem{fermi} E. Fermi, \emph{Zeitschrift für Physik.} \textbf{88} (1934)
161.
\bibitem{feynman} R. P. Feynman and M. Gell-Mann, \emph{Phys. Rev.} \textbf{109} (1958) 193.
\bibitem{higgs} P. W. Higgs, \emph{Phys. Lett.} \textbf{12} (1964) 132; F. Englert and R. Brout,
\emph{Phys. Rev. Lett.} (1964) 321; G. S. Gurallnik, C. R. Hagen,
and T. W. B. Kibble, \emph{Phys. Rev. Lett.} \textbf{13} (1964)
585; P. W. Higgs, \emph{Phys. Rev.} \textbf{145} (1966) 1156; T.
W. B. Kibble, \emph{Phys. Rev.} \textbf{155} (1967) 1554.
\bibitem{goldstone} Y. Nambu, \emph{Phys. Rev. Lett.} \textbf{4} (1960) 380; J. Goldstone and Nuovo Cimento
\textbf{19} (1961) 154; J. Goldstone, A. Salam, and S. Weinberg,
\emph{Phys. Rev.} \textbf{127} (1962) 965.
\bibitem{gell} M. Gell-Mann, \emph{Phys. Rev.} \textbf{92} (1953) 833; K. Nishijima and T. Nakano,
\emph{Prog. Theor. Phys.} \textbf{10} (1953) 581.
\bibitem{cern} UA1 Collobration, G. Armson et.al., \emph{Phys. Lett.} \textbf{B122} (1983)
103; UA2 Collobration, M. Banner et.al.,\emph{ Phys. Lett.}
\textbf{B122} (1983) 476.
\bibitem{ckm} Particle Data Group, \emph{Jour. of Phys.}
\textbf{G33} (2006) 1.
\bibitem{Pontecorvo} B. Pontecorvo, \emph{Zh. Eksp. Teor. Fiz.} \textbf{33} (1957) 549; Z. Maki, M. Nakagawa, and S. Sakata,
\emph{Prog. Theor. Phys.} \textbf{28} (1962) 870; B. Pontecorvo,
\emph{Sov. Phys. JETP} \textbf{26} (1968) 984.
\bibitem{illana} J. I. Illana, M. Jack, and T. Riemann, \emph{hep-ph/0001273} (2000); J. I. Illana and T. Riemann,
\emph{Phys. Rev.} \textbf{D63} 053004 (2001).
\bibitem{ccp} S. Weinberg, \emph{Rev. Mod. Phys.} \textbf{61} (1989) 1;
P. J. E. Peebles and B. Ratra, \emph{Rev. Mod. Phys.} \textbf{75}
(2003) 559; T. Padmanabhan, \emph{Phys. Rept.} \textbf{380} (2003)
235.
\bibitem{witten} E. Witten, \emph{hep-ph/ 0002297} (2000); P.
Binetray, \emph{hep-ph/ 0005037} (2000); P. Horawa and E. Witten,
\emph{Nucl. Phys.} \textbf{B 460} (1996) 506; P. Horawa and E.
Witten, \emph{Nucl. Phys.} \textbf{B 475} (1996) 94.
%
%

%
\bibitem{rubakov} V. A. Rubakov and M. E. Shaposhnikov, \emph{Phys. Lett.} \textbf{B 125} (1983) 136; \emph{Phys. Lett.}
\textbf{B 125} (1983) 139.
%
%
\bibitem{visser} M. Visser, \emph{Phys. Lett.} \textbf{B159} (1985) 22, \emph{hep-th/9910093.}
%
%
\bibitem{csaki1} C. Csaki, \emph{hep-ph/0404096} (2004).
\bibitem{csaki} C. Csaki, M. L. Graesser, and G. D. Gribs, \emph{Phys. Rev.} \textbf{D 63} (2001) 065002.
\bibitem{dirac} P. A. M. Dirac, \emph{General theory of
relativity} (Wiley, New York, 1975) Cambridge University Press,
Cambridge, 1996).
\bibitem{goldberger} W. D. Goldberger and M. B. Wise, \emph{Phys. Rev. Lett.} \textbf{D 83} (1999)
4922.
\bibitem{charmousis} C. Charmousis, R. Gregory, and V. A. Rubakov, \emph{Phys. Rev.} \textbf{D 62}
(2000) 067505.
\bibitem{das} P. K. Das, \emph{hep-ph 0407041} (2004); P. K. Das, \emph{Phys. Rev.} \textbf{D 72} (2005)
055009.
\bibitem{kribs} G. D. Kribs, \emph{eConf} \textbf{C010630} (2001) P317,
\emph{hep-ph/0110242} (2001).
\bibitem{han} T. Han, G. D. Kribs, and B. McElrath \emph{Phys. Rev.} \textbf{D 64} (2001)
076003.
\bibitem{cheung} K. Cheung, \emph{Phys. Rev.} \textbf{D 63} (2001)
056007.
\bibitem{cheung2} K. Cheung, \emph{hep-ph/0408200} (2004).
\bibitem{giudice} G. F. Giudice, R. Rattazzi, and J. D. Wells, \emph{Nucl. Phys.} \textbf{B 595} (2001)
250.
\bibitem{mahanta} U. Mahanta and A. Datta, \emph{Phys. Lett.} \textbf{B 483} (2000) 196.
%
\bibitem{Kingman} K. Cheung, C. S. Kim, J. Song,
{\it Phys. Rev.} {\bf D67} (2003) 075017.
%
\bibitem{iltan} E. Iltan and H. Sundu, \emph{Acta Phys.Slov.} \textbf{53} (2003)
17.
\bibitem{brooks} M. L. Brooks et. al., MEGA Collaboration, \emph{Phys. Rev. Lett.} \textbf{83} (1999)
1521.
\bibitem{iltan1} E. Iltan, \emph{Phys. Rev.} \textbf{D64} (2001) 115005; \emph{Phys. Rev.} \textbf{D64} (2001) 013013.
\bibitem{Makale} E. Iltan and B. Korutlu, \emph{hep-ph/0610147}
(2006).
\bibitem{MEG} S. Ritt for the MEG Collaboration, \emph{Nuc. Phys. B (Proc. Suppl.)} \textbf{162} (2006)
279.
\bibitem{MEGA} M. L. Brooks et. al., MEGA Collaboration, \emph{Phys. Rev. Lett.} \textbf{83} (1999) 1521.
\bibitem{KEKB} Donato Nicolo, MUEGAMMA Collaboration, \emph{Nucl. Instrum. Meth} \textbf{A503}
(2003) 287.
\bibitem{PEP1} K. Hayasaka et al., \emph{Phys. Lett.} \textbf{B613} (2005) 20.
\bibitem{PEP2} J.M. Roney and the BABAR Collaboration, \emph{Nucl. Phys. Proc. Suppl.} \textbf{144}
(2005) 155.
\bibitem{LHC} LHC/LC Study Group, \emph{Phys. Rept.} \textbf{426} (2006) 47.
\bibitem{spincon} S. M. Carroll, Lecture Notes on General
Relativity (1997).
\end{thebibliography}

\newpage
\begin{center}
\normalfont{\huge{\bfseries{\sc{Appendix A}}}} \vspace{1.75cm}
\end{center}
\renewcommand{\theequation}{A-\arabic{equation}}
\renewcommand{\thesection}{A-\arabic{section}}
\setcounter{equation}{0}
The gauge invariance is an important
concept in modern particle theories as it is the origin of all of
the known four fundamental forces described in Chapter \ref{SM}.
The basic method to provide gauge invariance is to ensure that
Lagrangian remains invariant under certain symmetry
transformations which reflect conservation laws in nature. By
applying these transformations, we end up with conserved physical
quantities. Since these conserved quantities should not depend on
position in space-time, theories of particle interactions have to
be invariant under local as well as global gauge transformations
explained below. The transformations could be written as
\begin{equation}
\psi\rightarrow U\psi,
\end{equation}
where U is unitarity. So, the symmetry is called $U(1)$ gauge
invariance.
\\ \\ \\
{\Large{\textbf{Global Gauge Transformations}}}
\\ \\
The expression for global gauge transformation (GGT) is
\begin{equation}\label{global}
\psi\rightarrow e^{i\theta}\psi,
\end{equation}
where $\theta$ is a real number. Thus, GGT represents an identical
operation at all points in space-time and causes a simple shift in
the phase of a fermion wave function. As a first step we can see
how it works in QED. Equation of motion for free fermions are
obtained from Dirac Lagrangian
\begin{equation}\label{Lfree}
L_{free}=\bar{\psi}(i\gamma^{\mu}\partial_{\mu}-m)\psi,
\end{equation}
where $\psi$ is the fermion spinor. It is clear that such a
Lagrangian remains invariant under GGT. So, the global phase
symmetry is just a statement of the fact that the laws of physics
are independent of the choice of phase convention.
\\ \\ \\
{\Large{\textbf{Local Gauge Transformations}}}
\\ \\
The expression for local gauge transformation (LGT) is
\begin{equation}
\psi\rightarrow e^{iq\theta(x)}\psi,
\end{equation}
where $\theta$ is a function of $x=(\textbf{x}, t)$. Thus, LGT
corresponds to choosing a convention to define the phase of the
fermion wavefunction, which is different at different points in
space-time. Going back to the Dirac Lagrangian defined in eq.
\ref{Lfree}, one can simply realize that it is not invariant under
this more demanding symmetry transformation. However, it comes as
a pleasant surprise that if we introduce another field, $A_{\mu}$,
a Lagrangian which exhibits local gauge symmetry can be obtained.
The required field must have infinite range, since there is no
limit to the distances over which the phase transformations done.
Hence, invariance of Lagrangian requires this new field to be
massless. In fact, this field is not other than the long-range
electromagnetic field: the photon. Therefore, we should modify our
Lagrangian to make it gauge invariant by replacing the normal
derivative $\partial_{\mu}$ with the covariant derivative
$D_{\mu}=\partial_{\mu}+iqA_{\mu}$. So, the Lagrangian in eq.
\ref{Lfree} reads
\begin{equation}
L=\bar{\psi}(i\gamma^{\mu}D_{\mu}-m)\psi=\bar{\psi}(i\gamma^{\mu}
\partial_{\mu}-m)\psi
-qA_{\mu}\bar{\psi}\gamma^{\mu}\psi=L_{free}-J^{\mu}A_{\mu}.
\end{equation}
It so happens that, the photon transforms under LGT as
$A_{\mu}\rightarrow A_{\mu}+\partial_{\mu}\theta(x)$ so that, the
changes in the Lagrangian resulting from the LGT is cancelled out
by the changes in $A_{\mu}$.

To see the complete picture, we should also add the gauge
invariant kinetic term to the QED Lagrangian
\begin{equation}
L_{QED}=L_{free}-J^{\mu}A_{\mu}-\frac{1}{4}F^{\mu\nu} F_{\mu\nu},
\end{equation}
where $F_{\mu\nu}$ is the electromagnetic field tensor defined as
\begin{equation}
F_{\mu\nu}=\partial_{\mu}A_{\nu}-\partial_{\nu}A_{\mu}.
\end{equation}
This Lagrangian expresses the interaction of Dirac fields with the
massless photon.

\newpage
\begin{center} \normalfont{\huge{\bfseries{\sc{Appendix B}}}}
\vspace{1.75cm}
\end{center}
\renewcommand{\theequation}{B-\arabic{equation}}
\renewcommand{\thesection}{B-\arabic{section}}
\setcounter{equation}{0}
{\Large{\textbf{The Einstein Equation in Another Form}}}
\\ \\
Here we show another form of the Einstein equation which we use in
the text. By using the equation
$G_{MN}=R_{MN}-\frac{1}{2}g_{MN}R=\kappa^{2}T_{MN}$ we obtain
$R_{MN}$ as
\begin{equation}\label{RMNson}
R_{MN}=\kappa^{2}T_{MN}+\frac{1}{2}g_{MN}R,
\end{equation}
and multiplying this equation by $g^{MN}$ from left we get
\begin{equation}
g^{MN}R_{MN}=\kappa^{2}g^{MN}T_{MN}+\frac{1}{2}g^{MN}g_{MN}R,
\end{equation}
which is equal to,
\begin{equation}
R=\kappa^{2}T+\frac{1}{2}d R=\frac{2\kappa^{2}}{2-d}T,
\end{equation}
where d is the number of dimensions. Substituting this into the
eq. \ref{RMNson} we obtain $R_{MN}$ as
\begin{eqnarray}
R_{MN}&=&\kappa^{2}T_{MN}+\frac{1}{2}g_{MN}\frac{2\kappa^{2}}{2-d}T\nonumber\\
&=&\kappa^{2}(T_{MN}+g_{MN}\frac{1}{2-d}T)\nonumber\\
&=&\kappa^{2}\tilde{T}_{MN} \label{RMNtildeT},
\end{eqnarray}
with
\begin{equation}
\tilde{T}_{MN}=T_{MN}-\frac{1}{3}g_{MN}T \label{tildeT},
\end{equation}
where $d=5$ in our case.
\\ \\ \\ \\
{\Large{\textbf{The Derivation of the Linearized Einstein
Equations}}}
\\ \\
Now, we consider the metric in eq. \ref{ansatzfluctuation} and
first we will derive the explicit form of the curvature $R_{MN}$
and its linearized form. The curvature $R_{\mu\nu}$ in terms of
connection coefficients can be written as
\begin{equation}
R_{\mu\nu}=\Gamma_{\mu K,\nu}^{ K} -\Gamma_{\mu\nu, K}^{
K}-\Gamma_{\mu\nu}^{ K} \Gamma_{ K M}^{ M}+\Gamma_{\mu M}^ {
K}\Gamma_{\nu K}^{ M}.
\end{equation}
Let us analyze this term by term. The first term is derived as
follows
\begin{eqnarray}
\Gamma_{\mu K,\nu}^{ K}&=&\{g^{ K L}\Gamma_{ L\mu K}\}_{,\nu}\nonumber\\
&=&\frac{1}{2}\{g^{ K L}[g_{ L\mu, K}+g_{ L K,\mu}-g_{\mu K,
L}]\}_{,\nu}
\nonumber\\
&=&\frac{1}{2}\{g^{ K \rho}[g_{\rho\mu, K}+g_{\rho K,\mu}-g_{\mu
K,\rho}]+
g^{ K 5}[g_{5 K,\mu}-g_{\mu K,5}]\}_{,\nu}\nonumber\\
&=&\frac{1}{2}\{g^{\xi \rho}[g_{\rho\mu,\xi}+g_{\rho \xi
,\mu}-g_{\mu \xi ,\rho}]+g^{55}
[g_{55,\mu}]\}_{,\nu}\nonumber\\
&=&\frac{1}{2}\{e^{2A+2F}\eta^{\xi \rho }[(e^{-2A-2F}\eta_{\rho \mu})_{,\xi}+(e^{-2A-2F}\eta_{\rho \xi })_{,\mu}-(e^{-2A-2F}\eta_{\mu \xi })_{,\rho }]\nonumber\\
&+&(1+G)^{-2}(1+G)^{2}_{,\mu}\}_{,\nu}\nonumber\\
&=&\frac{1}{2}\{e^{2A+2F}\eta^{\xi \rho }[(-2e^{-2A-2F}\partial_{\xi }F\eta_{\rho \mu})+(-2e^{-2A-2F}\partial_{\mu}F\eta_{\rho \xi })\nonumber\\
&-&(-2e^{-2A-2F}\partial_{\rho }F\eta_{\mu \xi })+(1+G)^{-2}2\partial_{\mu}G(1+G)\}_{,\nu}\nonumber\\
&=&\{-\eta^{\xi }_{\mu}\partial_{\xi }F- 4\partial_{\mu}F+ \eta^{\rho }_{\mu}\partial_{\rho }F+ \frac{\partial_{\mu}G}{1+G}\}_{,\nu}\nonumber\\
&=&\{-\partial_{\mu}F- 4\partial_{\mu}F+ \partial_{\mu}F+ \frac{\partial_{\mu}G}{1+G}\}_{,\nu}\nonumber\\
&=&\{-4\partial_{\mu}F+\frac{\partial_{\mu}G}{1+G}\}_{,\nu}\nonumber\\
&=&-4\partial_{\mu}\partial_{\nu}F+\frac{\partial_{\mu}\partial_{\nu}G}{1+G}-\frac{\partial_{\mu}G\partial_{\nu}G}{(1+G)^{2}},
\end{eqnarray}
and for small fluctuations we have
\begin{eqnarray}
\Gamma_{\mu K,\nu}^{ K}&\cong& -4\partial_{\mu}\partial_{\nu}F+
\partial_{\mu}\partial_{\nu}G(1-G)-\partial_{\mu}G\partial_{\nu}G(1-2G).
\end{eqnarray}
Then, the linearized form of the first term is obtained as
\begin{eqnarray}
\delta\Gamma_{\mu K,\nu}^{ K}=-4\partial_{\mu}\partial_{\nu}F+
\partial_{\mu}\partial_{\nu}G.
\end{eqnarray}
The second term is
\begin{eqnarray}
\Gamma_{\mu\nu, K}^{ K}&=&\{g^{ K L}\Gamma_{ L\mu\nu}\}_{, K}\nonumber\\
&=&\frac{1}{2}\{g^{ K L}[g_{ L\mu,\nu}+g_{ L\nu,\mu}-g_{\mu\nu, L}]\}_{, K}\nonumber\\
&=&\frac{1}{2}\{g^{ K \rho }[g_{\rho \mu,\nu}+g_{\rho \nu,\mu}-g_{\mu\nu,\rho }]+g^{ K 5}[-g_{\mu\nu,5}]\}_{, K}\nonumber\\
&=&\frac{1}{2}\{g^{\xi \rho }[g_{\rho \mu,\nu}+g_{\rho \nu,\mu}-g_{\mu\nu,\rho }]\}_{,\xi}+\frac{1}{2}\{g^{55}(-g_{\mu\nu,5})\}_{,5}\nonumber\\
&=&\frac{1}{2}\{e^{2A+2F}\eta^{\xi \rho }[(e^{-2A-2F}\eta_{\rho \mu})_{,\nu}+(e^{-2A-2F}\eta_{\rho \nu})_{,\mu}-(e^{-2A-2F}\eta_{\mu\nu})_{,\rho }]\}_{,\xi}\nonumber\\
&+&\frac{1}{2}\{(1+G)^{-2}(e^{-2A-2F}\eta_{\mu\nu})_{,5}\}_{,5}\nonumber\\
&=&\frac{1}{2}\{e^{2A+2F}\eta^{\xi \rho }[(-2e^{-2A-2F}\partial_{\nu}F\eta_{\rho \mu})+(-2e^{-2A-2F}\partial_{\mu}F\eta_{\rho \nu})\nonumber\\
&-&(-2e^{-2A-2F}\partial_{\rho }F\eta_{\mu\nu})\}_{,\xi}+\frac{1}{2}\{-2(1+G)^{-2}\partial_{5}(A+F)e^{-2A-2F}\eta_{\mu\nu}\}_{,5}\nonumber\\
&=&\{-\eta^{\xi }_{\mu}\partial_{\nu}F- \eta^{\xi }_{\nu}\partial_{\mu}F+ \eta^{\xi \rho }\eta_{\mu\nu}\partial_{\rho }F\}_{,\xi}- \{\frac{e^{-2A-2F}\eta_{\mu\nu}\partial_{5}(A+F)}{(1+G)^{2}}\}_{,5}\nonumber\\
&=&-\eta^{\xi }_{\mu}\partial_{\xi }\partial_{\nu}F- \eta^{\xi }_{\nu}\partial_{\xi }\partial_{\mu}F+ \eta^{\xi \rho }\eta_{\mu\nu}\partial_{\xi }\partial_{\rho }F\nonumber\\
&-& e^{-2A-2F}\eta_{\mu\nu}\{\frac{-2(A'+F')^{2}+A''+F''}{(1+G)^{2}}-\frac{2(A'+F')G'(1+G)}{(1+G)^{4}}\}\nonumber\\
&=&-\partial_{\mu}\partial_{\nu}F-\partial_{\nu}\partial_{\mu}F+\eta_{\mu\nu}\partial_{\xi }\partial^{\xi }F\nonumber\\
&-&\frac{e^{-2A-2F}\eta_{\mu\nu}}{(1+G)^{2}}[-2(A'+F')^{2}+A''+F''-\frac{2(A'+F')G'}{(1+G)}]\nonumber\\
&=&-2\partial_{\mu}\partial_{\nu}F+\eta_{\mu\nu}\Box F-\frac{e^{-2A-2F}\eta_{\mu\nu}}{(1+G)^{2}}\nonumber\\
&\times&[-2A'^{2}-2F'^{2}-4A'F'+A''+F''-
\frac{2(A'+F')G'}{(1+G)}],
\end{eqnarray}
and for small fluctuations we get
\begin{eqnarray}
\Gamma_{\mu K,\nu}^{
K}&\cong&-2\partial_{\mu}\partial_{\nu}F+\eta_{\mu\nu}\Box
F\nonumber\\
&-&e^{-2A}\eta_{\mu\nu}(1-2F)(1-2G)\nonumber\\
&\times&[A''+F''-2A'^{2}-2F'^{2}-4A'F'-2(A'+F')G'(1-G)].
\end{eqnarray}
The linearized form of the second term is obtained as
\begin{equation}
\begin{split}
\delta\Gamma_{\mu\nu, K}^{
K}&=-2\partial_{\mu}\partial_{\nu}F+\eta_{\mu\nu}\Box F-
e^{-2A}\eta_{\mu\nu}\\
&\times[F''-4A'F'-2A'G'-2FA''+4FA'^{2}-2GA''+4GA'^{2}].
\end{split}
\end{equation}
The third term is calculated as
\begin{eqnarray}
\Gamma_{\mu\nu}^{ K}\Gamma_{ K M}^{ M}&=&g^{ K L}\Gamma_{
L\mu\nu}g^{ M
S}\Gamma_{S K M}\nonumber\\
&=&\frac{1}{2}g^{ K L}[g_{ L\mu,\nu}+g_{ L\nu,\mu}-g_{\mu\nu,
L}]\frac{1}{2}g^{ M
S}[g_{S K, M}+g_{S M, K}-g_{ K M,S}]\nonumber\\
&=&\frac{1}{4}g^{ K \rho }[g_{\rho \mu,\nu}+g_{\rho
\nu,\mu}-g_{\mu\nu,\rho }]g^{ M
S}[g_{S K, M}+g_{S M, K}-g_{ K M,S}]\nonumber\\
&+&\frac{1}{4}g^{ K 5}[-g_{\mu\nu,5}]g^{ M S}[g_{S K, M}+g_{S M, K}-g_{ K M,S}]\nonumber\\
&=&\frac{1}{4}g^{\xi \rho }[g_{\rho \mu,\nu}+g_{\rho
\nu,\mu}-g_{\mu\nu,\rho }]g^{ M
S}[g_{S\xi , M}+g_{S M,\xi}-g_{\xi  M,S}]\nonumber\\
&+&\frac{1}{4}g^{55}[-g_{\mu\nu,5}]g^{ M S}[g_{S5, M}+g_{S M,5}-g_{5 M,S}]\nonumber\\
&=&\frac{1}{4}g^{\xi \rho }[g_{\rho \mu,\nu}+g_{\rho \nu,\mu}-g_{\mu\nu,\rho }]g^{\alpha S}[g_{S\xi ,\alpha }+g_{S\alpha ,\xi}-g_{\xi \alpha ,S}]\nonumber\\
&+&\frac{1}{4}g^{55}[-g_{\mu\nu,5}]g^{\alpha S}[g_{S5,\alpha }+g_{S\alpha ,5}]\nonumber\\
&+&\frac{1}{4}g^{\xi \rho }[g_{\rho \mu,\nu}+g_{\rho
\nu,\mu}-g_{\mu\nu,\rho }]g^{5
S}[g_{S\xi ,5}+g_{S5,\xi}]\nonumber\\
&+&\frac{1}{4}g^{55}[-g_{\mu\nu,5}]g^{5S}[g_{S5,5}+g_{S5,5}-g_{55,S}]\nonumber\\
&=&\frac{1}{4}g^{\xi \rho }[g_{\rho \mu,\nu}+g_{\rho \nu,\mu}-g_{\mu\nu,\rho }]g^{\alpha \beta }[g_{\beta \xi ,\alpha }+g_{\beta \alpha ,\xi}-g_{\xi \alpha ,\beta }]\nonumber\\
&+&\frac{1}{4}g^{55}[-g_{\mu\nu,5}]g^{\alpha \beta }[g_{\beta \alpha ,5}]\nonumber\\
&+&\frac{1}{4}g^{\xi \rho }[g_{\rho \mu,\nu}+g_{\rho \nu,\mu}-g_{\mu\nu,\rho }]g^{55}[g_{55,\xi}]\nonumber\\
&+&\frac{1}{4}g^{55}[-g_{\mu\nu,5}]g^{55}[g_{55,5}]\nonumber\\
&=&\frac{1}{4}e^{2A+2F}\eta^{\xi \rho }[(e^{-2A-2F}\eta_{\rho
\mu})_{,\nu}+
(e^{-2A-2F}\eta_{\rho \nu})_{,\mu}- (e^{-2A-2F}\eta_{\mu\nu})_{,\rho }]\nonumber\\
&\times& e^{2A+2F}\eta^{\alpha \beta }[(e^{-2A-2F}\eta_{\beta \xi })_{,\alpha }+ (e^{-2A-2F}\eta_{\beta \alpha })_{,\xi}-(e^{-2A-2F}\eta_{\xi \alpha })_{,\beta }]\nonumber\\
&+&\frac{1}{4}(1+G)^{-2}(e^{-2A-2F}\eta_{\mu\nu})_{,5} e^{2A+2F}\eta^{\alpha \beta }(e^{-2A-2F}\eta_{\beta \alpha })_{,5}\nonumber\\
&+&\frac{1}{4}e^{2A+2F}\eta^{\xi \rho }[(e^{-2A-2F}\eta_{\rho \mu})_{,\nu}+(e^{-2A-2F}\eta_{\rho \nu})_{,\mu}- (e^{-2A-2F}\eta_{\mu\nu})_{,\rho }]\nonumber\\
&\times&[(1+G)^{-2}(1+G)^{2}_{,\xi}]\nonumber\\
&+&\frac{1}{4}(1+G)^{-2}[(e^{-2A-2F}\eta_{\mu\nu})_{,5}](1+G)^{-2}(1+G)^{2}_{,5}\nonumber
\end{eqnarray}
\begin{eqnarray}
&=&\frac{1}{4}[-2\eta_{\mu}^{\xi }\partial_{\nu}F-2\eta_{\nu}^{\xi }\partial_{\mu}F+2\eta_{\mu\nu}\partial^{\xi }F]\times[-2\eta_{\xi }^{\alpha }\partial_{\alpha }F-8\partial_{\xi }F+2\eta_{\xi }^{\beta }\partial_{\beta }F]\nonumber\\
&+&\frac{1}{4}[\frac{16e^{-2A-2F}\eta_{\mu\nu}\partial_{5}(A+F)\partial_{5}(A+F)}{(1+G)^{2}}]\nonumber\\
&+&\frac{1}{4}[-2\eta_{\mu}^{\xi }\partial_{\nu}F-2\eta_{\nu}^{\xi }\partial_{\mu}F+2\eta_{\mu\nu}\partial^{\xi }F][2\frac{\partial_{\xi }G}{1+G}]\nonumber\\
&+&\frac{1}{4}[-4\frac{e^{-2A-2F}\eta_{\mu\nu}\partial_{5}(A+F)}{(1+G)^{2}}\frac{\partial_{5}G}{1+G}]\nonumber\\
&=& 8\partial_{\mu}F\partial_{\nu}F-4\eta_{\mu\nu}\partial_{\xi }F\partial^{\xi }F+\frac{4e^{-2A-2F}\eta_{\mu\nu}(A'+F')^{2}}{(1+G)^{2}}\nonumber\\
&-&\frac{\partial_{\mu}G\partial_{\nu}F}{1+G}-\frac{\partial_{\mu}F\partial_{\nu}G}{1+G}+\frac{\eta_{\mu\nu}\partial^{\xi }F\partial_{\xi }G}{1+G}-\frac{e^{-2A-2F}\eta_{\mu\nu}(A'+F')G'}{(1+G)^{3}}\nonumber\\
&=& 8\partial_{\mu}F\partial_{\nu}F-
\frac{\partial_{\mu}G\partial_{\nu}F}{1+G}-\frac{\partial_{\mu}F\partial_{\nu}G}{1+G}-\frac{e^{-2A-2F}\eta_{\mu\nu}}{(1+G)^{2}}\nonumber\\
&\times&[-4(A'+F')^{2}+\frac{(A'+F')G'}{1+G}]+\eta_{\mu\nu}[-4(\partial
F)^{2}+\frac{\partial^{\xi }F\partial_{\xi }G}{1+G}],
\end{eqnarray}
and for small fluctuations we have
\begin{eqnarray}
\Gamma_{\mu\nu}^{ K}\Gamma_{ K M}^{ M}&\cong&
8\partial_{\mu}F\partial_{\nu}F
-\partial_{\mu}G\partial_{\nu}F(1-G)-\partial_{\mu}F\partial_{\nu}G(1-G)\nonumber\\
&-&e^{-2A}\eta_{\mu\nu}(1-2F)(1-2G)\nonumber\\
&\times&[-4A'^{2}-4F'^{2}-8A'F'+(A'+F')G'(1-G)]\nonumber\\
&+&\eta_{\mu\nu}[-4(\partial F)^{2}+\partial^{\xi }F\partial_{\xi
}G(1-G)].
\end{eqnarray}
Then, the linearized form of the third term is obtained as
\begin{eqnarray}
\delta(\Gamma_{\mu\nu}^{ K}\Gamma_{ K M}^{
M})=-e^{-2A}\eta_{\mu\nu}[-8A'F'+A'G'+8FA'^{2}+8GA'^{2}].
\end{eqnarray}
Finally, we will derive the fourth term in the same manner as
\begin{eqnarray}
\Gamma_{\mu M}^{ K}\Gamma_{\nu K}^{ M}&=&g^{ K L}\Gamma_{ L\mu
M}g^{ M
S}\Gamma_{S\nu K}\nonumber\\
&=&\frac{1}{2}g^{ K L}[g_{ L\mu, M}+g_{ L M,\mu}-g_{\mu M,
L}]\frac{1}{2}g^{ M
S}[g_{S\nu, K}+g_{S K,\nu}-g_{\nu K,S}]\nonumber\\
&=&\frac{1}{4}g^{ K \rho }[g_{\rho \mu, M}+g_{\rho  M,\mu}-g_{\mu
M,\rho }]g^{ M
S}[g_{S\nu, K}+g_{S K,\nu}-g_{\nu K,S}]\nonumber\\
&+&\frac{1}{4}g^{ K 5}[g_{5 M,\mu}-g_{\mu M,5}]g^{ M
S}[g_{S\nu, K}+g_{S K,\nu}-g_{\nu K,S}]\nonumber\\
&=&\frac{1}{4}g^{\xi \rho }[g_{\rho \mu, M}+g_{\rho  M,\mu}-g_{\mu
M,\rho }]g^{ M
S}[g_{S\nu,\xi}+g_{S\xi ,\nu}-g_{\nu \xi ,S}]\nonumber\\
&+&\frac{1}{4}g^{55}[g_{5 M,\mu}-g_{\mu M,5}]g^{ M S}[g_{S\nu,5}+g_{S5,\nu}]\nonumber\\
&=&\frac{1}{4}g^{\xi \rho }[g_{\rho \mu,\alpha }+g_{\rho \alpha ,\mu}-g_{\mu \alpha ,\rho }]g^{\alpha S}[g_{S\nu,\xi}+g_{S\xi ,\nu}-g_{\nu \xi ,S}]\nonumber\\
&+&\frac{1}{4}g^{55}[-g_{\mu \alpha ,5}]g^{\alpha S}[g_{S\nu,5}+g_{S5,\nu}]\nonumber\\
&+&\frac{1}{4}g^{\xi \rho }[g_{\rho \mu,5}]g^{5S}[g_{S\nu,\xi}+g_{S\xi ,\nu}-g_{\nu \xi },s]\nonumber\\
&+&\frac{1}{4}g^{55}[g_{55,\mu}]g^{5S}[g_{S\nu,5}+g_{S5,\nu}]\nonumber\\
&=&\frac{1}{4}g^{\xi \rho }[g_{\rho \mu,\alpha }+g_{\rho \alpha ,\mu}-g_{\mu \alpha ,\rho }]g^{\alpha \beta }[g_{\beta \nu,\xi}+g_{\beta \xi ,\nu}-g_{\nu \xi ,\beta }]\nonumber\\
&+&\frac{1}{4}g^{55}[-g_{\mu \alpha ,5}]g^{\alpha \beta }[g_{\beta \nu,5}]\nonumber\\
&+&\frac{1}{4}g^{\xi \rho }[g_{\rho \mu,5}]g^{55}[-g_{\nu \xi ,5}]\nonumber\\
&+&\frac{1}{4}g^{55}[g_{55,\mu}]g^{55}[g_{55,\nu}]\nonumber\\
&=&\frac{1}{4}e^{2A+2F}\eta^{\xi \rho }[(e^{-2A-2F}\eta_{\rho
\mu})_{,\alpha }+
(e^{-2A-2F}\eta_{\rho \alpha })_{,\mu}-(e^{-2A-2F}\eta_{\mu \alpha })_{,\rho }]\nonumber\\
&\times& e^{2A+2F}\eta^{\alpha \beta }[(e^{-2A-2F}\eta_{\beta \nu})_{,\xi}+ (e^{-2A-2F}\eta_{\beta \xi })_{,\nu}-(e^{-2A-2F}\eta_{\nu \xi })_{,\beta }]\nonumber\\
&+&\frac{1}{4}(1+G)^{-2}(e^{-2A-2F}\eta_{\mu \alpha })_{,5} e^{2A+2F}\eta^{\alpha \beta }(e^{-2A-2F}\eta_{\beta \nu})_{,5}\nonumber\\
&+&\frac{1}{4}e^{2A+2F}\eta^{\xi \rho }(e^{-2A-2F}\eta_{\rho \mu})_{,5}(1+G)^{-2}(e^{-2A-2F}\eta_{\nu \xi })_{,5}\nonumber\\
&+&\frac{1}{4}(1+G)^{-2}(1+G)^{2}_{,\mu}(1+G)^{-2}(1+G)^{2}_{,\nu}\nonumber
\end{eqnarray}
\begin{eqnarray}
&=&\frac{1}{4}[-2\eta_{\mu}^{\xi}\partial_{\alpha }F-2\eta_{\alpha }^{\xi}\partial_{\mu}F+ 2\eta_{\mu \alpha }\partial^{\xi}F]\times[-2\eta_{\nu}^{\alpha }\partial_{\xi }F-2\eta_{\xi }^{\alpha }\partial_{\nu}F+2\eta_{\nu \xi }\partial^{\alpha }F]\nonumber\\
&+&\frac{1}{4}[\frac{4e^{-2A-2F}\eta_{\mu \alpha }\eta_{\nu}^{\alpha }\partial_{5}(A+F)\partial_{5}(A+F)}{(1+G)^{2}}]\nonumber\\
&+&\frac{1}{4}[\frac{4e^{-2A-2F}\eta_{\mu}^{\xi }\eta_{\nu \xi }\partial_{5}(A+F)\partial_{5}(A+F)}{(1+G)^{2}}]\nonumber\\
&+&\frac{1}{4}[\frac{4\partial_{\mu}G\partial_{\nu}G}{(1+G)^{2}}]\nonumber\\
&=&6\partial_{\mu}F\partial_{\nu}F-2\eta_{\mu\nu}(\partial
F)^{2}+\frac{2\eta_{\mu\nu}e^{-2A-2F}(A'+F')^{2}}{(1+G)^{2}}\nonumber\\
&+&\frac{\partial_{\mu}G\partial_{\nu}G}{(1+G)^{2}}.
\end{eqnarray}
Since we have small fluctuations, we can write
\begin{eqnarray}
\Gamma_{\mu M}^{ K}\Gamma_{\nu K}^{
M}&\cong&6\partial_{\mu}F\partial_{\nu}F-2\eta_{\mu\nu}(\partial
F)^{2}+\eta_{\mu\nu}e^{-2A}(1-2F)(1-2G)\nonumber\\
&\times&[2A'^{2}+2F'^{2}+4A'F']+\partial_{\mu}G\partial_{\nu}G(1-2G).
\end{eqnarray}
As a result, the linearized form of the fourth term is obtained as
\begin{eqnarray}
\delta(\Gamma_{\mu M}^{ K}\Gamma_{\nu K}^{
M})=\eta_{\mu\nu}e^{-2A}[4A'F'-4A'^{2}F-4A'^{2}G].
\end{eqnarray}
If we add these four terms, we get $\delta R_{\mu\nu}$ as
\begin{eqnarray}
\delta R_{\mu\nu}&=&-2\partial_{\mu}\partial_{\nu}F+
\partial_{\mu}\partial_{\nu}G-\eta_{\mu\nu}\Box
F+ e^{-2A}\eta_{\mu\nu}\nonumber\\
&\times&[F''-4A'F'-2A'G'-2FA''+4FA'^{2}-2GA''+4GA'^{2}\nonumber\\
&-&8A'F'+A'G'+8FA'^{2}+8GA'^{2}+ 4A'F'-4A'^{2}F-4A'^{2}G]\nonumber\\
&=&-2\partial_{\mu}\partial_{\nu}F+
\partial_{\mu}\partial_{\nu}G-\eta_{\mu\nu}\Box
F+ e^{-2A}\eta_{\mu\nu}\nonumber\\
&\times&[F''-8A'F'-A'G'-2FA''+8FA'^{2}-2GA''+8GA'^{2}].
\end{eqnarray}
Now, we will calculate the curvature $R_{\mu 5}$ which can be
written in terms of connection coefficients as
\begin{equation}
R_{\mu 5}=\Gamma_{\mu K,5}^{ K}-\Gamma_{\mu 5, K}^{ K}-\Gamma_{\mu
5}^{ K}\Gamma_{ K M}^{ M}+\Gamma_{\mu M}^{ K}\Gamma_{5
 K}^{ M}.
\end{equation}
Let us analyze the $R_{\mu 5}$ also term by term. The first term
is obtained as
\begin{eqnarray}
\Gamma_{\mu K,5}^{ K}&=&\{g^{ K L}\Gamma_{ L\mu K}\}_{,5}\nonumber\\
&=&\frac{1}{2}\{g^{ K L}[g_{ L\mu, K}+g_{ L K,\mu}-g_{\mu K, L}]\}_{,5}\nonumber\\
&=&\frac{1}{2}\{g^{ K \rho }[g_{\rho \mu, K}+g_{\rho  K,\mu}-g_{\mu K,\rho }]+g^{ K 5}[g_{5 K,\mu}-g_{\mu K,5}]\}_{,5}\nonumber\\
&=&\frac{1}{2}\{g^{\xi \rho }[g_{\rho \mu,\xi}+g_{\rho \xi ,\mu}-g_{\mu \xi ,\rho }]+g^{55}[g_{55,\mu}]\}_{,5}\nonumber\\
&=&\frac{1}{2}\{e^{2A+2F}\eta^{\xi \rho }[(e^{-2A-2F}\eta_{\rho \mu})_{,\xi}+(e^{-2A-2F}\eta_{\rho \xi })_{,\mu}-(e^{-2A-2F}\eta_{\mu \xi })_{,\rho }]\nonumber\\
&+&(1+G)^{-2}(1+G)^{2}_{,\mu}\}_{,5}\nonumber\\
&=&\frac{1}{2}\{e^{2A+2F}\eta^{\xi \rho }[(-2e^{-2A-2F}\partial_{\xi }F\eta_{\rho \mu})+(-2e^{-2A-2F}\partial_{\mu}F\eta_{\rho \xi })\nonumber\\
&-&(-2e^{-2A-2F}\partial_{\rho }F\eta_{\mu \xi })+(1+G)^{-2}2\partial_{\mu}G(1+G)\}_{,5}\nonumber\\
&=&\{-\eta^{\xi }_{\mu}\partial_{\xi }F- 4\partial_{\mu}F+ \eta^{\rho }_{\mu}\partial_{\rho }F+ \frac{\partial_{\mu}G}{1+G}\}_{,5}\nonumber\\
&=&\{-\partial_{\mu}F- 4\partial_{\mu}F+ \partial_{\mu}F+ \frac{\partial_{\mu}G}{1+G}\}_{,5}\nonumber\\
&=&\{-4\partial_{\mu}F+\frac{\partial_{\mu}G}{1+G}\}_{,5}\nonumber\\
&=&-4\partial_{\mu}F'+\frac{\partial_{\mu}G'}{1+G}-\frac{G'\partial_{\mu}G}{(1+G)^{2}}.
\end{eqnarray}
Since $G=2F$ we get
\begin{eqnarray}
\Gamma_{\mu K,5}^{
K}&=&-4\partial_{\mu}F'+\frac{2\partial_{\mu}F'}{1+2F}-
\frac{4F'\partial_{\mu}F}{(1+2F)^2},
\end{eqnarray}
and taking into account the small fluctuation we obtain
\begin{eqnarray}
\Gamma_{\mu K,5}^{
K}&\cong&-4\partial_{\mu}F'+2\partial_{\mu}F'(1-2F)-4F'\partial_{\mu}F(1-4F).
\end{eqnarray}
Then, the linearized form of the first term will be
\begin{eqnarray}
\delta\Gamma_{\mu K,5}^{ K}&=&-4\partial_{\mu}F'+2\partial_{\mu}F'\nonumber\\
&=&-2 \partial_{\mu}F'.
\end{eqnarray}
The second term can be calculated as follows
\begin{eqnarray}
\Gamma_{\mu 5, K}^{ K}&=&\{g^{ K L}\Gamma_{ L\mu 5}\}_{, K}\nonumber\\
&=&\frac{1}{2}\{g^{ K L}[g_{ L\mu,5}+g_{ L 5,\mu}\}_{, K}\nonumber\\
&=&\frac{1}{2}\{g^{ K \rho }[g_{\rho \mu,5}]+g^{ K 5}[g_{55,\mu}]\}_{, K}\nonumber\\
&=&\frac{1}{2}\{g^{\xi \rho }[g_{\rho \mu,5}]\}_{,\xi}+\frac{1}{2}\{g^{55}(g_{55,\mu})\}_{,5}\nonumber\\
&=&\frac{1}{2}\{e^{2A+2F}\eta^{\xi \rho }[(e^{-2A-2F}\eta_{\rho \mu})_{,5}]\}_{,\xi}\nonumber\\
&+& \frac{1}{2}\{(1+G)^{-2}(1+G)^{2}_{,\mu}\}_{,5}\nonumber\\
&=&\frac{1}{2}\{e^{2A+2F}\eta^{\xi
\rho}[(-2e^{-2A-2F}\partial_{5}(A+F)\eta_{\rho \mu})]\}
_{,\xi}\nonumber\\
&+&\frac{1}{2}\{(1+G)^{-2}2(1+G)\partial_{\mu}G\}_{,5}\nonumber\\
&=&\{-\eta^{\xi }_{\mu}\partial_{5}(A+F)\}_{,\xi}+ \{\frac{\partial_{\mu}G}{(1+G)}\}_{,5}\nonumber\\
&=&-\partial_{\mu}F'+\frac{\partial_{\mu}G'}{(1+G)}-\frac{G'\partial_{\mu}G}{(1+G)^{2}}.
\end{eqnarray}
Substituting $G=2F$ we have
\begin{eqnarray}
\Gamma_{\mu 5, K}^{
K}&=&-\partial_{\mu}F'+\frac{2\partial_{\mu}F'}{(1+2F)}
-\frac{4F'\partial_{\mu}F}{(1+2F)^{2}},
\end{eqnarray}
and due to the small fluctuation in $F$ it can be written as
\begin{eqnarray}
\Gamma_{\mu 5, K}^{
K}&\cong&-\partial_{\mu}F'+2\partial_{\mu}F'(1-2F)-4F'\partial_{\mu}F(1-4F).
\end{eqnarray}
We get the linearized form of the second term as
\begin{eqnarray}
\delta\Gamma_{\mu5, K}^{ K}&=&-\partial_{\mu}F'+2\partial_{\mu}F'\nonumber\\
&=&\partial_{\mu}F'.
\end{eqnarray}
The third term is
\begin{eqnarray}
\Gamma_{\mu 5}^{ K}\Gamma_{ K M}^{ M}&=&g^{ K L}\Gamma_{ L\mu
5}g^{ M
S}\Gamma_{S K M}\nonumber\\
&=&\frac{1}{2}g^{ K L}[g_{ L\mu,5}+g_{ L 5,\mu}]\frac{1}{2}g^{ M
S}[g_{S K, M}+g_{S M, K}-g_{ K M,S}]\nonumber\\
&=&\frac{1}{4}g^{ K \rho }[g_{\rho \mu,5}]g^{ M
S}[g_{S K, M}+g_{S M, K}-g_{ K M,S}]\nonumber\\
&+&\frac{1}{4}g^{ K 5}[g_{55,\mu}]g^{ M S}[g_{S K, M}+g_{S M, K}-g_{ K M,S}]\nonumber\\
&=&\frac{1}{4}g^{\xi \rho }[g_{\rho \mu,5}]g^{ M
S}[g_{S\xi , M}+g_{S M,\xi}-g_{\xi  M,S}]\nonumber\\
&+&\frac{1}{4}g^{55}[g_{55,\mu}]g^{ M S}[g_{S5, M}+g_{S M,5}-g_{5 M,S}]\nonumber\\
&=&\frac{1}{4}g^{\xi \rho }[g_{\rho \mu,5}]g^{\alpha S}[g_{S\xi ,\alpha }+g_{S\alpha ,\xi}-g_{\xi \alpha ,S}]\nonumber\\
&+&\frac{1}{4}g^{55}[g_{55,\mu}]g^{\alpha S}[g_{S5,\alpha }+g_{S\alpha ,5}]\nonumber\\
&+&\frac{1}{4}g^{\xi \rho }[g_{\rho \mu,5}]g^{5S}[g_{S\xi ,5}+g_{S5,\xi}]\nonumber\\
&+&\frac{1}{4}g^{55}[g_{55,\mu}]g^{5S}[g_{S5,5}+g_{S5,5}-g_{55,S}]\nonumber\\
&=&\frac{1}{4}g^{\xi \rho }[g_{\rho \mu,5}]g^{\alpha \beta }[g_{\beta \xi ,\alpha }+g_{\beta \alpha ,\xi}-g_{\xi \alpha ,\beta }]\nonumber\\
&+&\frac{1}{4}g^{55}[g_{55,\mu}]g^{\alpha \beta }[g_{\beta \alpha ,5}]\nonumber\\
&+&\frac{1}{4}g^{\xi \rho }[g_{\rho \mu,5}]g^{55}[g_{55,\xi}]\nonumber\\
&+&\frac{1}{4}g^{55}[g_{55,\mu}]g^{55}[g_{55,5}]\nonumber\\
&=&\frac{1}{4}e^{2A+2F}\eta^{\xi \rho }[(e^{-2A-2F}\eta_{\rho \mu})_{,5}]\nonumber\\
&\times&e^{2A+2F}\eta^{\alpha \beta }[(e^{-2A-2F}\eta_{\beta \xi })_{,\alpha }+ (e^{-2A-2F}\eta_{\beta \alpha })_{,\xi}-(e^{-2A-2F}\eta_{\xi \alpha })_{,\beta }]\nonumber\\
&+&\frac{1}{4}(1+G)^{-2}(1+G)^{2}_{,\mu}e^{2A+2F}\eta^{\alpha \beta }(e^{-2A-2F}\eta_{\beta \alpha })_{,5}\nonumber\\
&+&\frac{1}{4}e^{2A+2F}\eta^{\xi \rho }[(e^{-2A-2F}\eta_{\rho \mu})_{,5}(1+G)^{-2}(1+G)^{2}_{,\xi}]\nonumber\\
&+&\frac{1}{4}(1+G)^{-2}(1+G)^{2}_{,\mu}(1+G)^{-2}(1+G)^{2}_{,5}\nonumber
\end{eqnarray}
\begin{eqnarray}
&=&\frac{1}{4}[-2\eta_{\mu}^{\xi }\partial_{5}(A+F)]\times[-2\eta_{\xi }^{\alpha }\partial_{\alpha }F-8\partial_{\xi }F+2\eta_{\xi }^{\beta }\partial_{\beta }F]\nonumber\\
&+&\frac{1}{4}[\frac{-16\partial_{\mu}G\partial_{5}(A+F)}{(1+G)}]\nonumber\\
&+&\frac{1}{4}[-4\eta_{\mu}^{\xi }\partial_{5}(A+F)\frac{\partial_{\xi }G}{1+G}]\nonumber\\
&+&\frac{1}{4}[4\frac{\partial_{\mu}G}{(1+G)}\frac{\partial_{5}G}{1+G}]\nonumber\\
&=&4\partial_{\mu}F(A'+F')-\frac{4\partial_{\mu}G(A'+F')}{1+G}\nonumber\\
&-&\frac{\partial_{\mu}G(A'+F')}{1+G}+\frac{G'\partial_{\mu}G}{(1+G)^{2}}.
\end{eqnarray}
By using $G=2F$ we get
\begin{eqnarray}
\Gamma_{\mu5}^{ K}\Gamma_{ K M}^{ M}&=&4\partial_{\mu}F(A'+F')
-\frac{10\partial_{\mu}F(A'+F')}{1+2F}+\frac{4F'\partial_{\mu}F}{(1+2F)^{2}},
\end{eqnarray}
and for small fluctuations we have
\begin{eqnarray}
\Gamma_{\mu 5}^{ K}\Gamma_{ K M}^{ M}&\cong&4\partial_{\mu}F(A'+F')-10\partial_{\mu}F(A'+F')(1-2F)\nonumber\\
&+&4F'\partial_{\mu}F(1-4F).
\end{eqnarray}
Then, the third term in the linearized form is obtained as
\begin{eqnarray}
\delta(\Gamma_{\mu5}^{ K}\Gamma_{ K M}^{ M})&=&4A'\partial_{\mu}F-10A'\partial_{\mu}F\nonumber\\
&=&-6A'\partial_{\mu}F.
\end{eqnarray}
The last term is calculated as
\begin{eqnarray}
\Gamma_{\mu M}^{ K}\Gamma_{5 K}^{ M}&=&g^{ K L}\Gamma_{ L\mu M}g^{
M
S}\Gamma_{S5 K}\nonumber\\
&=&\frac{1}{2}g^{ K L}[g_{ L\mu, M}+g_{ L M,\mu}-g_{\mu M,
L}]\frac{1}{2}g^{ M
S}[g_{S5, K}+g_{S K,5}-g_{5 K,S}]\nonumber\\
&=&\frac{1}{4}g^{ K \rho }[g_{\rho \mu, M}+g_{\rho  M,\mu}-g_{\mu
M,\rho }]g^{ M
S}[g_{S5, K}+g_{S K,5}-g_{5 K,S}]\nonumber\\
&+&\frac{1}{4}g^{ K 5}[g_{5 M,\mu}-g_{\mu M,5}]g^{ M
S}[g_{S5, K}+g_{S K,5}-g_{5 K\,S}]\nonumber\\
&=&\frac{1}{4}g^{\xi \rho }[g_{\rho \mu, M}+g_{\rho  M,\mu}-g_{\mu
M,\rho }]g^{ M
S}[g_{S5,\xi}+g_{S\xi ,5}\nonumber\\
&+&\frac{1}{4}g^{55}[g_{5 M,\mu}-g_{\mu M,5}]g^{ M S}[g_{S5,5}+g_{S5,5}-g_{55,s}]\nonumber\\
&=&\frac{1}{4}g^{\xi \rho }[g_{\rho \mu,\alpha }+g_{\rho \alpha ,\mu}-g_{\mu \alpha ,\rho }]g^{\alpha S}[g_{S5,\xi}+g_{S\xi ,5}]\nonumber\\
&+&\frac{1}{4}g^{55}[-g_{\mu \alpha ,5}]g^{\alpha S}[2g_{S5,5}-g_{55,S}]\nonumber\\
&+&\frac{1}{4}g^{\xi \rho }[g_{\rho \mu,5}]g^{5S}[g_{S5,\xi}+g_{S\xi ,5}]\nonumber\\
&+&\frac{1}{4}g^{55}[g_{55,\mu}]g^{5S}[2g_{S5,5}-g_{55,S}]\nonumber\\
&=&\frac{1}{4}g^{\xi \rho }[g_{\rho \mu,\alpha }+g_{\rho \alpha ,\mu}-g_{\mu \alpha ,\rho }]g^{\alpha \beta }[g_{\beta \xi ,5}]\nonumber\\
&+&\frac{1}{4}g^{55}[-g_{\mu \alpha ,5}]g^{\alpha \beta }[-g_{55,\beta }]\nonumber\\
&+&\frac{1}{4}g^{\xi \rho }[g_{\rho \mu,5}]g^{55}[g_{55,\xi}]\nonumber\\
&+&\frac{1}{4}g^{55}[g_{55,\mu}]g^{55}[g_{55,5}]\nonumber\\
&=&\frac{1}{4}e^{2A+2F}\eta^{\xi \rho }[(e^{-2A-2F}\eta_{\rho
\mu})_{,\alpha }+
(e^{-2A-2F}\eta_{\rho \alpha })_{,\mu}-(e^{-2A-2F}\eta_{\mu \alpha })_{,\rho }]\nonumber\\
&\times& e^{2A+2F}\eta^{\alpha \beta }[(e^{-2A-2F}\eta_{\beta \xi })_{,5}]\nonumber\\
&+&\frac{1}{4}(1+G)^{-2}(e^{-2A-2F}\eta_{\mu \alpha })_{,5} e^{2A+2F}\eta^{\alpha \beta }(1+G)^{2}_{,\beta }\nonumber\\
&+&\frac{1}{4}e^{2A+2F}\eta^{\xi \rho }(e^{-2A-2F}\eta_{\rho \mu})_{,5}(1+G)^{-2}(1+G)^2_{,\xi}\nonumber\\
&+&\frac{1}{4}(1+G)^{-2}(1+G)^{2}_{,\mu}(1+G)^{-2}(1+G)^{2}_{,5}\nonumber
\end{eqnarray}
\begin{eqnarray}
&=&\frac{1}{4}[-2\eta_{\mu}^{\xi }\partial_{\alpha }F+2\eta_{\alpha }^{\xi }\partial_{\mu}F- 2\eta_{\mu \alpha }\partial^{\xi }F]\times[-2\eta_{\xi }^{\alpha }\partial_{5}(A+F)]\nonumber\\
&+&\frac{1}{4}[\frac{-4e^{-2A-2F}\eta_{\mu \alpha }e^{2A+2F}\eta^{\alpha \beta }\partial_{5}(A+F)\partial_{\beta }G}{(1+G)}]\nonumber\\
&+&\frac{1}{4}[\frac{-4\eta_{\mu}^{\xi }\partial_{5}(A+F)\partial_{\xi }G}{1+G}+\frac{4\partial_{\mu}G\partial_{5}G}{(1+G)^{2}}]\nonumber\\
&=&4\partial_{\mu}F(A'+F')-\frac{2(A'+F')\partial_{\mu}G}{1+G}+\frac{G'\partial_{\mu}G}{(1+G)^{2}}.
\end{eqnarray}
Substituting $G=2F$, for small fluctuations, we get
\begin{eqnarray}
\Gamma_{\mu M}^{ K}\Gamma_{5 K}^{ M}&=&4\partial_{\mu}F(A'+F')-
\frac{4(A'+F')\partial_{\mu}F}{1+2F}+\frac{4F'\partial_{\mu}F}{(1+2F)^{2}}\nonumber\\
&\cong&4\partial_{\mu}F(A'+F')-4(A'+F')\partial_{\mu}F(1-2F)\nonumber\\
&+&4F'\partial_{\mu}F(1-4F).
\end{eqnarray}
We obtain the linearized form of the fourth term as
\begin{eqnarray}
\delta(\Gamma_{\mu M}^{ K}\Gamma_{5 K}^{ M})&=&4A'\partial_{\mu}F-4A'\partial_{\mu}F\nonumber\\
&=&0.
\end{eqnarray}
If we add these four terms, the linearized form of $R_{\mu 5}$
reads
\begin{eqnarray}
\delta R_{\mu
5}&=&-2\partial_{\mu}F'-\partial_{\mu}F'+6A'\partial_{\mu}F+0\nonumber\\
&=&-3\partial_{\mu}F'+6A'\partial_{\mu}F.
\end{eqnarray}
Finally, we will derive the curvature $R_{55}$ and, in terms of
connection coefficients, it reads
\begin{equation}
R_{55}=\Gamma_{5 K,5}^{ K}-\Gamma_{55, K}^{ K}-\Gamma_{55}^{
K}\Gamma_{ K M}^{ M}+\Gamma_{5 M}^{ K}\Gamma_{5 K}^{ M}.
\end{equation}
Now, we start with the first term in $R_{55}$ which reads
\begin{eqnarray}
\Gamma_{5 K,5}^{ K}&=&\{g^{ K L}\Gamma_{ L5 K}\}_{,5}\nonumber\\
&=&\frac{1}{2}\{g^{ K L}[g_{ L5, K}+g_{ L K,5}-g_{5 K, L}]\}_{,5}\nonumber\\
&=&\frac{1}{2}\{g^{ K \rho }[g_{\rho  K,5}-g_{5 K,\rho }]+g^{ K 5}[g_{55, K}+g_{5 K,5}-g_{5 K,5}]\}_{,5}\nonumber\\
&=&\frac{1}{2}\{g^{\xi \rho }[g_{\rho \xi ,5}]+g^{55}[g_{55,5}]\}_{,5}\nonumber\\
&=&\frac{1}{2}\{e^{2A+2F}\eta^{\xi \rho }(e^{-2A-2F}\eta_{\rho \xi })_{,5}+(1+G)^{-2}(1+G)^{2}_{5}\}_{,5}\nonumber\\
&=&\frac{1}{2}\{-8\partial_{5}(A+F)+\frac{2\partial_{5}G}{1+G}\}_{5}\nonumber\\
&=&\{-4(A'+F')+\frac{G'}{1+G}\}_{,5}\nonumber\\
&=&-4(A''+F'')+\frac{G''}{1+G}-\frac{G'^{2}}{(1+G)^{2}}.
\end{eqnarray}
Since $G=2F$ we get
\begin{eqnarray}
\Gamma_{5 K,5}^{
K}&=&-4(A''+F'')+\frac{2F''}{1+2F}-\frac{4F'^{2}}{(1+2F)^{2}}.
\end{eqnarray}
For small fluctuations we have
\begin{eqnarray}
\Gamma_{5 K,5}^{ K}&\cong&-4(A''+F'')+2F''(1-2F)-4F'^{2}(1-4F).
\end{eqnarray}
Then, one can simply write the linearized form of the first term
as
\begin{eqnarray}
\delta\Gamma_{5 K,5}^{ K}&=&-4F''+2F''\nonumber\\
&=&-2F''.
\end{eqnarray}
The second term is
\begin{eqnarray}
\Gamma_{55, K}^{ K}&=&\{g^{ K L}\Gamma_{ L55}\}_{, K}\nonumber\\
&=&\frac{1}{2}\{g^{ K L}[g_{ L5,5}+g_{ L5,5}-g_{55, L}\}_{, K}\}\nonumber\\
&=&\frac{1}{2}\{g^{ K \rho }[-g_{55,\rho }]+g^{ K 5}[2g_{55,5}-g_{55,5}]\}_{, K}\nonumber\\
&=&\frac{1}{2}\{g^{\xi \rho }[-g_{55,\rho }]\}_{,\xi}+\frac{1}{2}\{g^{55}(g_{55,5})\}_{,5}\nonumber\\
&=&\frac{1}{2}\{e^{2A+2F}\eta^{\xi \rho }(1+G)^{2}_{,\rho }\}_{,\xi}+ \frac{1}{2}\{(1+G)^{-2}(1+G)^{2}_{,5}\}_{,5}\nonumber\\
&=&\frac{1}{2}\{e^{2A+2F}\eta^{\xi \rho }2\partial_{\rho }G(1+G)\}_{,\xi}+\frac{1}{2}\{\frac{G'}{1+G}\}_{,5}\nonumber\\
&=&e^{2A+2F}[2\partial_{\xi }F\partial^{\xi }G(1+G)+\partial_{\xi }\partial^{\xi }G(1+G)+\partial_{\xi }G\partial^{\xi }G]\nonumber\\
&+&\frac{G''}{1+G}-\frac{G'^{2}}{(1+G)^{2}}.
\end{eqnarray}
Substituting $G=2F$ we get
\begin{eqnarray}
\Gamma_{55, K}^{ K}&=&e^{2A+2F}[4(\partial F)^{2}(1+2F)+2\Box F(1+2F)+4(\partial F)^{2}]\nonumber\\
&+&\frac{2F''}{1+2F}-\frac{4F'^{2}}{(1+2F)^{2}},
\end{eqnarray}
and for small fluctuations we have
\begin{eqnarray}
\Gamma_{55, K}^{ K}&\cong&e^{2A}(1+2F)[4(\partial
F)^{2}(1+2F)+2\Box
F(1+2F)+4(\partial F)^{2}]\nonumber\\
&+&2F''(1-2F)-4F'^{2}(1-4F).
\end{eqnarray}
Then, the linearized form of the second term is obtained as
\begin{equation}
\delta\Gamma_{55, K}^{ K}=2e^{2A}\Box F+2F''.
\end{equation}
The third term in the curvature $R_{55}$ is
\begin{eqnarray}
\Gamma_{55}^{ K}\Gamma_{ K M}^{ M}&=&g^{ K L}\Gamma_{ L 55}g^{ M
S}\Gamma_{S K M}\nonumber\\
&=&\frac{1}{2}g^{ K L}[g_{ L5,5}+g_{ L 5,5}-g_{55,
L}]\frac{1}{2}g^{ M
S}[g_{S K, M}+g_{S M, K}-g_{ K M,S}]\nonumber\\
&=&\frac{1}{4}g^{ K \rho }[-g_{55,\rho }]g^{ M
S}[g_{S K, M}+g_{S M, K}-g_{ K M,S}]\nonumber\\
&+&\frac{1}{4}g^{ K 5}[2g_{55,5}-g_{55,5}]g^{ M S}[g_{S K, M}+g_{S M, K}-g_{ K M,S}]\nonumber\\
&=&\frac{1}{4}g^{\xi \rho }[-g_{55,\rho }]g^{ M
S}[g_{S\xi , M}+g_{S M,\xi}-g_{\xi  M,S}]\nonumber\\
&+&\frac{1}{4}g^{55}[g_{55,5}]g^{ M S}[g_{S5, M}+g_{S M,5}-g_{5 M,S}]\nonumber\\
&=&\frac{1}{4}g^{\xi \rho }[-g_{55,\rho }]g^{\alpha S}[g_{S\xi ,\alpha }+g_{S\alpha ,\xi}-g_{\xi \alpha ,S}]\nonumber\\
&+&\frac{1}{4}g^{55}[g_{55,5}]g^{\alpha S}[g_{S5,\alpha }+g_{S\alpha ,5}]\nonumber\\
&+&\frac{1}{4}g^{\xi \rho }[-g_{55,\rho }]g^{5S}[g_{S\xi ,5}+g_{S5,\xi}]\nonumber\\
&+&\frac{1}{4}g^{55}[g_{55,5}]g^{5S}[g_{S5,5}+g_{S5,5}-g_{55,S}]\nonumber\\
&=&\frac{1}{4}g^{\xi \rho }[-g_{55,\rho }]g^{\alpha \beta }[g_{\beta \xi ,\alpha }+g_{\beta \alpha ,\xi}-g_{\xi \alpha ,\beta }]\nonumber\\
&+&\frac{1}{4}g^{55}[g_{55,5}]g^{\alpha \beta }[g_{\beta \alpha ,5}]\nonumber\\
&+&\frac{1}{4}g^{\xi \rho }[-g_{55,\rho }]g^{55}[g_{55,\xi}]\nonumber\\
&+&\frac{1}{4}g^{55}[g_{55,5}]g^{55}[g_{55,5}]\nonumber\\
&=&\frac{1}{4}e^{2A+2F}\eta^{\xi \rho }(1+G)^{2}_{,\rho }e^{2A+2F}\eta^{\alpha \beta }\nonumber\\
&\times&[(e^{-2A-2F}\eta_{\beta \xi })_{,\alpha }+ (e^{-2A-2F}\eta_{\beta \alpha })_{,\xi}-(e^{-2A-2F}\eta_{\xi \alpha })_{,\beta }]\nonumber\\
&+&\frac{1}{4}(1+G)^{-2}(1+G)^{2}_{,5}e^{2A+2F}\eta^{\alpha \beta }(e^{-2A-2F}\eta_{\beta \alpha })_{,5}\nonumber\\
&+&\frac{1}{4}e^{2A+2F}\eta^{\xi \rho }[(1+G)^{2}_{,rho}(1+G)^{-2}(1+G)^{2}_{,\xi}]\nonumber\\
&+&\frac{1}{4}(1+G)^{-2}(1+G)^{2}_{,5}(1+G)^{-2}(1+G)^{2}_{,5}\nonumber
\end{eqnarray}
\begin{eqnarray}
&=&\frac{1}{4}[e^{2A+2F}2(1+G)\partial^{\xi }G]\times[-2\eta_{\xi }^{\alpha }\partial_{\alpha }F-8\partial_{\xi }F+2\eta_{\xi }^{\beta }\partial_{\beta }F]\nonumber\\
&+&\frac{1}{4}[\frac{-16G'\partial_{5}(A+F)}{(1+G)}]\nonumber\\
&+&\frac{1}{4}[e^{2A+2F}2(1+G)\partial^{\xi }G\frac{2\partial_{\xi }G}{1+G}]\nonumber\\
&+&\frac{1}{4}[4\frac{G'}{1+G}\frac{\partial_{5}G'}{1+G}]\nonumber\\
&=&-4e^{2A+2F}(1+G)\partial^{\xi }G\partial_{\xi }F-\frac{4(A'+F')G'}{(1+G)}\nonumber\\
&+&e^{2A+2F}\partial^{\xi }G\partial_{\xi
}G+\frac{G'^{2}}{(1+G)^{2}},
\end{eqnarray}
and by using $G=2F$, the third term is obtained to be equal to
\begin{eqnarray}
\Gamma_{55}^{ K}\Gamma_{ K M}^{ M}&=&-8e^{2A+2F}(1+2F)(\partial
F)^{2}-\frac{8(A'+F')F'}{1+2F}\nonumber\\
&+&4e^{2A+2F}(\partial F)^{2}+\frac{4F'^{2}}{(1+2F)^{2}},
\end{eqnarray}
and, considering the small fluctuation, we get
\begin{eqnarray}
\Gamma_{55}^{ K}\Gamma_{ K M}^{
M}&\cong&-8e^{2A}(1+2F)(1+2F)(\partial
F)^{2}-8(A'+F')F'(1-2F)\nonumber\\
&+&4e^{2A}(1+2F)(\partial F)^{2}+4F'^{2}(1-4F).
\end{eqnarray}
Then, the linearized form of the third term is obtained as
\begin{equation}
\delta(\Gamma_{55}^{ K}\Gamma_{ K M}^{ M})=-8A'F'.
\end{equation}
Finally, the fourth term is derived as follows:
\begin{eqnarray}
\Gamma_{5 M}^{ K}\Gamma_{5 K}^{ M}&=&g^{ K L}\Gamma_{ L5 M}g^{ M
S}\Gamma_{S5 K}\nonumber\\
&=&\frac{1}{2}g^{ K L}[g_{ L5, M}+g_{ L M,5}-g_{5 M,
L}]\frac{1}{2}g^{ M
S}[g_{S5, K}+g_{S K,5}-g_{5 K,S}]\nonumber\\
&=&\frac{1}{4}g^{ K \rho }[g_{\rho  M,5}-g_{5 M,\rho }]g^{ M
S}[g_{S5, K}+g_{S K,5}-g_{5 K,S}]\nonumber\\
&+&\frac{1}{4}g^{ K 5}[g_{55, M}]g^{ M
S}[g_{S5, K}+g_{S K,5}-g_{5 K\,S}]\nonumber\\
&=&\frac{1}{4}g^{\xi \rho }[g_{\rho  M,5}-g_{5 M,\rho }]g^{ M
S}[g_{S5,\xi}+g_{S\xi ,5}]\nonumber\\
&+&\frac{1}{4}g^{55}[g_{55, M}]g^{ M S}[g_{S5,5}+g_{S5,5}-g_{55,S}]\nonumber\\
&=&\frac{1}{4}g^{\xi \rho }[g_{\rho \alpha ,5}]g^{\alpha S}[g_{S5,\xi}+g_{S\xi ,5}]\nonumber\\
&+&\frac{1}{4}g^{55}[g_{55,\alpha }]g^{\alpha S}[2g_{S5,5}-g_{55,S}]\nonumber\\
&+&\frac{1}{4}g^{\xi \rho }[g_{55,\rho }]g^{5S}[g_{S5,\xi}+g_{S\xi ,5}]\nonumber\\
&+&\frac{1}{4}g^{55}[g_{55, M}]g^{5S}[2g_{S5,5}-g_{55,S}]\nonumber\\
&=&\frac{1}{4}g^{\xi \rho }[g_{\rho \alpha ,5}]g^{\alpha \beta }[g_{\beta \xi ,5}]\nonumber\\
&+&\frac{1}{4}g^{55}[g_{55,\alpha }]g^{\alpha \beta }[-g_{55,\beta }]\nonumber\\
&+&\frac{1}{4}g^{\xi \rho }[-g_{55,\rho }]g^{55}[g_{55,\xi}]\nonumber\\
&+&\frac{1}{4}g^{55}[g_{55,5}]g^{55}[g_{55,5}]\nonumber\\
&=&\frac{1}{4}e^{2A +2F}\eta^{\xi \rho }[(e^{-2A-2F}\eta_{\rho \alpha })_{,5}e^{2A+2F}\eta^{\beta \alpha }(e^{-2A-2F}\eta_{\beta \xi })_{,5}]\nonumber\\
&+&\frac{1}{4}(1+G)^{-2}(1+G)^{2}_{,\alpha }e^{2A+2F}\eta^{\alpha \beta }(1+G)^{2}_{,\beta }\nonumber\\
&+&\frac{1}{4}[e^{2A+2F}\eta^{\xi \rho }(1+G)^{2}_{,\rho }][(1+G)^{-2}(1+G)^2_{,\xi}\nonumber\\
&+&\frac{1}{4}(1+G)^{-2}(1+G)^{2}_{,5}(1+G)^{-2}(1+G)^{2}_{,5}\nonumber
\end{eqnarray}
\begin{eqnarray}
&=&\frac{1}{4}[-2\eta_{\alpha }^{\xi }\partial_{5}(A+F)][-2\eta_{\xi }^{\alpha }\partial_{5}(A+F)]\nonumber\\
&+&\frac{1}{4}[\frac{4e^{2A+2F}\eta^{\alpha \beta }(1+G)\partial_{\alpha }G\partial_{\beta }G}{1+G}]\nonumber\\
&+&\frac{1}{4}[\frac{4e^{2A+2F}\partial^{\xi }G\partial_{\xi }G(1+G)}{1+G}\nonumber\\
&+&\frac{1}{4}[4\frac{G'^{2}}{(1+G)^{2}}]\nonumber\\
&=&4(A'+F')^{2}+2e^{2A+2F}(\partial
G)^{2}+\frac{G'^{2}}{(1+G)^{2}}.
\end{eqnarray}
Substituting $G=2F$ we obtain
\begin{eqnarray}
\Gamma_{5 M}^{ K}\Gamma_{5 K}^{
M}&=&4(A'+F')^{2}+8e^{2A+2F}(\partial
F)^{2}+\frac{4F'^{2}}{(1+2F)^2},
\end{eqnarray}
and, with small fluctuations, we have
\begin{eqnarray}
\Gamma_{5 M}^{ K}\Gamma_{5 K}^{
M}&\cong&4(A'+F')^{2}+8e^{2A}(1+2F)(\partial F)^{2}+4F'^{2}(1-4F).
\end{eqnarray}
Then, the linearized form of the fourth term is obtained as
\begin{equation}
\delta(\Gamma_{5 M}^{ K}\Gamma_{5 K}^{ M})=8A'F'.
\end{equation}
As a final step we will add these four terms so that we get the
linearized form of $R_{55}$ as
\begin{equation}
\delta R_{55}=-4F''-2e^{2A}\Box F+16A'F'.
\end{equation}
%
At this stage, we will derive the linearized form of the source
term. We start with the part of the action (see eq. \ref{action})
including the source term and taking $\sqrt{g_{55}}$ as
$\sqrt{g_{55}}=1+2F$. The metric variation on this action gives
%
%
%
%
%
%
%
\begin{equation}\label{TMNN}
\begin{split}
T^{MN}&=\frac{1}{2}g^{MN}[\frac{1}{2}(\partial\phi)^2-V(\phi)]-\frac{1}{2}\partial^{M}\phi\partial^{N}\phi\\
&-\frac{1}{2(1+2F)}g^M_\mu g^N_\nu g^{\mu\nu}\sum_{i}
\lambda_{i}(\phi)\delta(y-y_{i}).
\end{split}
\end{equation}
Since $T=g_{MN}T^{MN}$ we get
\begin{equation}\label{TSon}
T=\frac{5}{2}[\frac{1}{2}(\partial\phi)^2-V(\phi)]-\frac{1}{2}(\partial\phi)^2-\frac{2}{1+2F}
\sum_{i} \lambda_{i}(\phi)\delta(y-y_{i}).
\end{equation}
Using the form of Einstein equation given in eq. \ref{RMNson}, and
the eqs. \ref{TMNN} and \ref{TSon} to get
\begin{equation}
\tilde{T}_{\mu\nu}=T_{\mu\nu}-\frac{1}{3}g_{\mu\nu}T,
\end{equation}
we obtain
\begin{equation}\label{tmunu}
\begin{split}
\tilde{T}_{\mu\nu}=&\frac{1}{2}g_{\mu\nu}[\frac{1}{2}(\partial\phi)^2-V(\phi)]
-\frac{1}{2}\partial_{\mu}\phi\partial_{\nu}\phi-\frac{1}{2(1+2F)}
g_{\mu\nu}\sum_{i}\lambda_{i}(\phi)\delta(y-y_{i})\\
&-\frac{5}{6}g_{\mu\nu}[\frac{1}{2}(\partial\phi)^2-V(\phi)]+\frac{1}{6}
g_{\mu\nu}(\partial\phi)^2+
\frac{2}{3(1+2F)}g_{\mu\nu}\sum_{i} \lambda_{i}(\phi)\delta(y-y_{i})\\
&=\frac{1}{3}g_{\mu\nu}V(\phi)-\frac{1}{2}\partial_{\mu}\phi\partial_{\nu}\phi+
\frac{1}{6(1+2F)}g_{\mu\nu}\sum_{i}
\lambda_{i}(\phi)\delta(y-y_{i}),
\end{split}
\end{equation}
where
\begin{equation}\label{phiVlambda}
\begin{split}
&\phi(x,y)=\phi_{0}(y)+\varphi(x,y),\\
&V(\phi)=V(\phi_{0})+\varphi V'(\phi_{0}),\\
& \lambda_{i}(\phi)= \lambda_{i}(\phi_{0})+\varphi
\lambda_{i}'(\phi_{0}),
\end{split}
\end{equation}
for small fluctuations. Substituting the equations in eq.
\ref{phiVlambda} into $\tilde{T}_{\mu\nu}$ we have
\begin{equation}
\begin{split}
\tilde{T}_{\mu\nu}&\cong\frac{1}{3}e^{-2A}(1-2F)\eta_{\mu\nu}
[V(\phi_{0})+\varphi
V'(\phi_{0})]-\frac{1}{2}\partial_{\mu}\varphi\partial_{\nu}
\varphi\\
&+\frac{1}{6}e^{-2A}(1-2F)^{2}\eta_{\mu\nu}\sum_{i}[
\lambda_{i}(\phi_{0})+\varphi
\lambda_{i}'(\phi_{0})]\delta(y-y_{i}).
\end{split}
\end{equation}
The linearized form of the source term $\tilde{T}_{\mu\nu}$ is
obtained as
\begin{equation}\label{deltaTmunuson}
\begin{split}
\delta\tilde{T}_{\mu\nu}=&\frac{1}{3}e^{-2A}\eta_{\mu\nu}
[\varphi V'(\phi_{0})-2V(\phi_{0})F]\\
&+\frac{1}{6}e^{-2A}\eta_{\mu\nu}\sum_{i}[\frac{\partial
\lambda_{i}(\phi_{0})}{\partial\phi}\varphi-4
\lambda_{i}(\phi_{0})F]\delta(y-y_{i}).
\end{split}
\end{equation}
Now, let us derive the equation for the source term
$\tilde{T}_{\mu5}$ by using
\begin{equation}
\tilde{T}_{\mu5}=T_{\mu5}-\frac{1}{3}g_{\mu5}T.
\end{equation}
Since $g_{\mu5}=0$ we get
\begin{equation}
\tilde{T}_{\mu5}=T_{\mu5}.
\end{equation}
Using the eq. \ref{TMNN} we obtain $\tilde{T}_{\mu5}$ as
\begin{equation}
\begin{split}
\tilde{T}_{\mu5}&=-\frac{1}{2}\partial_{\mu}\phi\partial_{5}\phi\\
&=-\frac{1}{2}[\partial_{\mu}(\phi_{0}(y)+\varphi(x,y))\partial_{5}(\phi_{0}(y)+\varphi(x,y))].
\end{split}
\end{equation}
Then, the linearized form of $\tilde{T}_{\mu5}$ reads
\begin{equation}
\delta\tilde{T}_{\mu5}=-\frac{1}{2}\phi_{0}'\partial_{\mu}\varphi.
\end{equation}
Finally, $T_{55}$ can be obtained by using the eq. \ref{TMNN}
\begin{equation}
T_{55}=-\frac{1}{2}(1+2F)^2[\frac{1}{2}(\partial\phi)^2-V(\phi)]
-\frac{1}{2}\partial_{5}\phi\partial_{5}\phi.
\end{equation}
Then, the source term $\tilde{T}_{55}$ becomes
\begin{equation}
\begin{split}
\tilde{T}_{55}&=T_{55}-\frac{1}{3}g_{55}T\\
&=-\frac{1}{2}(1+2F)^2[\frac{1}{2}(\partial\phi)^2-V(\phi)]-\frac{1}{2}
\partial_{5}\phi\partial_{5}\phi\\
&+\frac{1}{2}(1+2F)^2[\frac{1}{2}(\partial\phi)^2-\frac{5}{3}V(\phi)]\\
&-\frac{2}{3}(1+2F)\sum_{i} \lambda_{i}(\phi)\delta(y-y_{i}).
\end{split}
\end{equation}
Making the simplifications and substituting the eqs.
\ref{phiVlambda} we get
\begin{equation}
\begin{split}
\tilde{T}_{55}&\cong-\frac{1}{3}(1+4F+4F^2)[V(\phi_{0})+\varphi V'(\phi_{0})]\\
&-\frac{1}{2}\partial_{5}[\phi_{0}(y)+\varphi(x,y)]\partial_{5}[\phi_{0}(y)+\varphi(x,y)]\\
&-\frac{2}{3}(1+2F)\sum_{i}[ \lambda_{i}(\phi_{0})+\varphi
\lambda_{i}'(\phi_{0})]\delta(y-y_{i}).
\end{split}
\end{equation}
The linearized form is obtained as
\begin{equation}
\begin{split}
\delta\tilde{T}_{55}&=-\frac{4}{3}V(\phi_{0})F-\frac{1}{3}\varphi
V'(\phi_{0})-
\varphi'\phi_{0}'\\
&-\frac{2}{3}\sum_{i}[\frac{\partial
\lambda_{i}(\phi_{0})}{\partial\phi}\varphi+2
\lambda_{i}(\phi_{0})F]\delta(y-y_{i}).
\end{split}
\end{equation}

\newpage
\begin{center}
\normalfont{\huge{\bfseries{\sc{Appendix C}}}} \vspace{1.75cm}
\end{center}
\renewcommand{\theequation}{C-\arabic{equation}}
\setcounter{equation}{0}
In this Appendix we present the
formulation for the spin connection which we use in our
calculations (see \cite{spincon}). A natural basis for the tangent
space $T_{p}$ at a point p is given by the partial derivatives
with respect to the coordinates at that point,
$\hat{e}_{(\mu)}=\partial_{\mu}$. Similarly, a basis for the
cotangent space $T_{p}^{*}$ is given by of the coordinate
functions, $\hat{\theta}_{(\mu)}=dx^{\mu}$. Let us imagine that at
each point in the manifold there exists a set of orthonormal basis
vectors $\hat{e}_{(a)}$\footnote{indexed in Latin letter rather
than Greek, to remind us that they are not related to any
coordinate system.}. If the canonical form of the metric is
written $\eta_{ab}$, the inner product of our basis vectors should
be
\begin{equation}\label{ehataehatb}
\hat{e}_{(a)}\cdot\hat{e}_{(b)}=\eta_{ab}.
\end{equation}
Thus, in Lorentzian space $\eta_{ab}$ represents the Minkowski
metric, while in a space with positive definite metric it
represents Euclidean metric. We can express our old basis vectors
of tangent space in terms of the new ones as
\begin{equation}\label{hate}
\hat{e}_{(\mu)}=e_{\mu}^{a}\hat{e}_{(a)},
\end{equation}
where the components $e_{\mu}^{a}$ are called as vielbeins form an
invertible $n\times n$ matrix. Their inverse is denoted by
switching the indices to obtain $e_{a}^{\mu}$, which satisfy
\begin{equation}
e_{a}^{\mu}e_{\nu}^{a}=\delta_{\nu}^{\mu}\qquad;\qquad
e_{\mu}^{a}e_{b}^{\mu}=\delta_{b}^{a}.
\end{equation}
Multiplying the eq. \ref{hate} with $e_{a}^{\mu}$ from left
\begin{equation}
\hat{e}_{(a)}=e_{a}^{\mu}\hat{e}_{(\mu)},
\end{equation}
Then, in terms of the inverse vielbeins the eq. \ref{ehataehatb}
becomes
\begin{equation}
\begin{split}
\eta_{ab}&=\hat{e}_{(a)}\cdot\hat{e}_{(b)}\\
&=e_{a}^{\mu}\hat{e}_{(\mu)}\cdot e_{b}^{\nu}\hat{e}_{(\nu)}\\
&=g_{\mu\nu}e_{a}^{\mu}e_{b}^{\nu}.
\end{split}
\end{equation}
We see that the components of the metric tensor in the orthonormal
basis are just those of flat metric, $\eta_{ab}$. Multiplying this
with $e_{\mu}^{a}e_{\nu}^{b}$ we get
\begin{equation}
g_{\mu\nu}=e_{\mu}^{a}e_{\nu}^{b}\eta_{ab}.
\end{equation}
Thus, vielbeins are squareroot of the metric. Similarly in
$T_{p}^{*}$, $\theta^{(a)}$ are the orthonormal basis of
one-forms. Choosing
\begin{equation}
\theta^{(a)}\hat{e}_{(b)}=\delta_{b}^{a}=e^{a}_{\mu}e^{\mu}_{b},
\end{equation}
it is an immediate consequence that the orthonormal one-forms are
related to the cooridinate-based $\hat{\theta}_{(\mu)}$ by
\begin{equation}
\hat{\theta}_{(\mu)}=e_{a}^{\mu}\hat{\theta}_{(a)}.
\end{equation}
Similarly,
\begin{equation}
\hat{\theta}_{(a)}=e^{a}_{\mu}\hat{\theta}_{(a)}.
\end{equation}
Any vector $V$ written in the coordinate basis as
$V=V^{\mu}\hat{e}_{\mu}$ can be expressed as in terms of its
orthonormal basis as $V=V^{a}\hat{e}_{a}$. Since
$V^{\mu}\hat{e}_{\mu}=V^{a}\hat{e}_{a}$ we obtain a relation
between the sets of components as
\begin{equation}
V^{a}=e_{\mu}^{a}V^{\mu}.
\end{equation}
So the vielbeins allow us to pass from Latin to Greek indices and
back. In the same manner multi index tensor $V^{a}_{b}$ can be
written as
\begin{equation}
V^{a}_{\;\;\;b}=e_{\mu}^{a}V^{\mu}_{\;\;\;b}=e_{b}^{\nu}V^{a}_{\;\;\;\nu}=e_{\mu}^{a}e_{b}^{\nu}V^{\mu}_{\;\;\;\nu}.
\end{equation}
Nice property of tensors is that, we can go on to refer multi
index tensors in terms of mixed components.

Now, we have a set of basis vectors $\hat{e}_{a}$ and
$\theta{e}_{a}$ which are non-coordinate bases. Thus, they can be
changed independently of the coordinates provided the
orthonormality property defined in eq. \ref{ehataehatb} is
preserved. In Euclidean signature metric the transformations that
preserve orthonormality condition are orthogonal transformations
whereas in Lorentz signature metric they are Lorentz
transformations. We therefore consider changes of basis of the
form
\begin{equation}
\hat{e}_{a}\rightarrow
\hat{e}_{a'}=\Lambda_{a'}^{\;\;\;a}(x)\hat{e}_{a},
\end{equation}
where $\Lambda_{a'}^{a}(x)$ represent position dependent
transformations at each point in space which leave the canonical
form of metric unaltered such that
\begin{equation}
\Lambda_{a'}^{\;\;\;a}(x)\Lambda_{b'}^{\;\;\;b}(x)\eta_{ab}=\eta_{a'b'}.
\end{equation}
In flat space, we call these matrices inverse Lorentz
transformations. We also have ordinary Lorentz transformations,
$\Lambda_{a}^{a'}(x)$ to transform one-forms. So, we now have
freedom to perform a Lorentz transformation at every point in
space. These are called local Lorentz transformations (LLT).

The covariant derivative of a tensor is given by its partial
derivative plus the connection terms, one for each index. The
connection terms in our ordinary formalism involve the tensor and
connection coefficients $\Gamma_{\mu\nu}^{\lambda}$ whereas in
non-coordinate basis connection coefficients are replaced by spin
connection, denoted by $w_{\mu\;\;\;b}^{\;\;a}$. Each Latin index
gets a factor of the spin connection as
\begin{equation}
\nabla_{\mu}X_{\;\;\;b}^{a}=\partial_{\mu}X_{\;\;\;b}^{a}
+w_{\mu\;\;\;c}^{\;\;a}X_{\;\;\;b}^{c}-w_{\mu\;\;\;b}
^{\;\;c}X_{\;\;\;c}^{a}.
\end{equation}
Now let us try to find a relation between the spin connection, the
vielbeins and connection coefficients using the property that a
tensor should be independent of the way it is written. For
simplicity, we will consider the covariant derivative of a vector
$X$. Its covariant derivative in a purely coordinate basis is
given by
\begin{equation}\label{nablexordinary}
\begin{split}
\nabla X&=(\nabla_{\mu}X^{\nu})dx^{\mu}\otimes\partial_{\nu}\\
&=(\partial_{\mu}X^{\nu}+\Gamma_{\mu\lambda}^{\nu}X^
{\lambda})dx^{\mu}\otimes\partial_{\nu},
\end{split}
\end{equation}
and in a mixed basis it is written as
\begin{equation}
\begin{split}
\nabla X&=(\nabla_{\mu}X^{\nu})dx^{\mu}\otimes\hat{e}_{(a)}\\
&=(\partial_{\mu}X^{a}+w_{\mu\;\;\;b}^{\;\;a}X^{a})dx^{\mu}\otimes\hat{e}_{(a)}\\
&=(\partial_{\mu}(e_{\nu}^{a}X^{\nu})+w_{\mu\;\;\;b}^{\;\;a}e_{\lambda}^{b}X^{\lambda})dx^{\mu}\otimes\hat{e}_{(a)}\\
&=(e_{\nu}^{a}\partial_{\mu}X^{\nu}+X^{\nu}\partial_{\mu}e_{\nu}^{a}+w_{\mu\;\;\;b}^{\;\;a}e_{\lambda}^{b}X^{\lambda})dx^{\mu}\otimes
e_{a}^{\sigma}\partial_{\sigma}\\
&=e_{a}^{\sigma}(e_{\nu}^{a}\partial_{\mu}X^{\nu}+X^{\nu}\partial_{\mu}e_{\nu}^{a}+w_{\mu\;\;\;b}^{\;\;a}e_{\lambda}^{b}X^{\lambda})dx^{\mu}\otimes\partial_{\sigma}\\
&=(\delta_{\nu}^{\sigma}\partial_{\mu}X^{\nu}+e_{a}^{\sigma}X^{\nu}\partial_{\mu}e_{\nu}^{a}+e_{a}^{\sigma}w_{\mu\;\;\;b}^{\;\;a}e_{\lambda}^{b}X^{\lambda})dx^{\mu}\otimes\partial_{\sigma}\\
&=(\partial_{\mu}X^{\sigma}+e_{a}^{\sigma}X^{\nu}\partial_{\mu}e_{\nu}^{a}+e_{a}^{\sigma}w_{\mu\;\;\;b}^{\;\;a}e_{\lambda}^{b}X^{\lambda})dx^{\mu}\otimes\partial_{\sigma}.
\end{split}
\end{equation}
Let $\sigma\rightarrow\nu$ and $\nu\rightarrow\lambda$
\begin{equation}\label{nableX}
\nabla X=(\partial_{\mu}X^{\nu}+e_{a}^{\nu}
(\partial_{\mu}e_{\lambda}^{a})X^{\lambda}+
e_{a}^{\nu}e_{\lambda}^{b}w_{\mu\;\;\;b}
^{\;\;a}X^{\lambda})dx^{\mu}\otimes\partial_{\nu}.
\end{equation}
Comparing the eqs. \ref{nablexordinary} and \ref{nableX} one can
conclude that
\begin{equation}\label{gammamulambdanu}
\Gamma_{\mu\lambda}^{\nu}=e_{a}^{\nu}
(\partial_{\mu}e_{\lambda}^{a})X^{\lambda}
+e_{a}^{\nu}e_{\lambda}^{b}w_{\mu\;\;\;b}^{\;\;a}.
\end{equation}
Multiplying this with $e_{b}^{\lambda}e_{\nu}^{a}$ we get
\begin{equation}
w_{\mu\;\;\;b}^{\;\;a}=e_{\nu}^{a}e_{b}^
{\lambda}\Gamma_{\mu\lambda}^{\nu}-e_{b}^
{\lambda}\partial_{\mu}e_{\lambda}^{a}.
\end{equation}
Now let us look the covariant derivative of a vielbein
\begin{equation}
\nabla_{\mu}e_{\nu}^{a}=\partial_{\mu}
e_{\nu}^{a}+w_{\mu\;\;\;b}^{\;\;a}e_{\nu}^{b}
-\Gamma_{\mu\nu}^{\alpha}e_{\alpha}^{a}.
\end{equation}
Substituting the equation for $\Gamma_{\mu\lambda}^{\nu}$ into the
covariant derivative of the vielbein we get
\begin{equation}
\nabla_{\mu}e_{\nu}^{a}=0.
\end{equation}
\label{lastpage}
\end{document}